\documentclass[11pt, prd, aps, showpacs, nofootinbib]{revtex4-1}

\usepackage{amsmath,amssymb}
\usepackage{graphicx}
\usepackage{textcomp}
\usepackage{enumitem}
\usepackage{graphicx}
\usepackage{dsfont}
\usepackage{bm}
\usepackage{xcolor}
\usepackage{subfig}
\usepackage{caption}
\usepackage{appendix}
\usepackage{comment}
\usepackage{bbold}
\usepackage{cancel}

\newcommand{\be}{\begin{equation}}
\newcommand{\ee}{\end{equation}}
\newcommand{\beq}{\begin{equation}}
\newcommand{\eeq}{\end{equation}}
\newcommand{\bea}{\begin{eqnarray}}
\newcommand{\eea}{\end{eqnarray}}

\newcommand{\msbar}{\overline{\footnotesize\textrm{MS}}}

\newcommand{\nn}{\nonumber}

\newcommand{\<}{\langle}   
\renewcommand{\>}{\rangle}

\newcommand{\beqn}{\begin{eqnarray}}   
\newcommand{\eeqn}{\end{eqnarray}}

\newcommand{\pslash}{{\not{\hspace{-0.08cm}p}}}

\newcommand{\tr}{\mathrm{Tr}}

\newcommand{\QCD}{\mathrm{QCD}}
\newcommand{\QED}{\mathrm{QED}}

\newcommand{\ZO}{Z_{O}}

\newcommand{\LO}{\Lambda_{O}}

\newcommand{\GO}{\Gamma_{O}}

\newcommand{\RIMOMprime}{{RI\textquotesingle -MOM}}
\newcommand{\RIprime}{{RI\textquotesingle}}

\newcommand{\betazero}{{\beta}_0}
\newcommand{\betaone}{{\beta}_1}
\newcommand{\alphaem}{\alpha_{\mathrm{em}}}

\newcommand{\goi}{\raisebox{-0.3\totalheight}{\includegraphics[scale=.4]{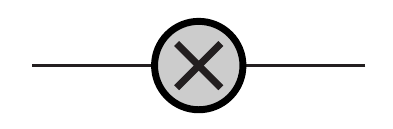}}}
\newcommand{\goip}{\raisebox{-0.3\totalheight}{\includegraphics[scale=.4]{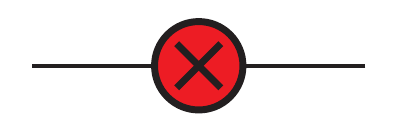}}}
\newcommand{\golself}{\raisebox{-0.0\totalheight}{\includegraphics[scale=.4]{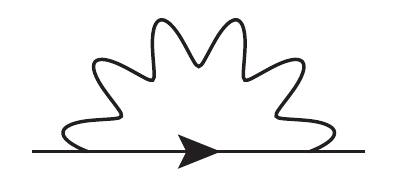}}}
\newcommand{\golltad}{\raisebox{-0.0\totalheight}{\includegraphics[scale=.4]{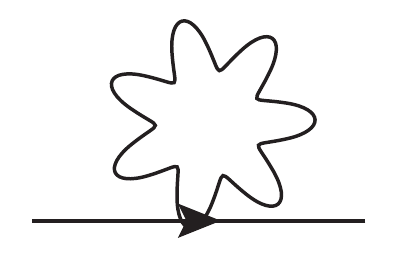}}}
\newcommand{\gexc}{\raisebox{-0.4\totalheight}{\includegraphics[scale=.7]{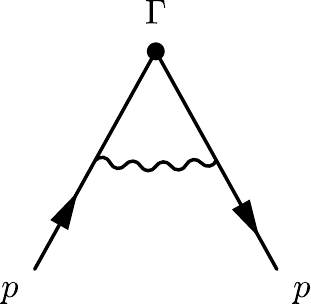}}}
\newcommand{\gin}{\raisebox{-0.4\totalheight}{\includegraphics[scale=.7]{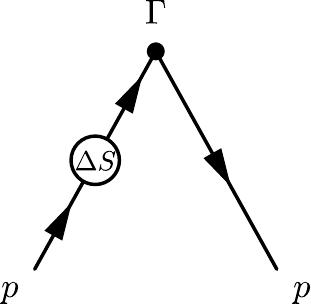}}}
\newcommand{\gout}{\raisebox{-0.4\totalheight}{\includegraphics[scale=.7]{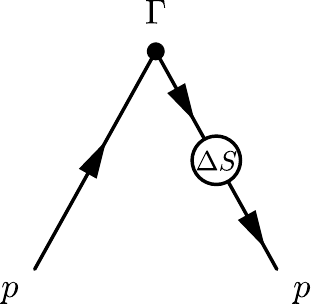}}}
\newcommand{\Gselfin}{\raisebox{-0.45\totalheight}{\includegraphics[scale=.7]{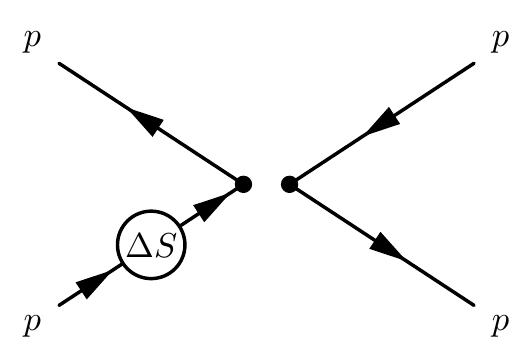}}}
\newcommand{\Gselfout}{\raisebox{-0.45\totalheight}{\includegraphics[scale=.7]{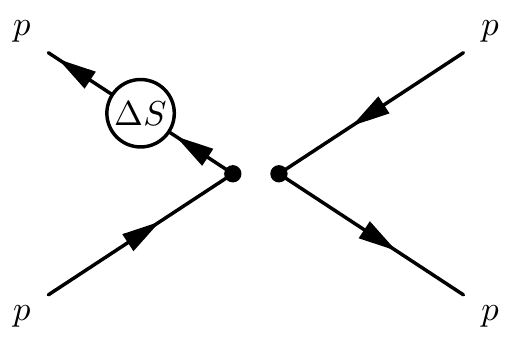}}}
\newcommand{\Gexch}{\raisebox{-0.45\totalheight}{\includegraphics[scale=.7]{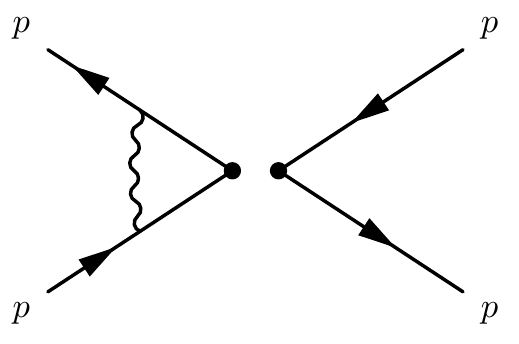}}}
\newcommand{\Gexchin}{\raisebox{-0.45\totalheight}{\includegraphics[scale=.7]{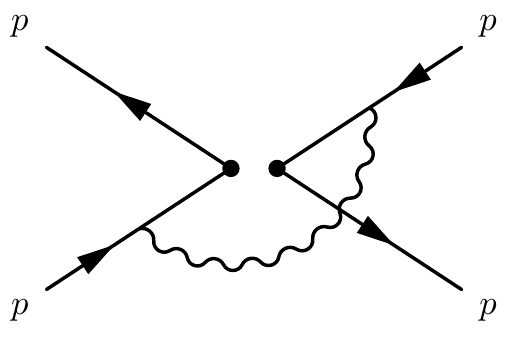}}}
\newcommand{\Gexchout}{\raisebox{-0.45\totalheight}{\includegraphics[scale=.7]{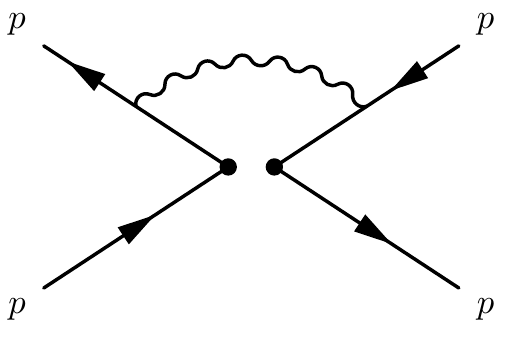}}}

\newcommand{\Romatre}{Dip.~di Matematica e Fisica, Universit\`a  Roma Tre and INFN, Sezione di Roma Tre,\\ Via della Vasca Navale 84, I-00146 Rome, Italy}
\newcommand{\RomatreINFN}{Istituto Nazionale di Fisica Nucleare, Sezione di Roma Tre,\\ Via della Vasca Navale 84, I-00146 Rome, Italy}
\newcommand{\soton}{Department of Physics and Astronomy, University of Southampton,\\ Southampton SO17 1BJ, UK}
\newcommand{\Romadue}{Dipartimento di Fisica and INFN, Universit\`a di Roma ``Tor Vergata",\\ Via della Ricerca Scientifica 1, I-00133 Roma, Italy}
\newcommand{\LaSapienza}{Dipartimento di Fisica and INFN Sezione di Roma La Sapienza,\\ Piazzale Aldo Moro 5, 00185 Roma, Italy}

\begin{document}

\title{\Large Light-meson leptonic decay rates in lattice QCD+QED}

\author{M.~Di Carlo}\affiliation{\LaSapienza}
\author{D.~Giusti}\affiliation{\Romatre}
\author{V.~Lubicz}\affiliation{\Romatre}
\author{G.~Martinelli}\affiliation{\LaSapienza}
\author{C.T.~Sachrajda}\affiliation{\soton}
\author{F.~Sanfilippo}\affiliation{\RomatreINFN}
\author{S.~Simula}\affiliation{\RomatreINFN}
\author{N.~Tantalo}\affiliation{\Romadue}


\begin{abstract} 
The leading electromagnetic (e.m.) and strong isospin-breaking corrections to the $\pi^+ \to \mu^+ \nu[\gamma]$ and $K^+ \to \mu^+ \nu[\gamma]$ leptonic decay rates are evaluated for the first time on the lattice. The results are obtained using gauge ensembles produced by the European Twisted Mass Collaboration with $N_f = 2 + 1 + 1$ dynamical quarks. 
The relative leading-order e.m.~and strong isospin-breaking corrections to the decay rates are 1.53(19)\% for $\pi_{\mu 2}$ decays and 0.24(10)\% for $K_{\mu 2}$ decays. Using the experimental values of the $\pi_{\mu 2}$ and $K_{\mu 2}$ decay rates and updated lattice QCD results for the pion and kaon decay constants in isosymmetric QCD, we find that the Cabibbo-Kobayashi-Maskawa matrix element $ \vert V_{us}\vert = 0.22538(46)$, reducing by a factor of about $1.8$ the corresponding uncertainty in the Particle Data Group review. Our calculation of $|V_{us}|$ allows also an accurate determination of the first-row CKM unitarity relation $\vert V_{ud}\vert^2 + \vert V_{us}\vert^2 + \vert V_{ub}\vert^2 = 0.99988(46)$. Theoretical developments in this paper include a detailed discussion of how QCD can be defined in the full QCD+QED theory and an improved renormalisation procedure in which the bare lattice operators are renormalised non-perturbatively into the (modified) Regularization Independent Momentum subtraction scheme and subsequently matched perturbatively at $O(\alphaem\alpha_s(M_W))$ into the W-regularisation scheme appropriate for these calculations. 
\end{abstract}


\maketitle

\newpage

\section{Introduction}
\label{sec:intro}

In flavour physics the determination of the elements of the Cabibbo-Kobayashi-Maskawa (CKM) matrix\,\cite{CKM}, which contains just 4 parameters, from a wide range of weak processes represents a crucial test of the limits of the Standard Model (SM) of particle physics.  Inconsistencies with theoretical  expectations would indeed  signal the existence of new physics beyond the SM and subsequently a detailed comparison of experimental measurements and theoretical predictions would provide a guide towards uncovering the underlying theory beyond the SM. For this to be possible non-perturbative hadronic effects need to be evaluated as precisely as possible and in this paper we report on progress in improving the precision of lattice computations of leptonic decay rates by including radiative corrections and strong isospin-breaking (IB) effects. A summary of our results has been presented in Ref.\,\cite{Giusti:2017dwk}; here we expand on the details of the calculation and include several improvements, most notably the renormalisation of the four-fermion weak operators in the combined QCD+QED theory (see Sec.\,\ref{sec:Wreg}). We also discuss in some detail how one might define the QCD component of the full (QCD+QED) theory (see Sec.\,\ref{sec:QEDQCD}). Although such a separate definition of QCD is not required in order to obtain results computed in the full theory, it is necessary if one wishes to talk about radiative (and strong IB) ``corrections"  to results obtained in QCD. For this we need to specify what we mean by QCD.

The extraction of the CKM elements from experimental data requires an accurate knowledge of a number of hadronic quantities and the main goal of large-scale QCD simulations on the lattice is the {\it ab initio} evaluation of the nonperturbative QCD effects in physical processes.  For several quantities relevant for flavour physics phenomenology, lattice QCD has recently reached the impressive level of precision of  ${\cal{O}}(1 \%)$ or even better .
Important examples are the ratio $f_K / f_\pi$ of kaon and pion leptonic decay constants and the $K_{\ell 3}$ vector form factor $f_+(0)$\,\cite{FLAG}, which play the central role in the accurate determination of the CKM entries $|V_{us} / V_{ud}|$ and $|V_{us}|$, respectively.  Such lattice computations are typically performed in the isospin symmetric limit of QCD, in which the up and down quarks are mass degenerate ($m_u = m_d$) and electromagnetic (e.m.) effects are neglected ($\alphaem = 0$).

Isospin breaking effects arise because of radiative corrections and because $m_u\neq m_d$; the latter contributions are usually referred to as strong isospin breaking effects. Since both $\alphaem$ and $(m_d - m_u) / \Lambda_{\mathrm{QCD}}$ are of ${\cal{O}}(1 \%)$, IB effects need to be included in lattice simulations to make further progress in flavour physics phenomenology, 
beyond the currently impressive precision obtained in isosymmetric QCD.

Since the electric charges of the up and down quarks are different, the presence of electromagnetism itself induces a difference in their masses, in addition to any explicit difference in the bare masses input into the action being simulated. The separation of IB effects into strong and e.m.~components therefore requires a convention. We discuss this in detail in Sec.\,\ref{sec:QEDQCD}, where we propose and advocate the use of hadronic schemes, based on taking a set of hadronic quantities, such as particle masses, which are computed with excellent precision in lattice simulations, to define QCD in the presence of electromagnetism. 

In recent years precise lattice results including e.m.~and strong IB effects have been obtained for the hadron spectrum; in particular for the mass splittings between charged and neutral pseudoscalar (P) mesons and baryons (see for example, Refs.\,\cite{deDivitiis:2013xla,Borsanyi:2014jba}).
The QED effects were included in lattice QCD simulations using the following two methods: 
\begin{itemize}
\item QED is added directly to the action and QED+QCD simulations are performed at a few values of the electric charge and the results extrapolated to the physical value of $\alphaem$ (see, e.g., Refs.\,\cite{Borsanyi:2014jba,Boyle:2017gzv,Hansen:2018zre});
\item the lattice path-integral is expanded in powers of the two {\it small} parameters $\alphaem$ and $(m_d - m_u) / \Lambda_{\mathrm{QCD}}$.  This is the RM123 approach of Refs.\,\cite{deDivitiis:2011eh,deDivitiis:2013xla,Giusti:2017dmp} which we follow in this paper. 
\end{itemize}

In practice, for all the relevant  phenomenological applications it is currently sufficient to work at first order in the small parameters $\alphaem$ and $(m_d - m_u) / \Lambda_{\textrm{QCD}}$. The attractive feature of the RM123 method is that it allows one naturally to work at first order in isospin breaking, computing the coefficients of the two small parameters directly.
Moreover, these coefficients can be determined from simulations of isosymmetric QCD.

The calculation of e.m.~and strong IB effects in the hadron spectrum has a very significant simplification in that there are no infrared (IR) divergences.
The same is not true when computing hadronic amplitudes, where e.m.~IR divergences are present and only cancel in well-defined, measurable physical quantities by summing diagrams containing real and virtual photons\,\cite{BN37}.
This is the case, for instance, for the leptonic $\pi_{\ell 2}$ and $K_{\ell 2}$ and semileptonic $K_{\ell 3}$ decay rates.  
The presence of IR divergences requires a new strategy beyond those developed for the calculation of IB effects in the hadron spectrum. 
Such a new strategy was proposed in Ref.\,\cite{Carrasco:2015xwa}, where the determination of the inclusive decay rate of a charged P meson into either a final $\ell^\pm \nu_\ell$ pair or a final $\ell^\pm \nu_\ell \gamma$ state was addressed.

The e.m.~corrections due to the exchange of a virtual photon and to the emission of a real one can be computed non-perturbatively, by numerical simulations, on a finite lattice with the corresponding uncertainties.
The exchange of a virtual photon depends on the structure of the decaying meson, since all momentum modes are included, and the corresponding amplitude must therefore be computed non-perturbatively.
On the other hand, the non-perturbative evaluation of the emission of a real photon is not strictly necessary\,\cite{Carrasco:2015xwa}.
Indeed, it is possible to compute the real emission amplitudes in perturbation theory by limiting the maximum energy of the emitted photon in the meson rest-frame, $\Delta E_\gamma$, to a value small enough so that the internal structure of the decaying meson is not resolved.
The IR divergences in the non-perturbative calculation of the corrections due to the exchange of a virtual photon are cancelled by the corrections due to the real photon emission even when the latter is computed perturbatively, because of the universality of the IR behaviour of the theory (i.e., the IR divergences do not depend on the structure of the decaying hadron).
Such a strategy, which requires an experimental cut on the energy of the real photon, makes the extraction of the relevant CKM element(s) cleaner.  

In the intermediate steps of the calculation it is necessary to introduce an IR regulator. 
In order to work with quantities that are finite when the IR regulator is removed, the inclusive rate $\Gamma(P^+ \to \ell^+ \nu_\ell [\gamma])$ is written as\,\cite{Carrasco:2015xwa}
\bea
     \Gamma(P^\pm \to \ell^\pm \nu_\ell [\gamma]) & = & \Gamma_0 + \Gamma_1^{\textrm{pt}}(\Delta E_\gamma) \nonumber \\
                                                                               & = & \displaystyle \lim_{L \to \infty} \left[ \Gamma_0(L) - \Gamma_0^{\textrm{pt}}(L) \right] + 
                                                                                         \displaystyle \lim_{\mu_\gamma \to 0} \left[ \Gamma_0^{\textrm{pt}}(\mu_\gamma) + 
                                                                                         \Gamma_1^{\textrm{pt}}(\Delta E_\gamma, \mu_\gamma) \right] ~ , 
     \label{eq:Gamma}
 \eea
where the subscripts $0, 1$ indicate the number of photons in the final state, while the superscript  {\footnotesize pt} denotes the point-like approximation of the decaying meson and $\mu_\gamma$ is an IR regulator.
In the first term on the r.h.s.~of Eq.\,(\ref{eq:Gamma}) the quantities $\Gamma_0(L)$ and $\Gamma_0^{\textrm{pt}}(L)$ are evaluated on the lattice.
Both have the same IR divergences which therefore cancel in the difference. 
We use the lattice size $L$ as the intermediate IR regulator by working in the QED$_\mathrm{L}$\,\cite{Hayakawa:2008an} formulation of QED on a finite volume (for a recent review on QED simulations in a finite box see Ref.\,\cite{Patella:2017fgk}).
The difference $\left[ \Gamma_0 - \Gamma_0^{\textrm{pt}} \right]$ is independent of the regulator as this is removed\,\cite{Lubicz:2016xro}. 
As already pointed out, since all momentum modes contribute to it, $\Gamma_0(L)$ depends on the structure of the decaying meson and must be computed non-perturbatively. 
The numerical determination of $\Gamma_0(L)$ for several lattice spacings, physical volumes and quark masses is indeed the focus of the present study. 

In the second term on the r.h.s.~of Eq.\,(\ref{eq:Gamma}) P is a point-like meson and both $\Gamma_0^{\textrm{pt}}(\mu_\gamma)$ and $\Gamma_1^{\textrm{pt}}(\Delta E_\gamma, \mu_\gamma)$ can be calculated directly in infinite volume in perturbation theory, using a photon mass $\mu_\gamma$ as the IR regulator. Each term is IR divergent, but the sum is convergent\,\cite{BN37} and independent of the IR regulator. In Refs.\,\cite{Carrasco:2015xwa} and \cite{Lubicz:2016xro} the explicit perturbative calculations of $\left[ \Gamma^{\textrm{pt}}_0+\Gamma^{\textrm{pt}}_1(\Delta E_\gamma) \right]$ and $\Gamma_0^{\textrm{pt}}(L)$ have been performed  with a small photon mass $\mu_\gamma$ or by using the finite volume respectively, as the IR cutoffs. 

In Ref.\,\cite{Giusti:2017dwk} we have calculated the e.m.~and IB corrections to the ratio of $K_{\mu2}$ and $\pi_{\mu2}$ decay rates of charged pions and kaons into muons \cite{Giusti:2017dwk}, using gauge ensembles generated by the European Twisted Mass Collaboration (ETMC) with $N_f = 2 + 1 + 1$ dynamical quarks\,\cite{Baron:2010bv,Baron:2011sf} in the quenched QED (qQED) approximation in which the charges of the sea quarks are set to 0. The ratio is 
less sensitive to various sources of uncertainty than the IB corrections to $\pi_{\mu 2}$ and $K_{\mu 2}$ decay rates separately.
In this paper, in addition to providing more details of the calculation than was possible in Ref.\,\cite{Giusti:2017dwk}, we do evaluate the e.m.~and strong IB corrections to the decay processes $\pi_{\mu 2}$ and $K_{\mu 2}$ separately.
Since the corresponding experimental rates are fully inclusive in the real photon energy, structure-dependent (SD) contributions to the real photon emission should be included, however
according to the Chiral Perturbation Theory (ChPT) predictions of Ref.\,\cite{Cirigliano:2007ga} these SD contributions are negligible for both kaon and pion decays into muons. The same is not true to the same extent for decays into  final-state electrons\,(see Ref.\,\cite{Carrasco:2015xwa}) and so in this paper we focus on decays into muons.
The SD contributions to $\Gamma_1$ are being investigated in an ongoing dedicated lattice study of light and heavy P-meson leptonic decays.

An important improvement presented in this paper is in the renormalisation of the effective weak Hamiltonian. To exploit the matching of the effective theory to the Standard Model performed in Ref.\,\cite{Sirlin:1981ie} it is particularly convenient to renormalise the weak Hamiltonian in the W-regularisation scheme.  The renormalisation is performed in two steps.
First of all, the lattice operators are renormalised non-perturbatively in the (modified) Regularization Independent Momentum subtraction (RI$^\prime$-MOM) scheme at $O(\alphaem)$ and to all orders in the strong coupling $\alpha_s$. Because of the breaking of chiral symmetry in the twisted mass formulation we have adopted in our study, this renormalisation includes the mixing with other four-fermion operators of different chirality. In the second step we perform the matching from the RI$^\prime$-MOM scheme to the W-regularisation scheme perturbatively. By calculating and including the two-loop anomalous dimension at $O(\alphaem\alpha_s)$, the residual truncation error of this matching is of $O(\alphaem\alpha_s(M_W))$, reduced from $O(\alphaem\alpha_s(1/a))$ in our earlier work\,\cite{Carrasco:2015xwa}.

The main results of the calculation are presented in Sec.\,\ref{sec:results} together with a detailed discussion of their implications. Here, we anticipate some key results:
After extrapolation of the data to the physical pion mass, and to the continuum and infinite-volume limits, the isospin-breaking corrections to the leptonic decay rates can be written in the form:
\bea
     \Gamma(\pi^\pm \to \mu^\pm \nu_\ell [\gamma]) 
     &=&(1.0153 \pm  0.0019)\,\Gamma^{(0)}(\pi^\pm \to \mu^\pm \nu_\ell), \label{eq:pion_result_final)2}\\[2mm]
      \Gamma(K^\pm \to \mu^\pm \nu_\ell [\gamma])  & =&(1.0024 \pm 0.0010) \Gamma^{(0)}(K^\pm \to \mu^\pm \nu_\ell)\,,
      \label{eq:kaon_result_final_2}
 \eea
where $\Gamma^{(0)}$ is the leptonic decay rate at tree level in the Gasser-Rusetsky-Scimemi (GRS) scheme which is a particular definition of QCD\,\cite{Gasser:2003hk} (see Sec.\,\ref{subsec:GRS} below). The corrections are  about 1.5\% for the pion decays and 0.2\% for the kaon decay, in line with na\"ive expectations. 
Taking the experimental value of the rate for the $K_{\mu2}$ decay, Eq.\,(\ref{eq:kaon_result_final_2}) together with $\Gamma^{(0)}(K^\pm \to \mu^\pm \nu_\ell)$ obtained using the lattice determination of the kaon decay constant  we obtain $|V_{us}| = 0.22567(42)$, in agreement with the latest estimate $|V_{us}| = 0.2253(7)$, recently updated by the Particle Data Group (PDG)\,\cite{PDG} but with better precision. 
Alternatively, by taking the ratio of $K_{\mu2}$ and $\pi_{\mu2}$ decay rates and the 
updated value $|V_{ud}| = 0.97420\,(21)$ from super-allowed nuclear beta decays\,\cite{Hardy:2016vhg}, we obtain $|V_{us}| = 0.22538(46)$. The unitarity of the first row of the CKM matrix is satisfied at the per-mille level; e.g.~taking the value of $V_{us} $ from the ratio of decay rates and $|V_{ub}| = 0.00413(49)$\,\cite{PDG},  we obtain $|V_{ud}|^2 + |V_{us}|^2 + |V_{ub}|^2 = 0.99988(46)$. See Sec.\,\ref{sec:results} for a more detailed discussion of our results and their implications.

The plan for the remainder of this paper is as follows.
A discussion of the relation between the ``{\it full}" QCD+QED theory, including e.m.~and strong IB~effects, and isosymmetric QCD without electromagnetism is given in Sec.\,\ref{sec:QEDQCD}. 
We discuss possible definitions of QCD in the full QCD+QED theory, and in particular we define and advocate hadronic schemes as well as the GRS scheme which is conventionally used\,\cite{Gasser:2003hk}.
In Sec.\,\ref{sec:master} we present the calculation of the relevant amplitudes using the RM123 approach. The renormalisation of the bare lattice operators necessary to obtain the effective weak Hamiltonian in 
the $W$-regularization scheme is performed in Sec.\,\ref{sec:Wreg}, while the subtraction of the universal IR-divergent finite volume effects (FVEs) is described in Sec.\,\ref{sec:FVE}. 
The lattice data for the e.m.~and strong IB corrections to the leptonic decay rates of pions and kaons are extrapolated to the physical pion mass, to the continuum and infinite volume limits in Sec.\,\ref{sec:results}. 
 Finally, Sec.\,\ref{sec:conclusions} contains our conclusions. There are four appendices.
The lattice framework and details of the simulation are presented in Appendix~\ref{sec:appA}. Appendix~\ref{sec:appB} contains a detailed discussion of the relation between observables in the full theory and in QCD, expanding on the material in Sec.\,\ref{sec:QEDQCD}.  An expanded discussion of the renormalisation of the effective weak Hamiltonian, including electromagnetic corrections, is presented in Appendices\,\ref{sec:appC}, which contains a general discussion of the non-perturbative renormalisation in the \RIMOMprime~scheme and \ref{sec:appD} in which issues specific to the twisted mass formulation are discussed.

\section{Defining QCD in the full theory (QCD+QED)}
\label{sec:QEDQCD}

Before presenting the detailed description of our calculation of leptonic decay rates, we believe that it is useful to discuss the relation between the ``{\it full}\," QCD+QED theory, that includes explicit e.m.~and strong isospin breaking effects, and QCD without electromagnetism (denoted in the following as the full theory and QCD, respectively). 

The action of the full theory can be schematically written as
\be
    \label{eq:fullaction}
    S^\textrm{full} = \frac{1}{g_s^2}S^{\mathrm{YM}}+S^A+ \sum_f\left\{S_f^\textrm{kin} + m_f \, S_f^m \right\}+\sum_\ell 
                              \left\{S^\textrm{kin}_\ell+m_\ell \, S_\ell^m\right\} ~ .
\ee
Here $g_s$ is the strong coupling constant, $S^\textrm{YM}$ is a discretisation of the gluon action, $S^A$ is the preferred discretisation of the Maxwell action of the photon, $S_f^\textrm{kin}$ is the kinetic term for the quark with flavour $f$, including the interaction with the gluon and photon fields, $m_f S_f^m = m_f \sum_x \overline{q}_f(x) q_f(x)$ is the mass term, $S^\textrm{kin}_\ell$ and $S^m_\ell$ are respectively the kinetic and mass terms for the lepton $\ell$ (for details see Appendix~\ref{sec:appB}). For fermion actions which break chiral symmetry, such as the Wilson action, a counterterm is needed to remove the critical mass and $m_f S_f^m$ has to be replaced with $m_f S_f^m + m_f^\textrm{cr} S_f^\textrm{cr}$. A mass counterterm is in principle needed also in the case of the lepton, but at leading order in $\alpha_{em}$ the lepton critical mass can be ignored.

At the level of precision to which we are currently working it is only the full theory, as defined in Eq.\,(\ref{eq:fullaction}), which is expected to reproduce physical results and that is therefore unambiguous. Nevertheless, a frequently asked question is what is the difference between the results for a physical quantity computed in the full theory and in pure QCD,  and how big are the strong isospin-breaking effects compared to the e.m.~corrections. We particularly wish to underline that in order to properly formulate such questions it is necessary to carefully define what is meant by QCD. It is naturally to be expected that in QCD alone physical quantities will not be reproduced with a precision of better than ${\cal{O}}(\alphaem) \simeq 1\%$ and this of course is the motivation for including QED. In order to define what is meant by QCD at this level of precision it is necessary to state the conditions which are used to determine the quark masses and the lattice spacing. The separation of the full theory into QCD and the rest is therefore prescription dependent.

In Ref.\,\cite{deDivitiis:2013xla} the subtle issue of a precise definition of QCD has been discussed by using the scheme originally proposed in Ref.\,\cite{Gasser:2003hk}, which we refer to as the GRS scheme and which has been widely used\,\cite{deDivitiis:2013xla,Giusti:2017dmp,Giusti:2017dwk}.
In the following and in Appendix~\ref{sec:appB} we present an extended and detailed discussion by introducing the hadronic schemes. Indeed, in light of the fact that hadron masses can nowadays be computed very precisely, we strongly suggest using hadronic schemes in future lattice calculations of QED radiative corrections. At the end of this section we discuss the connection with the GRS scheme that we had adopted at the time in which this calculation was started and that, for this reason, has been used in this work. A summary of the ideas discussed here has already been presented in Ref.\,\cite{Giusti:2018guw}. 

\subsection{Renormalisation of the full theory}
\label{subsec:renormalisationfull}

The main difference in the steps required to renormalise the full theory compared to the procedure in QCD is the presence of a massless photon and the corresponding finite-volume (FV) corrections which appear as inverse powers of $L$, where $L$ is the spatial extent of the lattice and the volume $V = L^3$.
By contrast, in QCD for leptonic and semileptonic decays the FV corrections are exponentially small in the volume. In the discussion below, if necessary, we imagine that the chiral Ward identities have been imposed to determine the critical masses $m_f^\textrm{cr}$\,\cite{Bochicchio:1985xa}. 

A possible strategy in principle is the following:
\begin{enumerate}
\item Fix the number of lattice points $N$, e.g.~$T = 2 aN$ and $L = aN$, where $T$ and $L$ are the temporal and spatial extents of the lattice and the lattice spacing $a$ will be determined later. (The specific choice $T = 2L$ is convenient for illustration but not necessary for the following argument.)
\item Using a four-flavour theory for illustration, we now need to determine the four physical bare quark masses, the bare electric charge and the lattice spacing. To this end we need to compute six quantities, e.g.~the five dimensionless ratios\footnote{An alternative procedure to determine the bare electric charge would be the evaluation of the hadronic corrections to a leptonic observable.} 
\bea
    &&
    R_1(aN; g_s, e, \textbf{m}) = \frac{aM_{\pi^+}}{aM_\Omega}(aN; g_s, e, \textbf{m}) ~ , 
    \nonumber \\
    &&
    R_2(aN; g_s, e, \textbf{m}) = \frac{aM_{K^0}}{aM_\Omega}(aN; g_s, e, \textbf{m}) 
    \nonumber \\
    &&
    R_3(aN; g_s, e, \textbf{m}) = \frac{aM_{D_s}}{aM_\Omega}(aN;g_s,e,\textbf{m}) ~ ,
    \nonumber \\
    &&
    R_4(aN; g_s, e, \textbf{m}) = \frac{aM_{K^+}-aM_{K^0}}{aM_\Omega}(aN; g_s, e, \textbf{m}) ~ ,     	
    \nonumber \\
    &&
    R_5(aN; g_s, e, \textbf{m}) = \frac{aM_{D^0}-aM_{D^+}}{aM_\Omega}(aN;g_s,e,\textbf{m}) ~ ,
    \label{eq:Ri}
\eea
as well as a dimensionful quantity, e.g.~the mass of the $\Omega$ baryon, computed in lattice units, from which the lattice spacing can be determined after extrapolation to the infinite volume limit (see below):
\be 
   R_0(aN; g_s, e, \textbf{m}) = \frac{aM_\Omega(aN; g_s, e, \textbf{m})}{M_\Omega^\textrm{phys}} ~ , 
    \label{eq:spacing}
\ee
where $M_\Omega^\textrm{phys} = 1.672$\,GeV is the physical value of the mass of the $\Omega$ baryon. For illustration we are considering the masses of QCD$+$QED stable pseudoscalar mesons in the numerators of the dimensionless ratios (\ref{eq:Ri}) and using $M_\Omega^{\mathrm{phys}}$ to determine the lattice spacing, but of course other quantities can be used instead. For example, in the four flavour theory that we are considering here one can in principle avoid potentially very noisy baryon observables by using one of the charmed mesons masses already considered above to set the scale. The choice of setting the scale with a charmed-meson observable could however, generate significant cutoff effects and reduce the sensitivity to the charm mass.  
In Eqs.\,(\ref{eq:Ri})\,-\,(\ref{eq:spacing}) we have used $a N$ instead of $L$ to highlight that the infinite-volume limit should be taken at fixed lattice spacing (see Eq.\,(\ref{eq:RiIV}) below).
The quantity $\textbf{m}$ represents the vector of bare quark masses $\textbf{m} \equiv \{m_u, m_d, m_s, m_c\}$. Note that in the RM123 strategy, since one works at first order in $\alphaem$, it is not necessary to impose a renormalisation condition to fix the e.m.~coupling\,\cite{deDivitiis:2013xla,Giusti:2017dmp}. In this case the electric charge can simply be fixed to the Thomson limit, i.e.~$e=\sqrt{4\pi / 137.036}$, and $R_5$ becomes a predictable quantity. For the remainder of this section, we assume that we are working to $O(\alpha_\mathrm{em})$ and only consider the four ratios $R_i$ (i=1,2,3,4) as well as $R_0$ when discussing the calibration of the lattices. Notice also that at first order in $\alphaem$ the $\pi^0$ cannot decay in two photons, so that it can also be used in the calibration procedure (see section\,\ref{sec:master} below). 
 
\item Up to this point the procedure is the standard one used in QCD simulations. The difference here is in the FV effects which behave as inverse powers of $L$. We therefore envisage extrapolating the ratios $R_i$ to the infinite-volume limit:
\be
    \label{eq:RiIV}
    R_i(g_s,e, \textbf{m}) \equiv \lim_{N \to \infty}R_i(aN; g_s, e, \textbf{m}) ~ , \qquad i = 0, 1, 2, 3, 4,5 ~ .
\ee 
\item For a given discretisation and choice of $g_s$, the {\it physical} bare quark masses, $\textbf{m}^\textrm{phys}(g_s)$, and the electric charge, $e^\textrm{phys}(g_s)$, are defined by requiring that the five ratios $R_{1, 2, 3, 4, 5}$ take their physical values
\be\label{eq:Riphys}
    R_i(g_s, e^\textrm{phys}(g_s), \textbf{m}^\textrm{phys}(g_s)) = R_i^\textrm{phys} ~ , \qquad i = 1, 2, 3, 4, 5 ~ .
\ee
In practice, of course, depending on the specific choice of the ratios $R_i$, this will require some extrapolations  of results obtained at different values of the bare quark masses and electric charge.

\item[5.] The lattice spacing $a$ at this value of $g_s$ can now be defined to be 
\be \label{eq:spacingfull}
    a(g_s) = R_0(g_s, e^\textrm{phys}(g_s), \textbf{m}^\textrm{phys}(g_s)) ~ .
\ee
Note that with such a procedure the bare parameters and the lattice spacing $a$ do not depend on the lattice volume.
\item[6.] At first order in isospin breaking, i.e.~${\cal{O}}(\alphaem, m_d - m_u)$, the renormalisation of the lepton masses is performed perturbatively, by requiring that the on-shell masses correspond to the physical ones. If one wishes to go beyond first order, when hadronic effects first enter, then the physical lepton masses should be added to the quantities used in the non-perturbative calibration. The bare lepton masses, together with the other parameters, should be chosen such that, in addition to satisfying the conditions in Eq.\,(\ref{eq:Ri}), the lepton-lepton correlators decay in time as $e^{-m_\ell t}$, where $m_\ell$ is the physical mass of the lepton $\ell$.
\end{enumerate}

In Eq.\,(\ref{eq:RiIV}) we have taken the infinite-volume limit of the computed hadron masses. By working in the QED$_\textrm{L}$ finite-volume formulation of QED, if for each hadron $H$  the FV corrections of order ${\cal{O}}(e^2/(M_HL)^3, e^4)$ can be neglected, then the extrapolation to the infinite-volume limit can be avoided by making use of the formula\,\cite{Hayakawa:2008an,Borsanyi:2014jba} (similar formulae also exist for other finite-volume formulations of the theory\,\cite{Lucini:2015hfa})
\be
    \label{eq:FVmass}
    \frac{aM_H(L; g_s, e, \textbf{m})}{aM_H(g_s,e,  \textbf{m})} = 1 - \kappa \, \alphaem \, e_H^2 \left\{\frac1{2L \, M_H(g_s,e, \textbf{m})} +
                                                                                                      \frac1{L^2 \, M^2_H(g_s, e, \textbf{m})} \right\} ~ ,
\ee
where $e_H$ is the charge of the hadron $H$ and $\kappa = 2.837297\,(1)$ is a known universal constant (independent of the structure of the hadron $H$). Equation~(\ref{eq:FVmass}) can be used to determine the infinite-volume mass of the hadron $H$ from the value measured on the finite-volume $L^3$, up to corrections of order of ${\cal{O}}(e^2/(m_H L)^3, e^4)$. (In any case, even if one wishes to study the behaviour with $L$ by performing simulations at different volumes, the subtraction of the universal $O(e^2/(M_HL))$ and $O(e^2/(M_HL)^2)$ terms using Eq.\,(\ref{eq:FVmass}) is a useful starting point; the residual leading behaviour of hadronic masses is then of $O(e^2/(M_HL)^3, e^4)$.)

\subsection{Defining observables in QCD}
\label{sec:QCDQCD}

The procedure discussed in section \ref{subsec:renormalisationfull} provides a full framework with which to perform lattice simulations of QCD together with isospin-breaking effects including radiative corrections. Nevertheless, one may wish to ask how different are the results for some physical quantities in the full theory (QCD+QED) and in QCD alone. We stress again that, under the assumption that isospin breaking effects are not negligible, QCD by itself is an unphysical theory and requires a definition. Different prescriptions are possible and, of course, lead to different results in QCD. In this section we propose and advocate hadronic schemes, based on the nonperturbative evaluation of a set of hadronic masses in lattice simulations and contrast this with schemes based on equating the renormalised strong coupling and masses in some renormalisation scheme and at a particular renormalisation scale which have been used previously.

We recall that the QCD action is given by
\be 
    S^\textrm{QCD} = \frac{1}{g_0^2}S^{\mathrm{YM}} + \sum_f \left\{S_{f, 0}^\textrm{kin} + m_{f, 0} S_f^m \right\} ~ , 
    \label{eq:QCDaction}
\ee
where the kinetic term only includes the gluon links and the subscripts 0 indicate that the bare coupling and masses are different from those in the full theory of Eq.\,(\ref{eq:fullaction}). Indeed the two theories have different dynamics that, in turn, generate a different pattern of ultraviolet divergences. The difference in the bare parameters of the two theories, for all schemes used to define QCD, can in fact be ascribed to the necessity of reabsorbing the different ultraviolet singularities.
In what follows we present two different approaches to making the choice of the parameters $g_0$ and $m_{f, 0}$. Explicit details of the lattice action, discretised using the Wilson formulation for the fermions for illustration, are presented in Appendix~\ref{subsec:actions}.

\subsubsection{Defining observables in QCD: hadronic schemes}
\label{subsec:hadronicschemes}

In hadronic schemes we choose a value of $g_0$ and determine the bare quark masses $\textbf{m}_0^\textrm{phys}$ and the lattice spacing $a_0$ imposing the same conditions as for the full theory for the ratios $R_{0,\dots,4}$ evaluated at vanishing electric charge, i.e.~following steps 1\,-\,5 in Sec.\,\ref{subsec:renormalisationfull} without imposing any constraint on the ratio $R_5$. We repeat that, for 
illustration we define the bare quark masses and lattice spacing using the five ratios $R_i$, but other hadronic quantities could be used instead, both in the full theory and in QCD.
These parameters differ by terms of order ${\cal{O}}(\alphaem)$ from those in the full theory. For this discussion, we make the natural and convenient choice $g_0 = g_s$. (In order to make the perturbative expansion in Eq.\,(\ref{eq:Of_full}) the difference $g_s - g_0$ should be less than ${\cal{O}}(\alphaem)$.) With this choice, the lattice spacings in QCD ($a_0$) and in the full theory ($a$) are therefore given by
\be
    \label{eq:latticespacings}
    a_0 = \frac{\langle a_0M_\Omega \rangle^\textrm{QCD}}{M_\Omega^\textrm{phys}} \quad \textrm{and} \quad 
    a = \frac{\langle aM_\Omega \rangle^\textrm{full}}{M_\Omega^\textrm{phys}} \equiv a_0 (1 + \delta a) ~ .
\ee

To illustrate the procedure imagine that we wish to calculate an observable $O$ of mass dimension 1, for example the mass of a hadron which has not been used for the calibration. The generalisation to other cases is straightforward and presented in Appendix~\ref{sec:appB}. At a fixed value of $g_s = g_0$, we denote the best estimate of the observable $O$, which is the one obtained in the full theory, by $O^\textrm{phys}$, and that obtained in QCD as defined above by $O^\textrm{QCD}$:
\be
    O^{\textrm{phys}} \equiv \frac{\langle a O\rangle^{\textrm{full}}}{a}\quad\textrm{and} \quad
    O^{\textrm{QCD}} \equiv \frac{\langle a_0 O\rangle^{\textrm{QCD}}}{a_0} ~ .
\ee
We \underline{define} the difference of the two as being due to QED effects, $\delta O^\textrm{QED}\equiv O^\textrm{phys} - O^\textrm{QCD}$. There are 3 contributions to $\delta O^\textrm{QED}$:
\begin{enumerate}
\item The first contribution comes from the fact that the covariant derivatives in the kinetic terms in \,(\ref{eq:fullaction}) and Eq.\,(\ref{eq:QCDaction}) are different. This generates the diagrams in the correlation functions which contain the explicit exchange of virtual photons.
\item The second contribution comes from the fact that the bare quark masses appearing in Eq.\,(\ref{eq:fullaction}) and Eq.\,(\ref{eq:QCDaction}) are different. The corresponding quark-mass counterterms must therefore be inserted into the correlation functions used to determine $O^\textrm{phys}$. We stress that the need to include quark-mass counterterms is generic and arises from the requirement that the conditions being used to determine the quark masses must be satisfied both in the full theory and in QCD (for the hadronic scheme being used for illustration we impose that the conditions in Eq.\,(\ref{eq:Riphys}) are satisfied in both theories).
 
\item Finally we must account for the difference in the lattice spacings $\delta a=a-a_0$ in the full theory and QCD. 
\end{enumerate}
Combining these contributions we arrive at
\be
    \label{eq:Ophys}
    O^\textrm{phys} = O^\textrm{QCD} + \frac{\langle a_0 \,\delta O\rangle^\textrm{QCD}}{a_0} - 
                                  \frac{\delta a}{a_0^2} \langle a_0\,\! O\rangle^\textrm{QCD} ~ ,
\ee
where we have combined the contributions to the correlation functions from the exchange of virtual photons and from the insertion of the mass counterterms into $\langle a_0 \delta O\rangle^\textrm{QCD}$.

The detailed derivation of Eq.\,(\ref{eq:Ophys}) is presented in Appendix~\ref{sec:appB} but some further comments may be helpful here. The first term on the right-hand side is one that can be calculated within QCD  alone. It has a well defined continuum limit as does the sum of all the  terms in Eq.\,(\ref{eq:Ophys}).  This term allows us to define  what is the difference between QCD (defined as above) and the full theory in the hadronic scheme: $\delta O^{\mathrm{QED}} = O^\textrm{phys} - O^\textrm{QCD}$.  

An important feature of the RM123 approach which we follow in the numerical study presented below, is that the ${\cal{O}}(\alphaem)$ terms are computed explicitly and so we do not have to take the difference between numerical calculations performed in the full theory and in QCD. Each of the terms on the right-hand side of Eq.\,(\ref{eq:Ophys}) is calculated directly. We now explain the procedure in some more detail by assuming that terms of order ${\cal{O}}(\alphaem^2)$ are negligible (the extension to higher orders in $\alphaem$ is straightforward). 
\begin{enumerate}
\item Correlation functions corresponding to diagrams with the exchange of a virtual photon and to the insertion of the mass counterterms
are already of ${\cal{O}}(\alphaem)$ and are calculated directly in QCD. The term proportional to the time separation in the correlation functions gives us the mass shift $\delta M_{H_i}$ ($i = 1, 2, 3, 4$) and $\delta M_\Omega$ for the five masses (or mass differences) in the ratios $R_i$ ($i=1,2,3,4$) in Eq.\,(\ref{eq:Ri}); 

\item\label{item:masscounterterms} In the hadronic scheme being used for illustration, we impose the condition that the four ratios $R_i = m_{H_i} / m_\Omega$ are the same in QCD   and in the full theory. This corresponds to requiring that
\be
    \frac{\delta M_{H_i}}{M_{H_i}} - \frac{\delta M_\Omega}{M_\Omega} = 0 \qquad (i = 1, 2, 3, 4) ~ .
    \label{eq:condition1}
\ee
The QED contribution to the left-hand side is different from zero (and also ultraviolet divergent) and 
we require the terms proportional to the counterterms to cancel this contribution. We therefore (in principle) scan the values of the four mass counterterms $\delta m_f = m_f - m_{f, 0}$ ($f = u, d, s, c$) until the four conditions (\ref{eq:condition1}) are satisfied. Also in this case no subtraction of results obtained in the full theory and in QCD is necessary.

\item Finally we determine the difference $\delta a \equiv a - a_0$ in the lattice spacing. Having determined the bare masses using item \ref{item:masscounterterms}, we can calculate the shift in the $\Omega$ mass, $\delta M_\Omega$ due to both QED and the mass counterterms and use Eq.\,(\ref{eq:latticespacings}). Since $a\delta M_\Omega$ is calculated directly, there is again no subtraction.
\end{enumerate} 

We have devoted a considerable discussion to the definition of the isospin-breaking effects due to electromagnetism, $\delta O^\textrm{QED}$. Having done this, the subsequent definition of the strong isospin breaking effects is straightforward. To do this however, we need to define the isosymmetric theory (labelled by ``ISO") by imposing appropriate conditions to determine the bare quark masses and the lattice spacing. Since $m_u=m_d$, in the $N_f=2+1+1$ theory we need to determine only three quark masses and hence we only need three conditions, e.g. we can use the ratios $R_{1,2,3}$ in Eq.\,({\ref{eq:Ri}) to determine the physical bare quark masses. For the determination of the lattice spacing we have two options. The simplest one is to work in a mass--independent scheme and set the lattice spacing in the isosymmetric theory, $a_0^\textrm{ISO}$, equal to the one of QCD with $m_u\neq m_d$, i.e. $a_0^\textrm{ISO}=a_0$. Notice that this choice is fully consistent with renormalisation because the ultraviolet divergences of the theories that we are considering do not depend on the quark masses. Note however, that they \emph{do} depend instead on the electric charge. The other option is that we set the lattice spacing in the isosymmetric theory by using  $R_0$ in Eq.\,(\ref{eq:spacingfull}). The difference between the two options is due to cutoff effects that disappear once the continuum limit is taken consistently. The strong isospin breaking correction $\delta O^\textrm{SIB}$ to the observable $O$ can now be defined by
\be
    \delta O^\textrm{SIB} = O^\textrm{QCD} - O^\textrm{ISO} ~ ,
\ee
where $O^\textrm{ISO} = \frac{\langle a_0^\textrm{ISO} O\rangle^\textrm{ISO}}{a_0^\textrm{ISO}}$ is the value of the observable obtained in isosymmetric QCD. With these definitions we have the natural relation $O^\textrm{phys} = O^\textrm{ISO} + \delta O^\textrm{QED} + \delta O^\textrm{SIB}$. We underline however that $\delta O^\textrm{SIB}$ depends on the quantities used for calibration, both in 4-flavour QCD and in isosymmetric QCD.

\subsubsection{Defining QCD: the GRS scheme}
\label{subsec:GRS}

A different prescription, called the GRS scheme, was proposed in Ref.\,\cite{Gasser:2003hk} to relate the bare quark masses and bare coupling of QCD ($m_{f,0}$ and $g_0$) to those in the full theory ($m_f$ and $g_s$). This prescription has been adopted in Refs.\,\cite{deDivitiis:2013xla,Giusti:2017dmp,Giusti:2017dwk}. In the GRS approach, instead of determining the bare parameters of QCD by requiring that the chosen hadronic masses in QCD are equal to their physical values, one imposes that the renormalised parameters in a given short--distance scheme (e.g.~the $\msbar$ scheme) and at a given scale are equal in the full and QCD theories. 

A consistent procedure is the following:

\begin{enumerate}

\item The full theory is renormalised by using a physical hadronic scheme as discussed in subsection\,
\ref{subsec:renormalisationfull}. This means that for each chosen value of $g_s$ we know the corresponding physical value of the bare electric charge $e^\textrm{phys}(g_s)$ and of the lattice spacing $a(g_s)$.

\item The renormalisation constants (RCs) of the strong coupling constant and of the quark masses are computed in a short--distance mass--independent scheme both in the full theory and in the theory at vanishing electric charge. 

\item In order to set the bare parameters of QCD at a given value of the lattice spacing we now chose a matching scale $\mu$ and impose that the renormalised strong coupling constant and the renormalised quark masses are the same as in the full theory. In practice we might want to simulate QCD at the same values of the lattice spacing used in the full theory simulations. In this case the matching conditions are
\bea
    g(\mu) & = & Z_g(0, g_0, a(g_s)\mu) g_0 = Z_g(e^\textrm{phys}(g_s), g_s, a(g_s)\mu) g_s = \widehat{g}(\mu) \nonumber \\
    m_f(\mu) & = & Z_{m_f}(0, g_0, a(g_s)\mu) m_{f,0}(g_0) = Z_{m_f}(e^\textrm{phys}(g_s), g_s, a(g_s)\mu) m_f(g_s) = \widehat{m}_f(\mu) ~ ,
    \label{eq:GRSdef}
\eea
where ~ $\widehat{}$ ~ indicates quantities in the QCD+QED theory.
Notice that quarks with the same electric charge have the same RC, e.g.~$Z_{m_u}(e, g_s, \mu) = Z_{m_c}(e, g_s, \mu)$, and that the quark mass RC at vanishing electric charge is flavour independent, $Z_{m_f}(0, g_0, \mu) = Z_{m}(g_0, \mu)$.

\item In order to define isosymmetric QCD by using this approach, the bare up--down quark mass is determined from
\be
    Z_{m}(g_0, a(g_s)\mu)\, m_{ud,0}(g_0) = \frac{\widehat{m}_u(\mu) + \widehat{m}_d(\mu)}{2} ~ .
\ee

\end{enumerate}

Some remarks are in order at this point. The GRS scheme is a short--distance matching procedure that can also be used to match the theories at unphysical values of the renormalised electric charge and/or quark masses with the physical theory. 

By following the procedure outlined above one can perform lattice simulations of the full theory and of (isosymmetric) QCD at the same value of the lattice spacing but, consequently, at different values of the bare strong coupling constant. This is different from the strategy outlined in the previous subsection where, by using hadronic schemes, it was more natural to chose the same value of the bare strong coupling at the price of having two different lattice spacings. The absence of the lattice spacing counterterm (see Eq.\,(\ref{eq:Ophys}) above) in the GRS scheme is compensated from the presence of the counterterm $(1/g_0^2 - 1/g_s^2)S^{\mathrm{YM}}$ originating from the difference of the bare strong coupling constants in the two theories.

A remark of some practical relevance concerns the possibility of implementing hadronically the GRS scheme. To this end, note that in the GRS scheme the dimensionless ratios $R_{i}$ will not be equal to the corresponding physical values and the difference can be parametrized as follows 
\be
    R_i^{\textrm{QCD-GRS}} = R_i^{\textrm{phys}}(1 + \epsilon_i^{\textrm{GRS}}),
    \label{eq:RiGRS}
\ee
where the $\epsilon_i^{\textrm{GRS}}$ are order ${\cal{O}}(\alphaem)$ and depend on the chosen matching scheme and also on the chosen matching scale. Once the $\epsilon_i^{\textrm{GRS}}$ (and hence the $R_i^{\textrm{QCD-GRS}}$) are known, for example from a particularly accurate lattice simulation, then they can be used in other lattice computations. The bare quark masses are then determined by requiring that the $R_i$ in (isosymmetric) QCD reproduce $R_i^{\textrm{QCD-GRS}}$ as given by Eq.(\ref{eq:RiGRS}), and, at this stage, the GRS scheme can be considered to be a hadronic one as it is defined in terms of non-perturbatively computed quantities (in this case meson masses). We stress however that this requires prior knowledge of the $\epsilon_i^{\textrm{GRS}}$.

Of course other schemes are also possible. In general, the $\epsilon_i$ provide a unifying language to discuss the different schemes for the definition of (isosymmetric) QCD in the presence of electromagnetism; in physical hadronic schemes the $\epsilon_i=0$ while in the GRS and other schemes they are of order ${\cal{O}}(\alphaem)$.
For later use, we make the simple observation that two schemes can be considered to be equivalent in practice if the $\epsilon_i$ in the two schemes are equal within the precision of the computations.

Although the GRS scheme is perfectly legitimate, we advocate the use of physical hadronic schemes in future lattice calculations. For lattice simulations of physical quantities, a non-perturbative calibration of the lattice is necessary in general, but the renormalisation required for the GRS conditions in Eq.\,(\ref{eq:GRSdef}) is not generally necessary (except perhaps for the determination of the renormalised coupling and quark masses themselves).
Now that hadronic masses are calculated with excellent precision in lattice simulations and their values are well known from experimental measurements, it is natural to use hadronic schemes. By contrast, the renormalised couplings and masses are derived quantities which are not measured directly in experiments. In spite of this, as explained above, at the time that our computation was started we chose to use the GRS scheme. Of course the physical results in the full theory do not depend on this choice.

\section{Evaluation of the amplitudes}
\label{sec:master}

At first order in $\alphaem$ and $(m_d - m_u) / \Lambda_{\mathrm{QCD}}$ the inclusive decay rate (\ref{eq:Gamma}) can be written as
 \be
      \Gamma(P^\pm \to \ell^\pm \bar{\nu}_\ell [\gamma]) =  \Gamma^{\mathrm{QCD}}\cdot \left[ 1 + \overline{\delta R}_P \right] + 
                                                                                             {\cal{O}}\left[ \alphaem^2, (m_d - m_u)^2, \alphaem (m_d - m_u) \right] ~ ,
      \label{eq:master0}
 \ee
where $\Gamma^{\mathrm{QCD}}$ is the tree-level decay rate given by
\be
     \Gamma^{\mathrm{QCD}} = \frac{G_F^2}{8 \pi} |V_{q_1 q_2}|^2 m_\ell^2 \left( 1 - \frac{m_\ell^2}{M_P^{(0)\,2}} \right)^{\!\!2} 
                                    f_P^{(0)\,2}  M_{P}^{(0)} ~ ,
     \label{eq:GammaQCD}
 \ee
and $M_P^{(0)}$ and $f_P^{(0)}$ are the mass and decay constant of the charged P-meson mass defined in isosymmetric QCD in the chosen scheme. 

The decay constant $f_P^{(0)}$ is defined in terms of the matrix element of the QCD axial current $A_P^{(0)}$ (in the continuum) as
 \be
      A_P^{(0)} \equiv \langle 0 | \bar{q}_2 \gamma_0 \gamma_5 q_1 | P^{(0)} \rangle \equiv f_P^{(0)} M_P^{(0)} \,,
      \label{eq:AP0}
 \ee
where the initial state meson $P^{(0)}$ is at rest.
The decay rate is obtained from the insertion of the lowest-order effective Hamiltonian
\be
     {\cal H}_W = \frac{G_F}{\sqrt{2}} \, V_{q_1 q_2}^\ast ~ O_1 = \frac{G_F}{\sqrt{2}} \,V_{q_1 q_2}^\ast \,
                          \big( \bar{q_2}\gamma_\mu (1-\gamma_5) {q_1} \big) \, \big(\bar{\nu}_{\ell} \gamma^\mu (1-\gamma_5)\ell \big) \, , 
    \label{eq:DeltaL0}
\ee
as depicted in the Feynman diagram of Fig.~\ref{fig:DiagLO}, where the decay of a charged kaon is shown as an example.  
At lowest order in $\alphaem$ the two full dots in the figure represent the two currents in the bare four-fermion operator 
\be
O_1=\big( \bar{q_2}\gamma_\mu (1-\gamma_5) {q_1} \big) \, \big(\bar{\nu}_{\ell} \gamma^\mu (1-\gamma_5)\ell \big)\,,
    \label{eq:O1bare}
\ee
whereas at order $\alphaem$ they will denote the insertion of the renormalised operator in the W-regularisation as defined in Sec.\,\ref{sec:Wreg}.  

\begin{figure}[t]
\includegraphics[width=0.40\hsize]{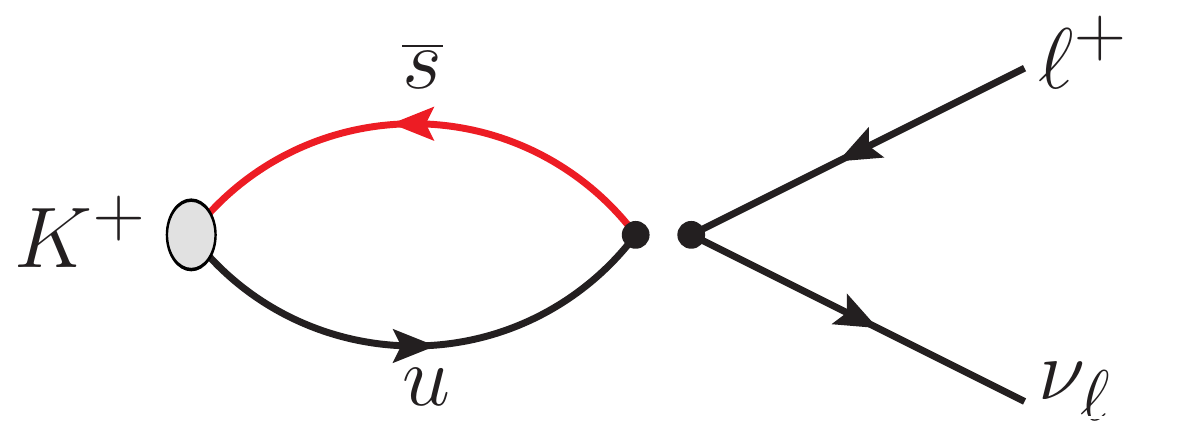}
\caption{\it \footnotesize Feynman diagram for the process $K^+ \to \ell^ + \nu_\ell$. In the effective theory the interaction is given by a local four-fermion operator denoted by the two full dots in the figure.\hspace*{\fill}}
\label{fig:DiagLO}
\end{figure}

In order to compare our results for the e.m.~and strong IB corrections to those obtained in Ref.\,\cite{Cirigliano:2011tm} and adopted by the PDG \cite{PDG,Rosner:2015wva} however, we will use a modified expression: 
 \be
      \Gamma(P^\pm \to \ell^\pm \bar{\nu}_\ell [\gamma]) =  \Gamma^{(0)} \cdot \left[ 1 + \delta R_P \right] + 
                                                                                             {\cal{O}}\left[ \alphaem^2, (m_d - m_u)^2, \alphaem (m_d - m_u) \right] ~ ,
      \label{eq:master}
 \ee
where $\Gamma^{(0)}$ is given by
 \be
     \Gamma^{(0)} = \frac{G_F^2}{8 \pi} |V_{q_1 q_2}|^2 m_\ell^2 \left( 1 - \frac{m_\ell^2}{M_P^2} \right)^2 
                                \left[ f_P^{(0)} \right]^2 M_{P} ~ ,
     \label{eq:Gamma0}
 \ee
and $M_P$ is the physical mass of the charged P-meson including both e.m.~and leading-order strong IB corrections.

The quantity $\delta R_P$ encodes both the e.m.~and the strong IB leading-order corrections to the tree-level decay rate. 
Its value depends on the prescription used for the separation between the QED and QCD corrections, while the quantity  
\be
    \mathcal{F}_P^2 \equiv \frac{\Gamma(P^\pm \to \ell^\pm \bar{\nu}_\ell [\gamma])}
                                          {\frac{G_F^2}{8 \pi} |V_{q_1 q_2}|^2 m_\ell^2 \left( 1 - \frac{m_\ell^2}{M_P^2} \right)^2 M_{P}}
                                         = \left[ f_P^{(0)} \right]^2 (1 + \delta R_P)
\ee
is prescription independent\,\cite{Gasser:2010wz} to all orders in both $\alphaem$ and $(m_d - m_u)$.

The quantity $\mathcal{F}_\pi$ may be used to set the lattice scale instead of the $\Omega$ baryon mass. The physical value $\mathcal{F}_\pi^\textrm{phys}$ can be obtained by taking the experimental pion decay rate $\Gamma(\pi^- \to \mu^- \bar{\nu}_\mu [\gamma]) = 3.8408 (7) \cdot 10^7$ s$^{-1}$ from the PDG~\cite{PDG} and the result for $|V_{ud}| = 0.97420 (21)$ determined accurately from super-allowed $\beta$-decays in Ref.\,\cite{Hardy:2016vhg}. 
Consequently, one may replace $M_\Omega$ with $\mathcal{F}_\pi$ (as the denominator of the ratios $R_{1,\dots,4}$ in Eqs.\,(\ref{eq:Ri})), $M_{\pi^+}$ with $M_{\pi^0}$ in the ratio $R_1$ (when working at leading order in $\alphaem$) and set the electron charge directly to its Thomson's limit (instead of using the ratio $R_5$), namely
\bea
    &&
    R_1(aN; g_s, e, \textbf{m}) = \frac{aM_{\pi^0}}{a\mathcal{F}_\pi}(aN; g_s, e, \textbf{m}) \, , 
    \nonumber \\
    &&
    R_2(aN; g_s, e, \textbf{m}) = \frac{aM_{K^0}}{a\mathcal{F}_\pi}(aN; g_s, e, \textbf{m}) 
    \nonumber \\
    &&
    R_3(aN; g_s, e, \textbf{m}) = \frac{aM_{D_s}}{a\mathcal{F}_\pi}(aN;g_s,e,\textbf{m}) ~ ,
    \nonumber \\
    &&
    R_4(aN; g_s, e, \textbf{m}) = \frac{aM_{K^+}-aM_{K^0}}{a\mathcal{F}_\pi}(aN; g_s, e, \textbf{m}) ~ .
    \label{eq:Ri_FLAG}
\eea

Note that for the present study we were unable to use $M_\Omega$ to determine the lattice spacing because the corresponding baryon correlators were unavailable. The choice of using $\mathcal{F}_\pi$ instead to set the scale clearly prevents us from being able to predict the value of $\vert V_{ud} \vert$. 
This is one of the reasons why we advocate the use of hadronic schemes with hadron masses as experimental inputs for future lattice calculations. 
However, as already explained above, in this work we renormalise the QCD theory using the same set of hadronic inputs adopted in our quark-mass analysis in Ref.\,\cite{Carrasco:2014cwa}, since we started the present calculations using the RM123 method on previously generated isosymmetric QCD gauge configurations from ETMC (see Appendix\,\ref{sec:appA}). 
The bare parameters of these QCD gauge ensembles were fixed in Ref.\,\cite{Carrasco:2014cwa} by using the hadronic scheme corresponding to $M_\pi^{(0),\textrm{FLAG}} = 134.98$\,MeV, $M_K^{(0),\textrm{FLAG}} = 494.2(3)$\,MeV and $f_\pi^{(0),\textrm{FLAG}} = 130.41(20)$\,MeV, while $M_{D_s}^{(0)}$ was chosen to be equal to the experimental $D_s^+$-meson mass, $M_{D_s^+} = 1969.0(1.4)$\,MeV\,\cite{PDG}. 
Note that in the absence of QED radiative corrections $\mathcal{F}_\pi$ reduces to the conventional definition of the pion decay constant $f_\pi^{(0)}$. 
The superscript $\textrm{FLAG}$ has been used because the chosen values of three out of the four hadronic inputs had been suggested in the previous editions of the FLAG review\,\cite{FLAG}. 
For this reason we refer to the scheme defined from these inputs as the FLAG scheme. 

We have calculated the same input parameters (\ref{eq:Ri_FLAG}) used in the FLAG scheme also in the GRS scheme (corresponding to the $\msbar$ scheme at $\mu = 2$ GeV) obtaining\footnote{These values differ slightly from those obtained in Ref.\,\cite{Giusti:2017dmp}, since we have now included the non-factorisable corrections of order ${\cal{O}}(\alphaem \alpha_s^n)$ (with $n \geq 1$) to the mass renormalisation constant (see the coefficient $Z_m^\mathrm{fact}$ in Eq.\,(\ref{eq:Zf}) and in Table~\ref{tab:factorisation} below). We take the opportunity to update Eqs.\,(8), (10), (14) and (15) of Ref.\,\cite{Giusti:2017dmp} with $\epsilon_{\pi^0} = 0.01 (4)$, $\epsilon_{K^0} = 0.01 (2)$, $\delta M_{D^+} + \delta M_{D^0} = 1.7 (1.0)$ MeV and $\delta M_{D_S^+} = 2.3 (4)$ MeV.}: $M_\pi^{(0),\mathrm{GRS}} = 135.0(2)$\,MeV, $M_K^{(0),\mathrm{GRS}} = 494.6(1)$\,MeV, $M_{D_s}^{(0),\mathrm{GRS}} = 1966.7(1.5)$\,MeV and $f_\pi^{(0),\mathrm{GRS}} = 130.65(12)$\,MeV (see Eq.\,(\ref{eq:fpi0}) in Sec.\,\ref{sec:results} below). 
Therefore, the values of the inputs determined in the GRS scheme differ at most by $\sim 0.15 \%$ from the corresponding values adopted in Ref.\,\cite{Carrasco:2014cwa} for the isosymmetric QCD theory and the differences are at the level of our statistical precision. 
Thus, the result of our analysis of the scheme dependence can be summarized by the conclusion that the FLAG and GRS schemes can be considered to be equivalent at the current level of precision. 
Nevertheless, we have used the results of this analysis to estimate the systematic error on our final determinations of the isospin breaking corrections $\delta R_P$ induced by residual scheme uncertainties (see the discussion at the end of Sec.\,\ref{sec:results}). 

In light of this quantitative analysis, given the numerical equivalence of the two schemes at the current level of precision, in the rest of the paper we shall compare our results obtained in the GRS scheme with the results obtained by other groups using the FLAG scheme and we shall not use superscripts to distinguish between the two schemes.

The correction $\delta R_P$, defined in Eq.\,(\ref{eq:master}), is given by (see Ref.\,\cite{Carrasco:2015xwa})
\be
    \delta R_P = \frac{\alphaem}{\pi} \mbox{log}\left( \frac{M_Z^2}{M_W^2} \right) + 2 \frac{\delta A_P}{A_P^{(0)}} - 
                        2 \frac{\delta M_P}{M_P^{(0)}} + \delta \Gamma_P^{\mathrm{(pt)}}(\Delta E_\gamma) ~ ,
    \label{eq:RPS} 
\ee
where 
\begin{enumerate}

\item[i)] the term containing $\mbox{log}(M^2_Z/M^2_W)$ comes from the short-distance matching  between the full theory (the Standard Model) and the effective theory  in the $W$-regularisation\,\cite{Sirlin:1981ie};

\item[ii)] the quantity $\delta \Gamma_P^{\mathrm{(pt)}}(\Delta E_\gamma)$ represents the ${\cal{O}}(\alphaem)$ correction to the tree-level decay rate for a point-like meson (see Eq.\,(\ref{eq:Gamma})), which can be read off from Eq.\,(51) of Ref.\,\cite{Carrasco:2015xwa}. The cut-off on the final-state photon's energy, $\Delta E_\gamma$, must be sufficiently small for the point like-approximation to be valid;

\item[iii)] $\delta A_P$ is the e.m.~and strong IB correction to the decay amplitude $P\to\ell\nu$ with the corresponding correction to the amplitude with a point-like meson subtracted (this subtraction term is added back in the term $\delta \Gamma_P^{\mathrm{(pt)}}(\Delta E_\gamma)$, see Eq.\,(\ref{eq:Gamma})). 

\item[iv)] $\delta M_P$ are the e.m.~and strong IB corrections to the mass of the P-meson.  The correction proportional to $2\, \delta M_P/M_P^{(0)}$ is present because of the definition of $f_P^{(0)}$ in terms of the amplitude  and of the meson mass in Eq.\,(\ref{eq:AP0}).  

\end{enumerate}
Since we adopt the qQED approximation, which neglects the effects of the sea-quark electric charges, the calculation of $\delta A_{P}$ and $\delta M_{P}$ only requires the evaluation of the connected diagrams. These are shown in Figs.~\ref{fig:DiagLO} - \ref{fig:diagrams_4} for the case of $K_{\ell 2}$ decays. At ${\cal{O}}(\alphaem)$ the diagram in Fig.~\ref{fig:DiagLO} corresponds to the insertion of the operator renormalised in the W-renormalisation scheme. 

\begin{figure}[htb!]
\subfloat[]{\includegraphics[scale=0.45]{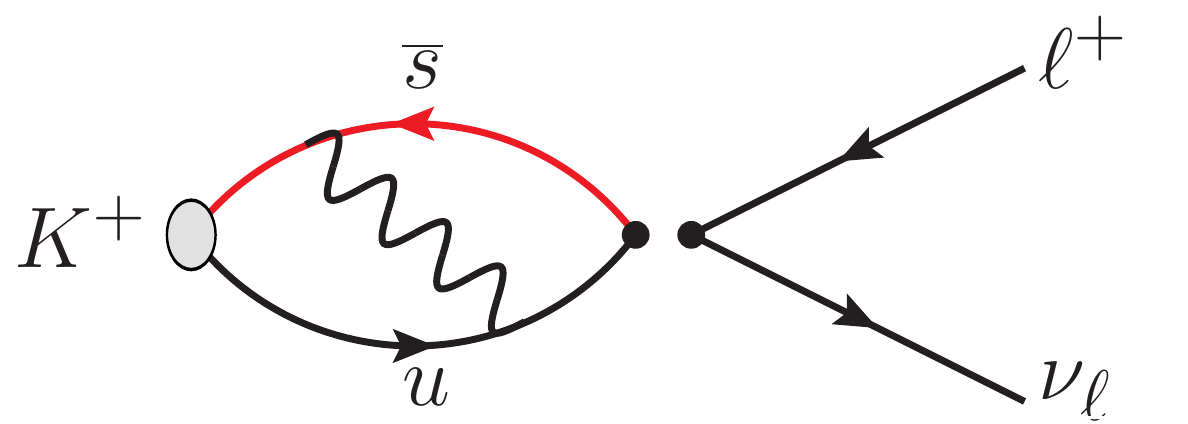}}

\subfloat[]{\raisebox{0.20\totalheight}{\includegraphics[scale=0.45]{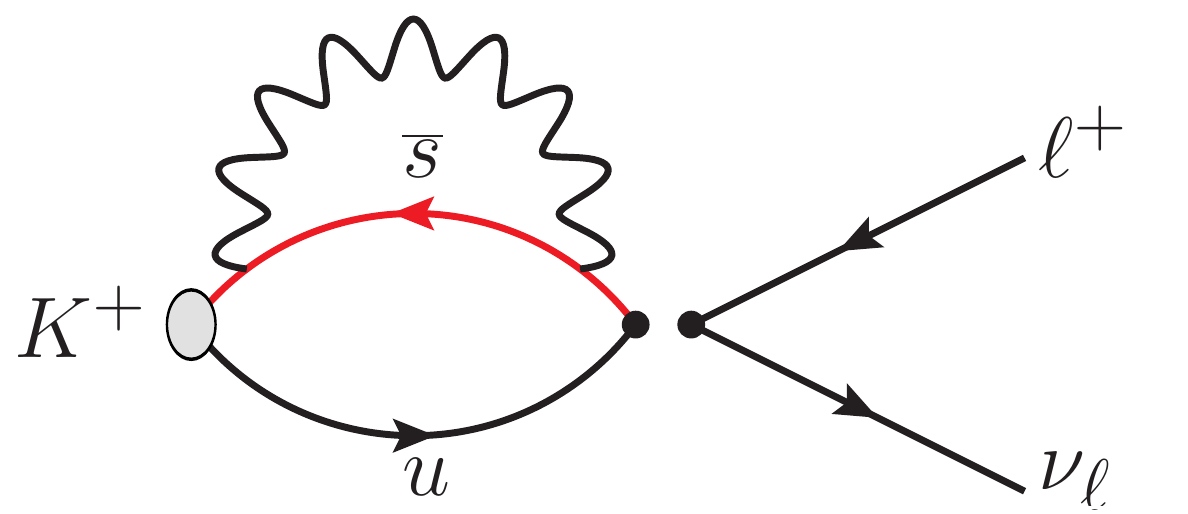}}}~
\subfloat[]{\includegraphics[scale=0.45]{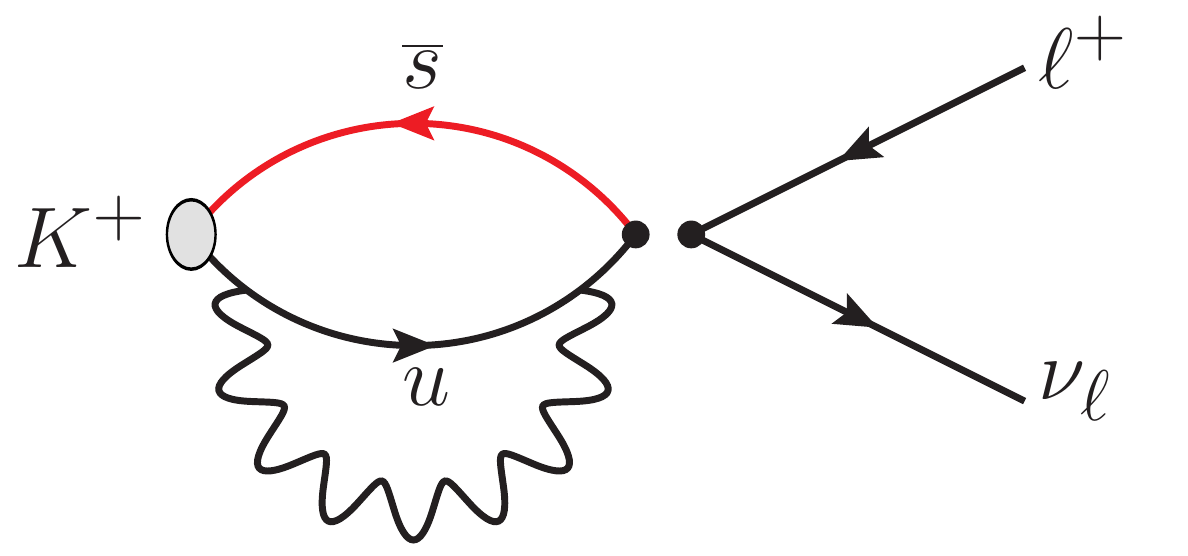}}

\subfloat[]{\raisebox{0.20\totalheight}{\includegraphics[scale=0.45]{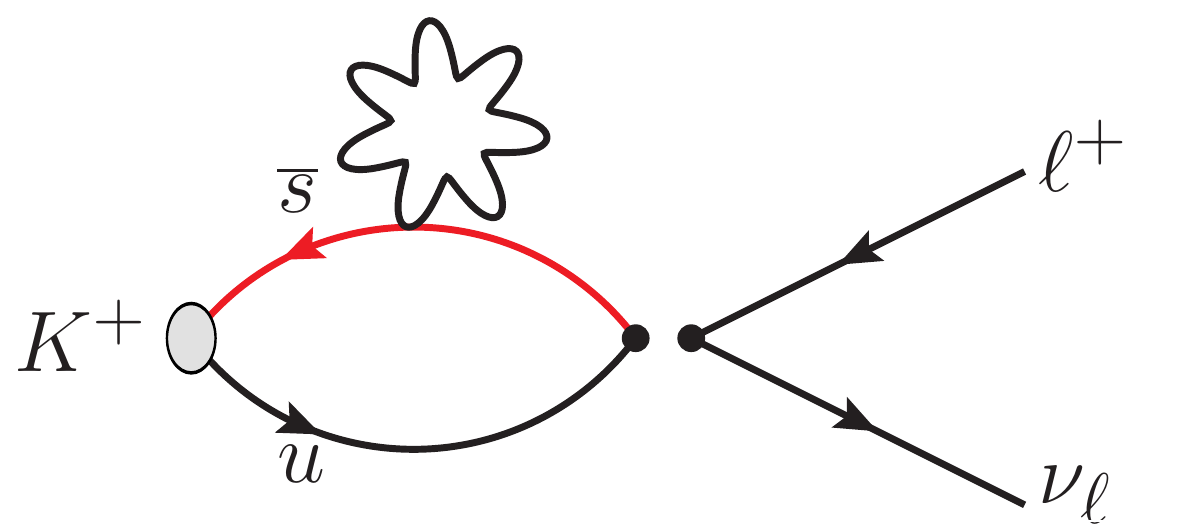}}}~
\subfloat[]{\includegraphics[scale=0.45]{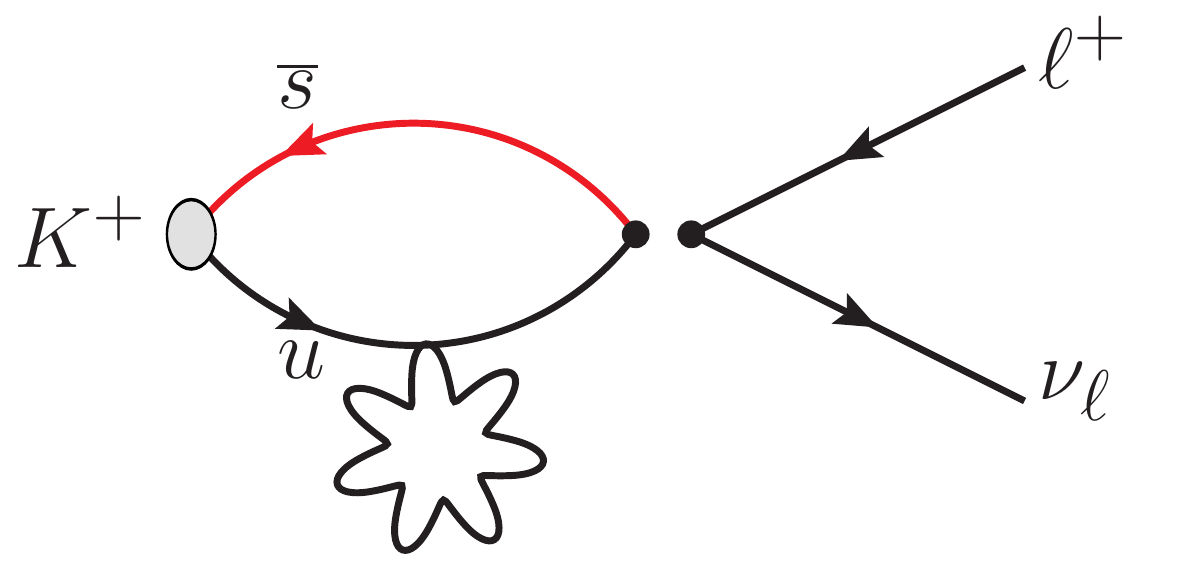}}
\caption{\it \footnotesize Connected diagrams contributing at ${\cal{O}}(\alphaem)$ to the $K^+ \to \ell^+ \nu_\ell$ decay amplitude, in which the photon is attached to quark lines: (a) exchange, (b, c) self-energy and (d, e) tadpole diagrams. The labels are introduced to identify the individual diagrams when describing their evaluation in the text.\hspace*{\fill}}
\label{fig:diagrams_1}
\end{figure}
\begin{figure}[htb!]
\subfloat[]{\includegraphics[scale=0.45]{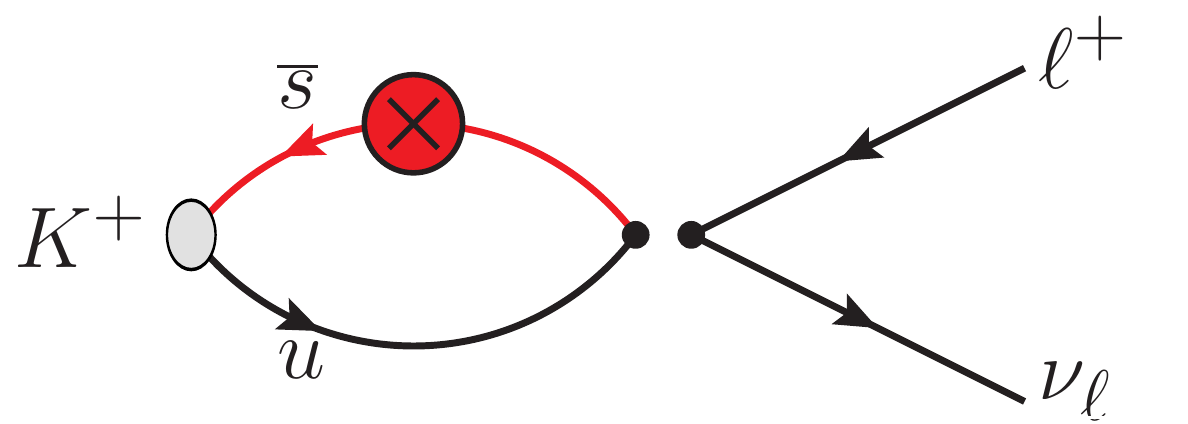}}~
\subfloat[]{\includegraphics[scale=0.45]{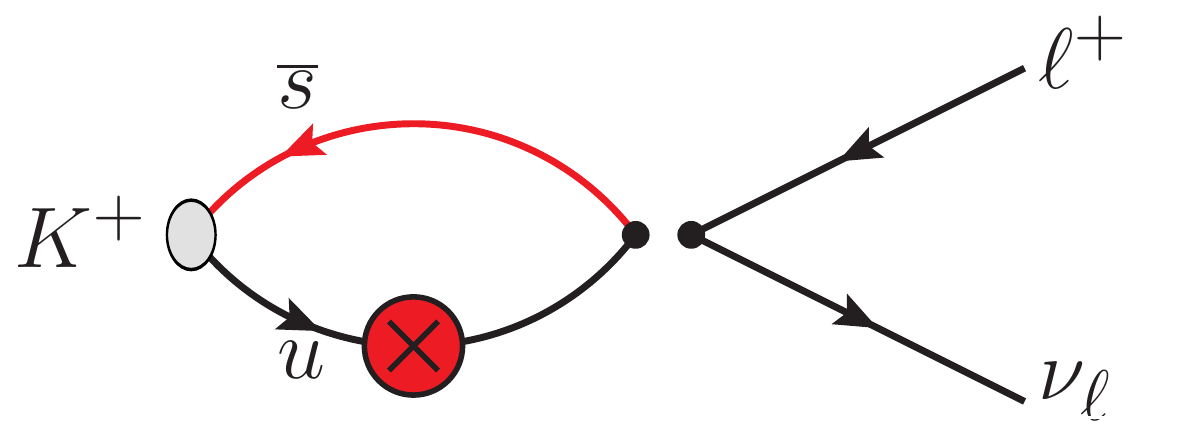}}
\caption{\it \footnotesize Connected diagrams contributing at ${\cal{O}}(\alphaem)$ to the $K^+ \to \ell^+ \nu_\ell$ decay amplitude corresponding to the insertion of the pseudoscalar density related to the e.m.~shift of the critical mass, $\delta m_f^{\rm crit}$, determined in Ref.\,\cite{Giusti:2017dmp}.\hspace*{\fill}}
\label{fig:diagrams_2}
\end{figure}
\begin{figure}[htb!]
\subfloat[]{\includegraphics[scale=0.50]{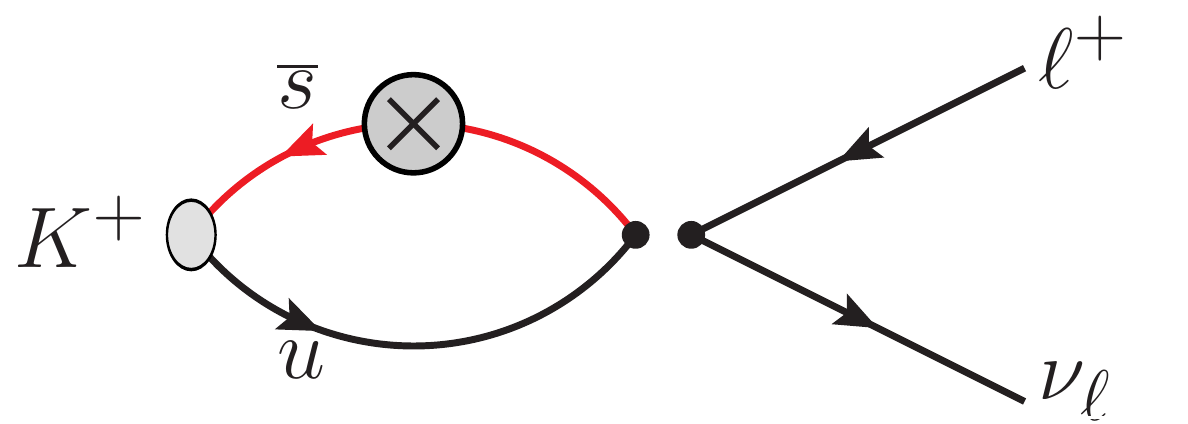}}~
\subfloat[]{\includegraphics[scale=0.50]{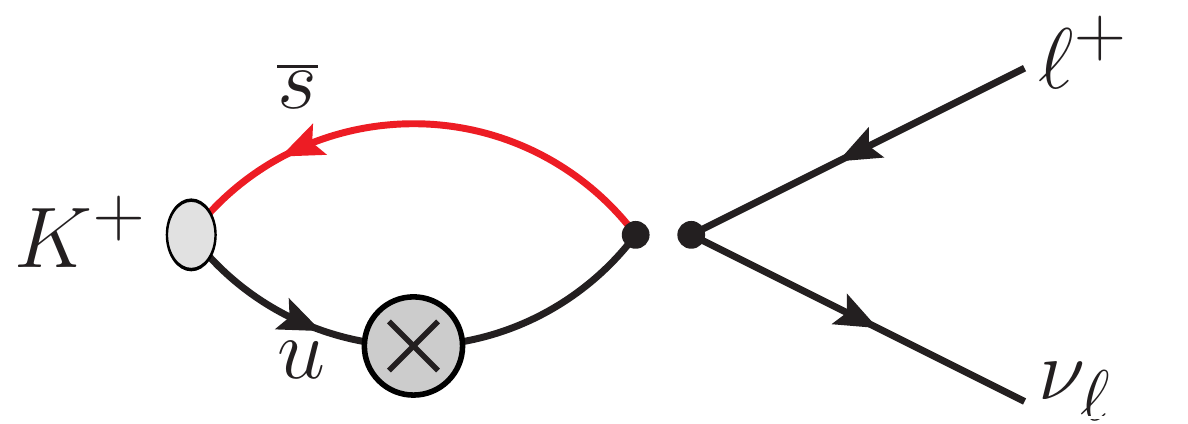}}
\caption{\it \footnotesize Connected diagrams contributing at ${\cal{O}}(\alphaem)$ and ${\cal{O}}(m_d - m_u)$ to the $K^+ \to \ell^+ \nu_\ell$ decay amplitude related to the insertion of the scalar density (see Ref.\,\cite{Giusti:2017dmp}).\hspace*{\fill}}
\label{fig:diagrams_3}
\end{figure}
\begin{figure}[htb!]
\subfloat[]{\includegraphics[scale=0.45]{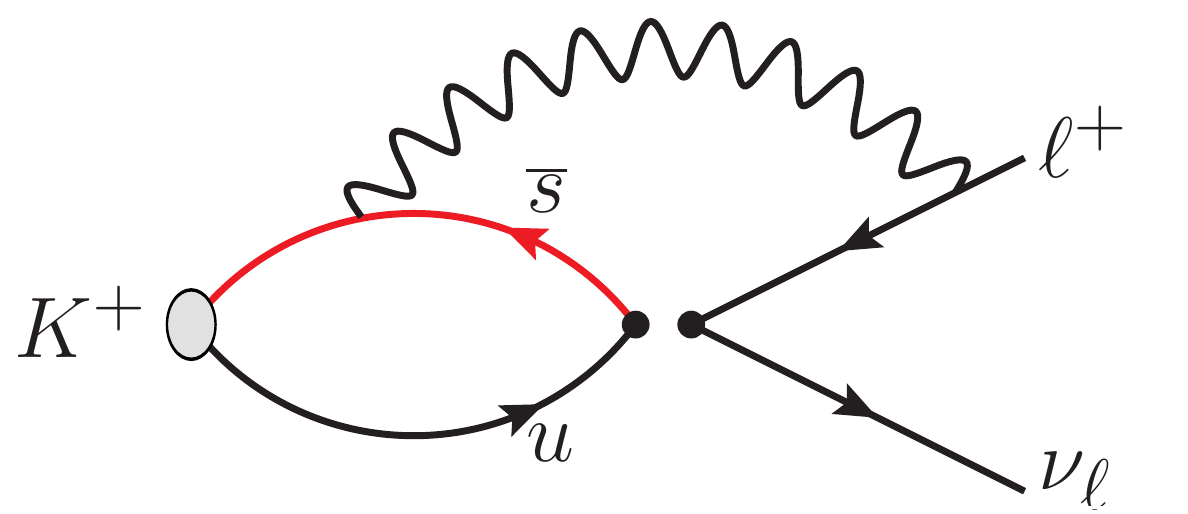}}~
\subfloat[]{\includegraphics[scale=0.45]{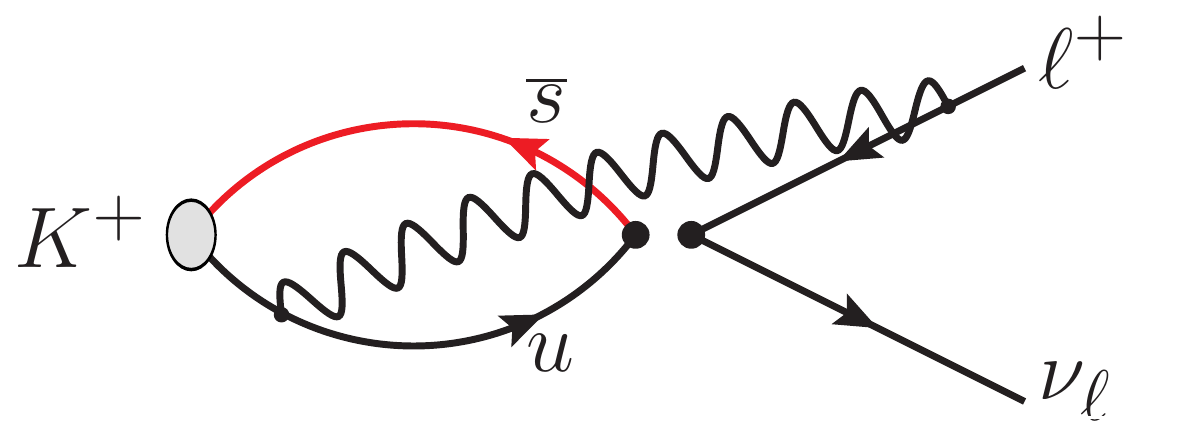}}~
\subfloat[]{\includegraphics[scale=0.45]{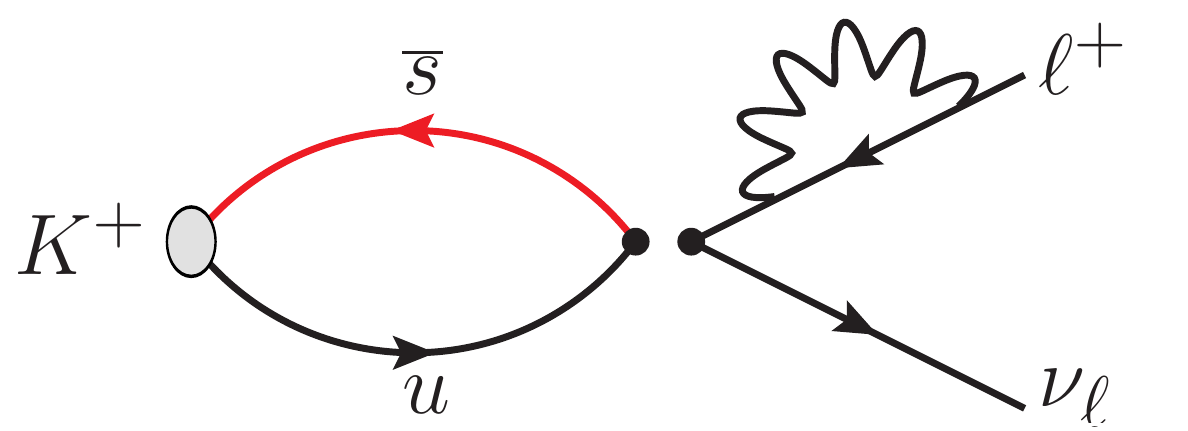}}
\caption{\it \footnotesize Connected diagrams contributing at ${\cal{O}}(\alphaem)$ to the $K^+ \to \ell^+ \nu_\ell$ decay amplitude corresponding to photon exchanges involving the final-state lepton.\hspace*{\fill}}
\label{fig:diagrams_4}
\end{figure}

In Eq.\,(\ref{eq:RPS}) $\delta A_P$ and $\delta M_P$ contain both the e.m.~and the strong IB leading-order corrections
 \bea
      \label{eq:deltaAPS}
      \delta A_P & = & \delta A_P^{W} + \delta A_P^{\mathrm{SIB}} + \sum_{i = J, T, P , S} \delta A_P^i + \delta A_P^\ell + 
                                 \delta A_P^{\ell,\rm{self}} ~  , \\
      \label{eq:deltaMPS}
      \delta M_P & = & \delta M_P^{\mathrm{SIB}} + \sum_{i = J, T, P , S} \delta M_P^i ~  ,
 \eea
where $\delta A_P^{W}$ is the e.m.~correction from both the matching of the four-fermion lattice weak operator to the W-renormalisation scheme and from the mixing with several bare lattice four-fermion operators generated by the breaking of chiral symmetry with the twisted-mass fermion action which we are using. Both the matching and the mixing will be discussed and calculated in Sec.\,\ref{sec:Wreg}. As already pointed out, the renormalised operator, defined in the W-renormalisation scheme, is inserted in the diagram of Fig.~\ref{fig:DiagLO}. As for the diagrams of Figs.~\ref{fig:diagrams_1}\,-\,\ref{fig:diagrams_4}, which are already of order ${\cal{O}}(\alphaem)$ and ${{\cal{O}}((m_d - m_u) / \Lambda_{\textrm{QCD}})}$, it is sufficient to insert the weak current operator renormalised in QCD only.

In Eqs.\,(\ref{eq:deltaAPS}) and (\ref{eq:deltaMPS}) the quantity $\delta A_P^{\mathrm{SIB}}$ ($\delta M_P^{\mathrm{SIB}}$) represents the strong IB corrections proportional to $m_d-m_u$ and to the diagram of Fig.\,\ref{fig:diagrams_3}(b), while the other terms are QED corrections coming from the insertions of the e.m.~current and tadpole operators, of the pseudoscalar and scalar densities (see Refs.\,\cite{deDivitiis:2013xla,deDivitiis:2011eh}).
The term $\delta A_P^J$ ($\delta M_P^J$) is generated by the diagrams of Fig.~\ref{fig:diagrams_1}(a-c), $\delta A_P^T$ ($\delta M_P^T$) by the diagrams of Fig.~\ref{fig:diagrams_1}(d-e), $\delta A_P^P$ ($\delta M_P^P$) by the diagrams of Fig.~\ref{fig:diagrams_2}(a-b) and $\delta A_P^S$ ($\delta M_P^S$) by the diagrams of Fig.~\ref{fig:diagrams_3}(a-b).
The term $\delta A_P^{\ell}$ corresponds to the exchange of a photon between the quarks and the final-state lepton and arises from the diagrams in Fig.\,\ref{fig:diagrams_4}(a-b). The term $\delta A_P^{\ell,\rm{self}}$ corresponds to the 
contribution to the amplitude from the lepton's wave function renormalisation; it arises from the self-energy diagram of Fig.~\ref{fig:diagrams_4}(c). The contribution of this term cancels out in the difference $\Gamma_0(L) - \Gamma_0^{\textrm{pt}}(L)$ and could be therefore omitted, as explained in the following section. The different insertions of the scalar density encode the strong IB effects together with the counter terms necessary to fix the masses of the quarks. The insertion of  the pseudoscalar density is peculiar to twisted mass quarks and would be absent in standard Wilson (improved) formulations of QCD. 

In the following subsection we discuss the calculation of all the diagrams that do not involve the photon attached to the charged lepton line. The determination of the contributions $\delta A_P^\ell$ and $\delta A_P^{\ell,\rm{self}}$ will be described later in subsection \ref{sec:crossed}.

\subsection{Quark-quark photon exchange diagrams and scalar and pseudoscalar insertions}
\label{sec:noncrossed}

The terms $\delta A_P^i$ and $\delta M_P^i$ ($i = J, T, P , S$) can be extracted from the following correlators:
 \bea
     \label{eq:deltaC_J}
     \delta C_P^J(t) & = & 4 \pi \alphaem \frac{1}{2} \sum_{\vec{x}, y_1, y_2} \langle 0| T \left \{ J^\rho_W(0) ~ 
                                        j^{\mathrm{em}}_\mu(y_1) j^{\mathrm{em}}_\nu(y_2) ~  \phi_P^\dagger(\vec{x}, - t) \right \} | 0 \rangle 
                                        \Delta_{\mu \nu}^{\mathrm{em}}(y_1, y_2) \frac{p^\rho_P}{M_P} ~ , ~ \\
    \label{eq:deltaC_T}
     \delta C_P^T(t) & = & 4 \pi \alphaem \sum_{\vec{x}, y} \langle 0| T \left \{ J^\rho_W(0) ~ T_\mu^{\mathrm{em}}(y) ~ 
                                        \phi_P^\dagger(\vec{x}, - t) \right \} | 0 \rangle \Delta_{\mu \mu}^{\mathrm{em}}(y,y) 
                                        \frac{p^\rho_P}{M_P} ~ , ~ \\
     \label{eq:deltaC_P}
     \delta C_P^P(t) & = & 4 \pi \alphaem \sum_{f = f_1, f_2} \delta m_f^{\rm crit} \cdot \sum_{\vec{x}, y} 
                                           \langle 0| T \left \{ J^\rho_W(0) ~ i \overline{q}_f(y) \gamma_5  q_f(y) ~ 
                                           \phi_P^\dagger(\vec{x}, - t) \right \} | 0 \rangle \frac{p^\rho_P}{M_P} ~ , ~ \\
     \label{eq:deltaC_S}
     \delta C_P^S(t) & = & - 4 \pi \alphaem \sum_{f = f_1, f_2} m_f \frac{{\cal{Z}}_m^f}{Z_m^{(0)}} \cdot \sum_{\vec{x}, y} 
                                        \langle 0| T \left \{ J^\rho_W(0) ~  \left[ \overline{q}_f(y) q_f(y) \right] ~ 
                                        \phi_P^\dagger(\vec{x}, - t) \right \} | 0 \rangle \frac{p^\rho_P}{M_P} ~ , ~ 
 \eea
where $\Delta_{\mu \nu}^{\mathrm{em}}(y_1, y_2)$ is the photon propagator, $J^\rho_W(x)$ is the local version of the hadronic $(V-A)$ weak current renormalised in QCD only\footnote{In our maximally twisted-mass setup, in which the Wilson $r$-parameters $r_{f_1}$ and $r_{f_2}$ are always chosen to be opposite $r_{f_1} =  - r_{f_2}$ (see Appendix~\ref{sec:appA}), the vector (axial) weak current in the physical basis renormalises multiplicatively with the RC $Z_A$ ($Z_V$) of the axial (vector) current for Wilson-like fermions, i.e.~$Z_V^{(0)} = Z_A$ and $Z_A^{(0)} = Z_V$ (see Appendix~\ref{sec:appD}).}
 \be
     J^\rho_W(x) = \overline{q}_{f_2}(x) \gamma^\rho \left[ Z_V^{(0)} - Z_A^{(0)} \gamma^5 \right] q_{f_1}(x) ~ ,
     \label{eq:V-A}
  \ee
$j^{\mathrm{em}}_\mu$ is the (lattice) conserved e.m.~current\footnote{The use of the conserved e.m.~current guarantees the absence of additional contact terms in the product $j^{\mathrm{em}}_\mu(y_1) j^{\mathrm{em}}_\nu(y_2)$.}
 \bea
     j^{\mathrm{em}}_\mu(y) & = & \sum_f e_f ~ \frac{1}{2} \left[ \bar{q}_f(y) (\gamma_\mu - i \tau^3 \gamma_5 ) 
                                                     U_\mu(y) q_f(y + a \hat{\mu}) \right. \nonumber \\
                                           & + & \left. \bar{q}_f(y + a \hat{\mu}) (\gamma_\mu + i \tau^3 \gamma_5 ) U_\mu^\dagger(y) q_f(y) \right]
     \label{eq:Jmu}
 \eea
and $T_\mu^{\mathrm{em}}$ is the tadpole operator
 \bea
     T_\mu^{\mathrm{em}}(y) & = & \sum_f e_f^2 ~ \frac{1}{2} \left[ \bar{q}_f(y) (\gamma_\mu - i \tau^3 \gamma_5 ) 
                                                       U_\mu(y) q_f(y+ a \hat{\mu}) \right. \nonumber \\
                                              & - & \left. \bar{q}_f(y + a \hat{\mu}) (\gamma_\mu + i \tau^3 \gamma_5 ) U_\mu^\dagger(y) 
                                                       q_f(y) \right] ~ .
     \label{eq:tadpole}
 \eea
In Eqs.\,(\ref{eq:deltaC_J})\,-\,(\ref{eq:deltaC_S}) $\phi^\dagger_P(\vec{x}, - t) = i \overline{q}_{f_1}(\vec{x}, - t) \gamma_5 q_{f_2}(\vec{x}, - t)$ is the interpolating field for a P-meson composed by two valence quarks $f_1$ and $f_2$ with charges $e_1 e$ and $e_2 e$. The Wilson $r$-parameters $r_{f_1}$ and $r_{f_2}$ are always chosen to be opposite $r_{f_1}=  - r_{f_2}$ (see Appendix~\ref{sec:appA}).
We have also chosen to place the weak current at the origin and to create the P-meson at a negative time $- t$, where $t$ and $T - t$ are sufficiently large to suppress the contributions from heavier states and from the backward propagating P-meson (this latter condition may be convenient but is not necessary). 
In Eq.\,(\ref{eq:deltaC_S}) $Z_m^{(0)}$ is the mass RC in pure QCD, which for our maximally twisted-mass setup is given by $Z_m^{(0)} = 1 / Z_P^{(0)}$, where $Z_P^{(0)}$ is the RC of the pseudoscalar density determined in Ref.\,\cite{Carrasco:2014cwa}.
The quantity ${\cal{Z}}^f_m$ is related to the e.m.~correction to the mass RC
\be
    Z_m^\mathrm{QCD+QED} = \left( 1 -  \frac{\alphaem}{4\pi} {\cal{Z}}_m^f \right)  Z_m^{(0)} + {\cal{O}}(\alphaem^m \alpha_s^n) 
                                                  \qquad \qquad (m > 1, ~ n \geq 0) 
\ee
and can be written in the form
\be
     {\cal{Z}}_m^f = {\cal{Z}}_\mathrm{QED}^f ~ Z_m^\mathrm{fact} ~ ,
     \label{eq:Zf}
\ee
where ${\cal{Z}}_\mathrm{QED}^f$ is the pure QED contribution at leading order in $\alphaem$, given in the $\msbar$ scheme at a renormalisation scale $\mu$ by\,\cite{Martinelli:1982mw,Aoki:1998ar}
 \be
     {\cal{Z}}_\mathrm{QED}^f(\overline{\rm MS}, \mu) = e_f^2  ~ \bigl( 6 \, \mbox{log}(a \mu) - 22.5954 \bigl) ~ ,
      \label{eq:Zf_em}  
 \ee
where $e_f$ is the fractional charge of the quark $q_f$ and $Z_m^\mathrm{fact}$ takes into account all the corrections of order ${\cal{O}}(\alpha_s^n)$ with $n \geq 1$.

The quantity $Z_m^\mathrm{fact}$ is computed non-perturbatively in section \ref{sec:Wreg} and represents the QCD corrections to the ``naive factorisation'' approximation ${\cal{Z}}_m^f = {\cal{Z}}_\mathrm{QED}^f$ (i.e.~$Z_m^\mathrm{fact} = 1$) introduced in Refs.\,\cite{Giusti:2017dmp,Giusti:2017jof}.

Analogously, the term $\left[ \delta A_P \right]^{\mathrm{SIB}}$ and $\left[ \delta M_P \right]^{\mathrm{SIB}}$ can be extracted from the correlator
 \be
      \delta C_P^{\mathrm{SIB}}(t) = - \sum_{f = f_1, f_2} \frac{\widehat{m}_f -  m_f}{Z_m^{(0)}} \cdot \sum_{\vec{x}, y} 
                                                        \langle 0| T \left \{ J^\rho_w(0) ~  \left[ \overline{q}_f(y) q_f(y) \right] ~ 
                                                        \phi_P^\dagger(\vec{x}, - t) \right \} | 0 \rangle \frac{p^\rho_P}{M_P} ~ , ~
      \label{eq:deltaC_QCD}
 \ee
where, following the notation of Ref.\,\cite{Giusti:2017dmp}, we indicate with $\widehat{m}_f$ and $m_f$ the renomalized masses of the quark with flavour $f$ in the full theory and in isosymmetric QCD only, respectively.
We stress again that the separation between QCD and QED corrections is prescription dependent and in this work we adopt the GRS prescription of Refs.\,\cite{deDivitiis:2013xla,Giusti:2017dmp,Giusti:2017dwk}, where
\bea
    \widehat{m}_u(\overline{\rm MS}, 2\,\mbox{\rm GeV})+\widehat{m}_d(\overline{\rm MS}, 2\,\mbox{\rm GeV}) &=& 
          2 \widehat{m}_{ud}(\overline{\rm MS}, 2\,\mbox{\rm GeV}) = 2 m_{ud}(\overline{\rm MS}, 2\,\mbox{\rm GeV}) \, ,  \nonumber \\ 
    \widehat{m}_s(\overline{\rm MS}, 2\,\mbox{\rm GeV}) =  m_s(\overline{\rm MS}, 2\,\mbox{\rm GeV})\,,&\mbox{}&    \widehat{m}_c(\overline{\rm MS}, 2\,\mbox{\rm GeV}) =  m_c(\overline{\rm MS}, 2\,\mbox{\rm GeV})\,. 
\eea
Thus, in Eq.\,(\ref{eq:deltaC_QCD}), the only relevant quark mass difference is $\widehat{m}_d -  m_{ud} = - (\widehat{m}_u -  m_{ud})$, whose value in the $(\overline{\rm MS}, 2\,\mbox{GeV})$ scheme was found to be equal to $1.19\,(9)$ MeV~\cite{Giusti:2017dmp} using as inputs the experimental values of the charged and neutral kaon masses.

 Following Ref.\,\cite{deDivitiis:2013xla} we form the ratio of $\delta C_P^i(t)$ with the corresponding tree-level correlator
 \be
      C_P^{(0)}(t) = \sum_{\vec{x}} \langle 0| T \left \{ J_W^\rho(0) \phi_P^\dagger(\vec{x}, -t) \right \} | 0 \rangle \frac{p_P^\rho}{M_P} 
      \label{eq:C0}
 \ee
and at large time distances $t$ we obtain ($i = J, T, P, S, QCD$)
 \bea
      \frac{\delta C_P^i(t)}{C_P^{(0)}(t)} &~& _{\overrightarrow{ t \gg a, (T-t) \gg a }} ~ \frac{\delta [G_P^i A_P^i]}{G_P^{(0)} A_P^{(0)}} + \nonumber \\
      && \frac{\delta M_P^i}{M_P^{(0)}} \left[ M_P^{(0)} \left( \frac{T}{2} - t \right)
            \frac{e^{-M_P^{(0)} t} + e^{-M_P^{(0)} (T-t)}}{e^{-M_P^{(0)} t} - e^{-M_P^{(0)} (T-t)}} - 1 - M_P^{(0)} \frac{T}{2}  \right] 
      \label{eq:ratio_i}
 \eea
where
\be
     G_P^{(0)} \equiv \langle 0 | \phi_P(0) | P^{(0)} \rangle
     \label{eq:ZP0}
\ee
is the coupling of the interpolating field of the P-meson with its ground-state in isosymmetric QCD. 
The term proportional to $\delta M_P^i$ in the r.h.s.~of Eq.\,(\ref{eq:ratio_i}) is related to the e.m.~and strong IB corrections of the meson mass.

The function in the square brackets on the r.h.s.~of Eq.\,(\ref{eq:ratio_i}) is an almost linear function of~$t$.
Thus, the correction to the P-meson mass, $\delta M_P^i$, can be extracted from the slope of the ratio $\delta C_P^i(t) / C_P^{(0)}(t)$ and the quantity $\delta [G_P^i A_P^i]$ from its intercept. 
As explained in Ref.\,\cite{Carrasco:2015xwa}, in order to obtain the quantity $\delta A_P^i$ the correction $\delta G_P^i$ is  separately determined by evaluating a correlator similar to those of Eqs.\,(\ref{eq:deltaC_J})\,-\,(\ref{eq:deltaC_S}), in which the weak operator $J_W^\rho p_P^\rho / M_P$ is replaced by the P-meson interpolating field $\phi_P$.

For illustration, in Fig.~\ref{fig:ratio_i} we show the ratios $C_P^i$ for the charged kaon ($P=K$) obtained from the ensemble D20.48 (see Appendix~\ref{sec:appA}). 
The top panel contains the ratio
$\delta C_{K}^{\mathrm{SIB}}(t) / C_{K}^{(0)}(t)$, the ratio $\delta C_{K}^J(t) / C_{K}^{(0)}(t)$ is shown in the middle panel and the ratios $\delta C_{K}^T(t) / C_{K}^{(0)}(t)$ and $\delta C_K^P(t) / C_K^{(0)}(t)$ are presented in the bottom panel. 

\begin{figure}[htb!]
\centering
\includegraphics[scale=0.65]{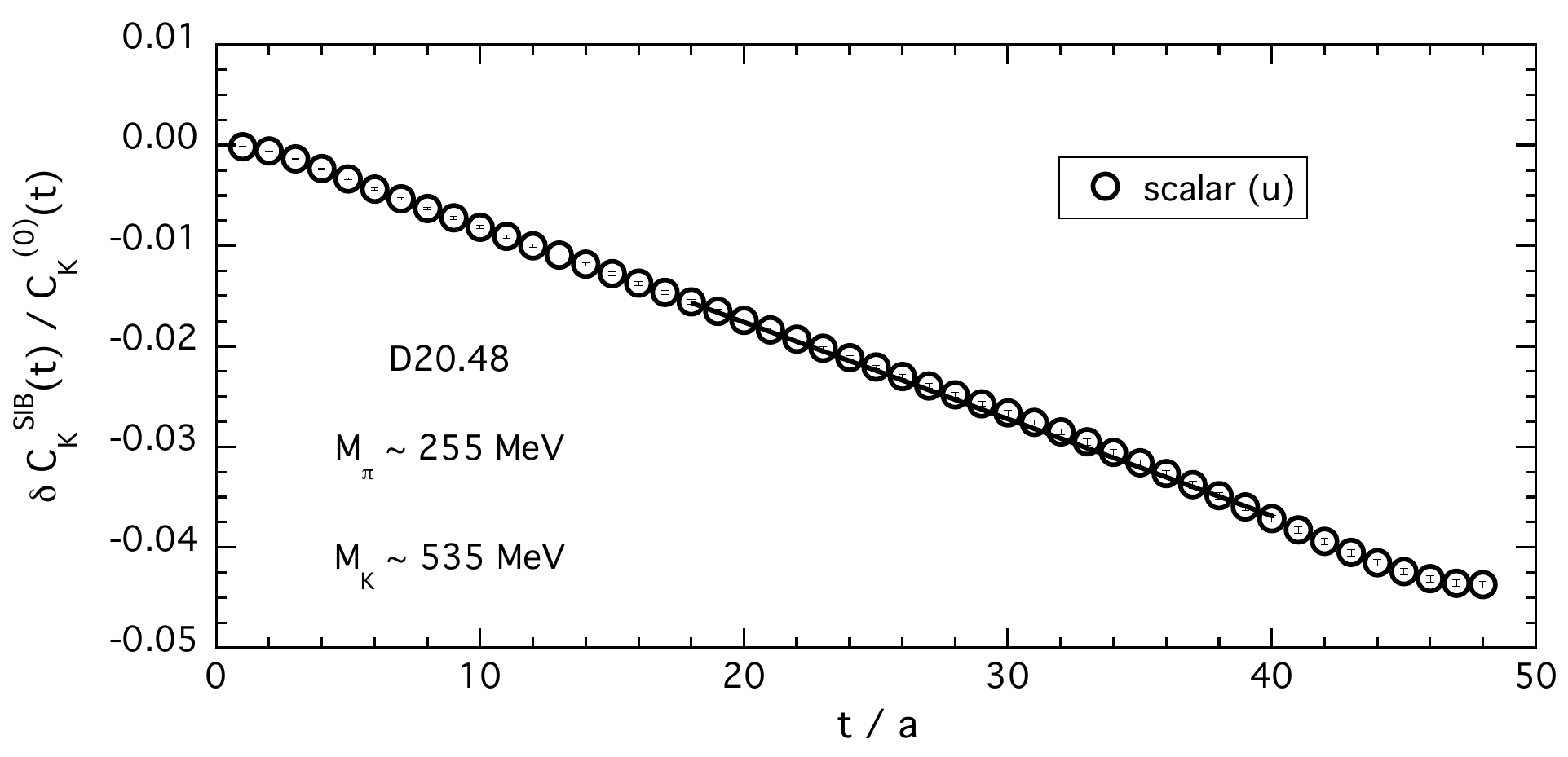}
\includegraphics[scale=0.65]{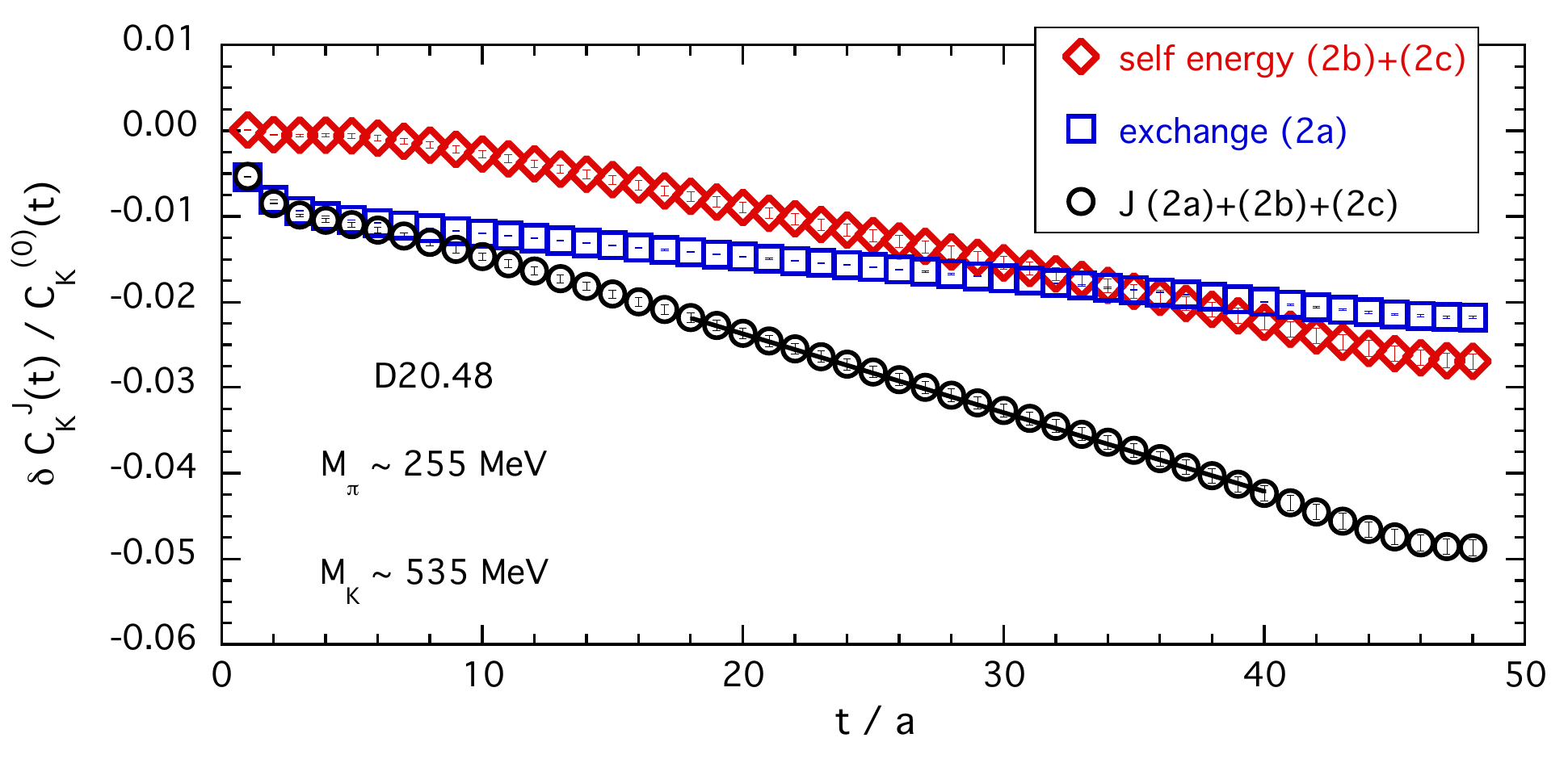}
\includegraphics[scale=0.65]{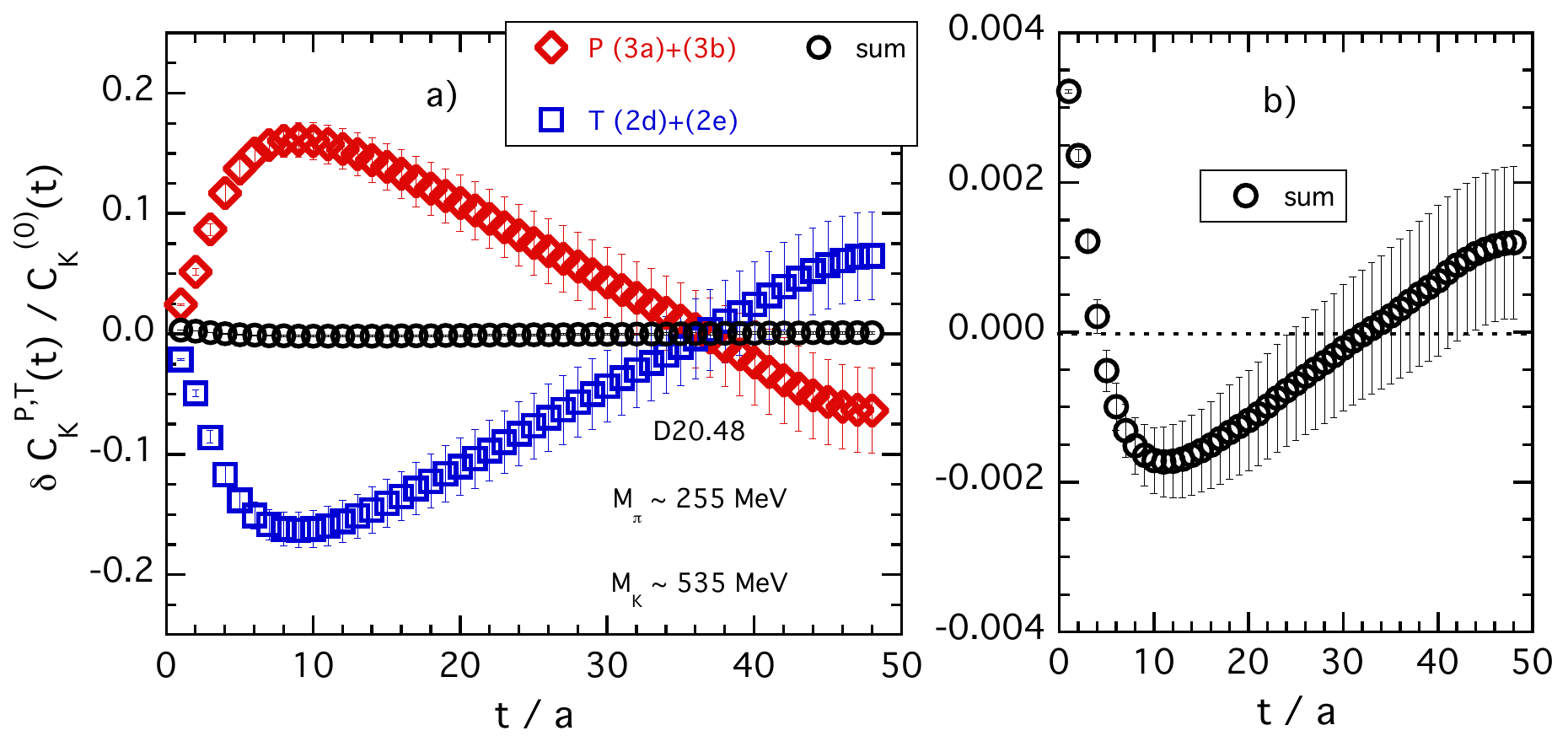}
\vspace{-0.50cm}
\caption{\it \footnotesize Top panel: The strong IB correction $\delta C_K^{\mathrm{SIB}}(t) / C_K^{(0)}(t)$ for the charged kaon obtained on the ensemble D20.48 (see Appendix~\ref{sec:appA}). The solid line is the ``linear" fit (\protect\ref{eq:ratio_i}) applied in the time interval where the ground-state is dominant. Middle panel: contributions of the exchange (\ref{fig:diagrams_1}a) and self-energy (\ref{fig:diagrams_1}b)+(\ref{fig:diagrams_1}c) diagrams. The circles represent the sum (\ref{fig:diagrams_1}a)+(\ref{fig:diagrams_1}b)+(\ref{fig:diagrams_1}c), i.e.~the ratio $\delta C_K^J(t) / C_K^{(0)}(t)$. Bottom panel: contributions of the tadpole operator $\delta C_K^T(t) / C_K^{(0)}(t)$, i.e.~diagrams (\ref{fig:diagrams_1}d)+(\ref{fig:diagrams_1}e), and of the e.m.~shift of the critical mass $\delta C_K^P(t) / C_K^{(0)}(t)$, i.e.~diagrams (\ref{fig:diagrams_2}a)+(\ref{fig:diagrams_2}b). The sum $\delta [C_K^T(t) + C_K^P(t) ] / C_K^{(0)}(t)$, shown by the circles, is non vanishing and it is determined quite precisely (see the right-hand plot where it is presented on an expanded scale). Errors are statistical only.\hspace*{\fill}}
\label{fig:ratio_i}
\end{figure}
We find: i) the contributions $\delta C_K^T(t) / C_P^{(0)}(t)$ and $\delta C_K^P(t) / C_P^{(0)}(t)$ are separately large, but strongly correlated, 
since the tadpole insertion dominates the values of the e.m.~shift of the critical mass $\delta m_f^{\mathrm{crit}}$ (see Ref.\,\cite{Giusti:2017dmp}). In the chiral limit they would cancel, but at finite masses the sum is small and linear in $t$. Because of the correlations it can nevertheless, be determined quite precisely (see the bottom right-hand plot of Fig.\,\ref{fig:ratio_i}) where the sum is presented on an expanded scale.
ii) the time dependence of the ratio $\delta C_K^J(t) / C_K^{(0)}(t)$ is almost linear in the time interval where the ground state is dominant. 

\subsection{Crossed diagrams and lepton self-energy}
\label{sec:crossed}

The evaluation of the diagrams~\ref{fig:diagrams_4}(a)\,-\,(b), corresponding to the term $\delta A_P^\ell$ in Eq.\,(\ref{eq:deltaAPS}), can be obtained by studying the correlator~\cite{Carrasco:2015xwa} 
 \bea
     \delta C_P^\ell(t) & = & - 4 \pi \alphaem \sum_{\vec{x}, x_1, x_2} \langle 0|\,T\left\{ J_W^\rho(0) j_\mu^{\mathrm{em}}(x_1) 
                                           \phi_P^\dagger(\vec{x}, -t) \right\} |0 \rangle \Delta_{\mu \nu}^{\mathrm{em}}(x_1, x_2) 
                                           e^{E_\ell t_2 - i \vec{p}_\ell \cdot \vec{x}_2} \nonumber \\
                                & \cdot & \overline{u}(p_\nu) \gamma_\rho (1 - \gamma_5) S^\ell(0, x_2) \gamma_\nu 
                                                v(p_\ell) \left[ \overline{v}(p_\ell) \gamma_\sigma (1 - \gamma_5) u(p_\nu) 
                                                \frac{p_P^\sigma}{M_P} \right] ~ ,
     \label{eq:Cql}
 \eea
where $S^\ell(0, x_2)$ stands for the free twisted-mass propagator of the charged lepton.  
For the numerical analysis we have found it to be convenient to saturate the Dirac indices by inserting on the r.h.s.~of Eq.\,(\ref{eq:Cql}) the factor $\left[ \overline{v}(p_\ell) \gamma_\sigma (1 - \gamma_5) u(p_\nu) \right]$, which represents the lowest order ``conjugate" leptonic ($V - A$) amplitude, and to sum over the lepton polarizations.
In this way we are able to study the time behaviour of the single function $\delta C_P^\ell(t)$.

The corresponding correlator at lowest order ($O(\alpha_\mathrm{em}^0)$) is
 \be
     C_P^{\ell(0)}(t) = \sum_{\vec{x}} \langle 0|\,T\left\{ J_W^\rho(0) \phi_P^\dagger(\vec{x}, -t) \right\} |0 \rangle ~   
                                 \overline{u}(p_\nu) \gamma_\rho (1 - \gamma_5) v(p_\ell) 
                                 \left[ \overline{v}(p_\ell) \gamma_\sigma (1 - \gamma_5) u(p_\nu) 
                                 \frac{p_P^\sigma}{M_P} \right] \, .
     \label{eq:Cql_tree}
 \ee
In Eqs.\,(\ref{eq:Cql}) and (\ref{eq:Cql_tree}) the contraction between the weak hadronic current $J_W^\rho(0)$ [see Eq.\,(\ref{eq:V-A})] and its leptonic ($V - A$) counterpart gives rise to two terms corresponding to the product of either the temporal or spatial components of these two weak currents, which are odd and even under time reversal, respectively. 
Thus, on a lattice with finite time extension $T$, for $t \gg a$ and $(T - t) \gg a$ one has
 \be
      \delta C_P^\ell(t) ~ _{\overrightarrow{ t \gg a, (T-t) \gg a }} ~ \frac{G_P^{(0)}}{2 M_P^{(0)}} \sum_{j =0}^4 \delta A_P^{\ell,j} ~ X_P^{\ell,j} ~ 
                                  \left[ e^{-M_P^{(0)} t} + s_j e^{- M_P^{(0)} (T - t)} \right] ~ ,
      \label{eq:crossed_larget}
 \ee
where $s_0 = -1$, $s_{1,2,3} = 1$ and 
 \be
     \label{eq:XPell}
     X_P^{\ell,j} = \mbox{Tr}\left[  \gamma_j (1 - \gamma_5) \ell  \overline{\ell} \gamma_0 (1 - \gamma_5) \nu \overline{\nu} \right]
    \ee
is the relevant leptonic trace evaluated on the lattice using for the charged lepton the free twisted-mass propagator and for the neutrino the free Wilson propagator in the P-meson rest frame [${p_P^\sigma = (M_P, \vec{0})}$]. 

Similarly, for the lowest-order correlator one has
 \be
      C_P^{\ell(0)}(t) ~ _{\overrightarrow{ t \gg a, (T-t) \gg a }} ~ \frac{G_P^{(0)}}{2 M_P^{(0)}} A_P^{(0)} ~ X_P^{\ell,0} ~ 
                               \left[ e^{-M_P^{(0)} t} - e^{- M_P^{(0)} (T - t)} \right] ~ ,
      \label{eq:LO_larget}
 \ee
where $A_P^{(0)}$ is the renormalised axial amplitude evaluated on the lattice in isosymmetric QCD in the P-meson rest frame, namely 
 \be
     Z_A^{(0)} \langle 0 | \bar{q}_2 \gamma_j \gamma_5 q_1 | P^{(0)} \rangle = \delta_{j, 0} A_P^{(0)} ~ .
     \label{eq:AP0_ren}
 \ee

The effect of the different signs of the backward-propagating signal in Eq.\,(\ref{eq:crossed_larget}) can be removed by introducing the following new correlators:
\bea
    \label{eq:backward_delta}
     \delta \overline{C}_P^\ell(t) & \equiv & \frac{1}{2} \left\{ \delta C_P^\ell(t) + \frac{\delta C_P^\ell(t-1) - 
                                                                  \delta C_P^\ell(t+1)}{e^{M_P^{(0)} } - e^{-M_P^{(0)}}} \right\} ~ 
                                                                  _{\overrightarrow{ t \gg a, (T-t) \gg a }} ~ \delta A_P^\ell \, X_P^{\ell,0}  
                                                                  \frac{G_P^{(0)}}{2 M_P^{(0)}} e^{-M_P^{(0)} t} ~ , \nonumber \\
     \label{eq:backward_tree}
     \overline{C}_P^{\ell(0)}(t) & \equiv & \frac{1}{2} \left\{ C_P^{\ell(0)}(t) + \frac{C_P^{\ell(0)}(t-1) - 
                                                              C_P^{\ell(0)}(t+1)}{e^{M_P^{(0)} } - e^{-M_P^{(0)}}} \right\} ~ 
                                                              _{\overrightarrow{ t \gg a, (T-t) \gg a }} ~  A_P^{(0)} X_P^{\ell,0} \frac{G_P^{(0)}}{2 M_P^{(0)}} 
                                                              e^{-M_P^{(0)} t} ~ , ~ \qquad
\eea
 where
 \be
      \delta A_P^\ell  = \frac{1}{X_P^{\ell,0}} \sum_{j =0}^4 \delta A_P^{\ell,j} ~ X_P^{\ell,j} ~ .
      \label{eq:deltaAP_ell}
 \ee
Thus, the quantity $\delta A_P^\ell / A_P^{(0)}$ can be extracted from the plateau of the ratio $\delta \overline{C}_P^\ell(t) / \overline{C}_P^{\ell(0)}(t)$ at large time separations, viz.
  \be
      \frac{\delta \overline{C}_P^\ell(t)}{\overline{C}_P^{\ell(0)}(t)} ~ _{\overrightarrow{ t \gg a, (T-t) \gg a }} ~ \frac{\delta A_P^\ell}{A_P^{(0)}} ~ .
      \label{eq:deltaR}
 \ee 
Note that the diagrams in Fig.~\ref{fig:diagrams_4}(a)\,-\,(b) do not contribute to the electromagnetic corrections to the masses of the mesons and therefore the ratio (\ref{eq:deltaR}) has no slope in $t$ in contrast to the ratios (\ref{eq:ratio_i}).
Moreover, the explicit calculation of $X_P^{\ell,j}$ on the lattice is not required.

In terms of the lattice momenta $a\widetilde{p}_\ell$  and $a\overline{p}_\ell$, defined as
 \bea
      a\widetilde{p}_\ell & = & \sqrt{ \sum_{k=1,2,3} \mbox{sin}^2\left( ap_{\ell k} \right)} ~ , \\
      a\overline{p}_\ell & = & 2 ~ \sqrt{ \sum_{k=1,2,3} \mbox{sin}^2\left( \frac{ap_{\ell k}}{2} \right)} ~ , 
      \label{eq:ptilde}
  \eea
the energy-momentum dispersion relations for the charged lepton and the neutrino in the P-meson rest frame are given by 
   \bea
      a\widetilde{E}_\ell  & = & 2 ~ \mbox{arcsinh}\left[ \frac{1}{2} \sqrt{ \frac{a^2 m_\ell^2 + 
                                             a^2 \widetilde{p}_\ell^2 + a^4 \overline{p}_\ell^4 / 4}
                                             {1 + a^2 \overline{p}_\ell^2 / 2 } } \right]  ~ , \\[2mm]
      a\widetilde{E}_\nu & = & \mbox{arcsinh}(a\widetilde{p}_\ell)  \, .  
      \label{eq:kinem}
 \eea
The 3-momentum of the final-state lepton  $\vec {p}_\ell$ ($\vec p_\nu = -\vec p_\ell$) must be chosen to satisfy the equation 
 \be
      \widetilde{E}_\ell + \widetilde{E}_\nu =  M_P^{(0)} \, .
      \label{eq:p_lepton}
 \ee
Thus, for any given simulated P-meson mass $M_P^{(0)}$, the 3-momentum $\vec {p}_\ell = | \vec {p}_\ell | (1,1,1)$, is calculated from Eq.\,(\ref{eq:p_lepton}) and is injected on the lattice using non-periodic boundary conditions \cite{Bedaque:2004kc,deDivitiis:2004kq} for the lepton field.
A simple calculation yields
\be
      X_P^{\ell,0} = \mbox{Tr}\left[  \gamma_0 (1 - \gamma_5) \ell  \overline{\ell} \gamma_0 (1 - \gamma_5) 
                             \nu \overline{\nu} \right] = 8 a\widetilde{p}_\ell \left[ \mbox{sinh}(a\widetilde{E}_\ell) - a\widetilde{p}_\ell \right] ~ .
      \label{eq:lepton_trace} 
\ee

In Fig.\,\ref{fig:backward} we show the correlators $C_\pi^{\mu(0)}(t)$, $\delta C_\pi^\mu(t)$, $\overline{C}_\pi^{\mu(0)}(t)$ and $\delta \overline{C}_\pi^\mu(t)$ for $\pi_{\mu 2}$ decays, multiplied by the ground-state exponential. These were obtained on the gauge ensembles A40.24 and D30.48 of Appendix~\ref{sec:appA}.
\begin{figure}[htb!]
\centering
\includegraphics[scale=0.85]{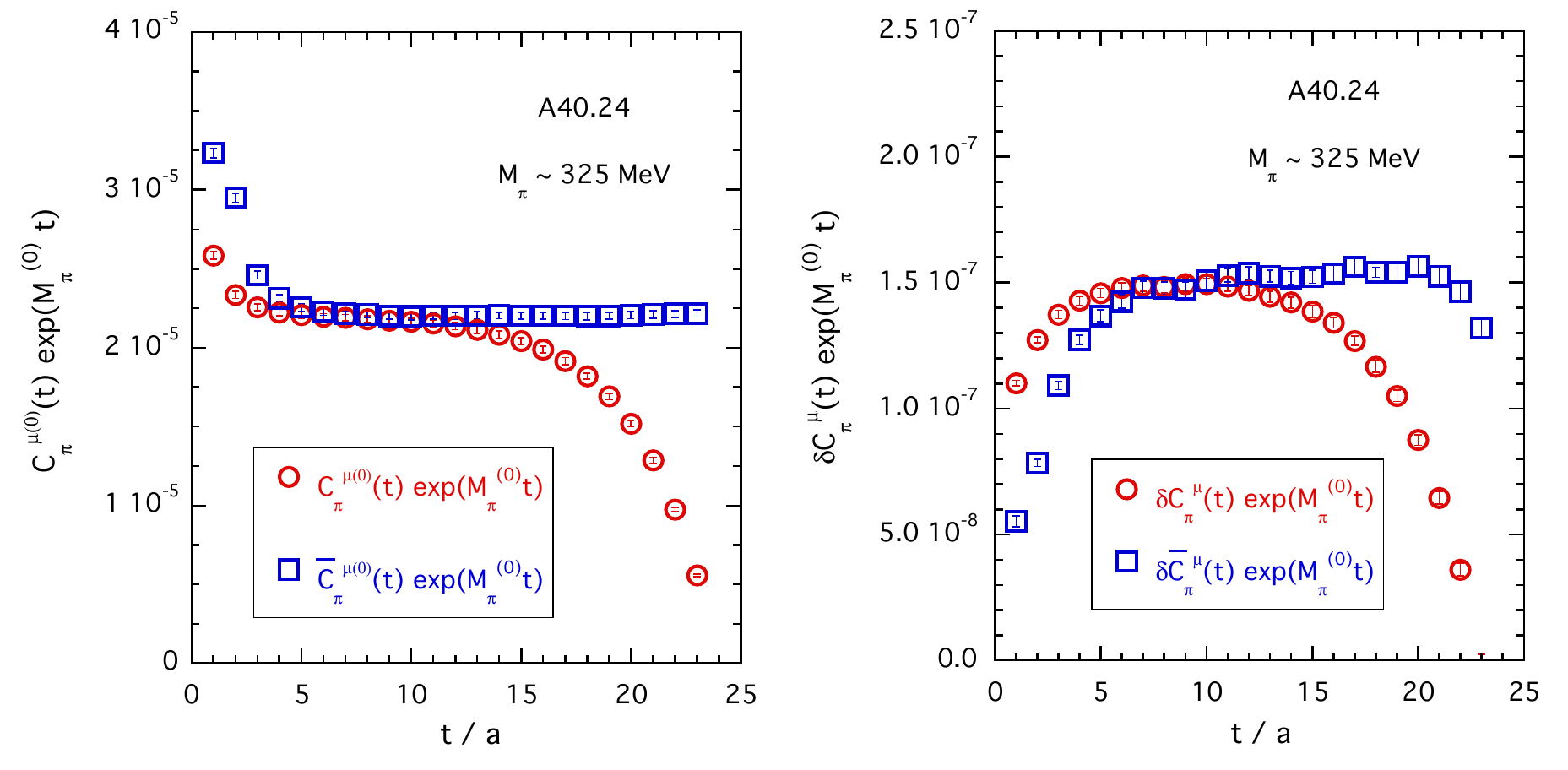}
\includegraphics[scale=0.85]{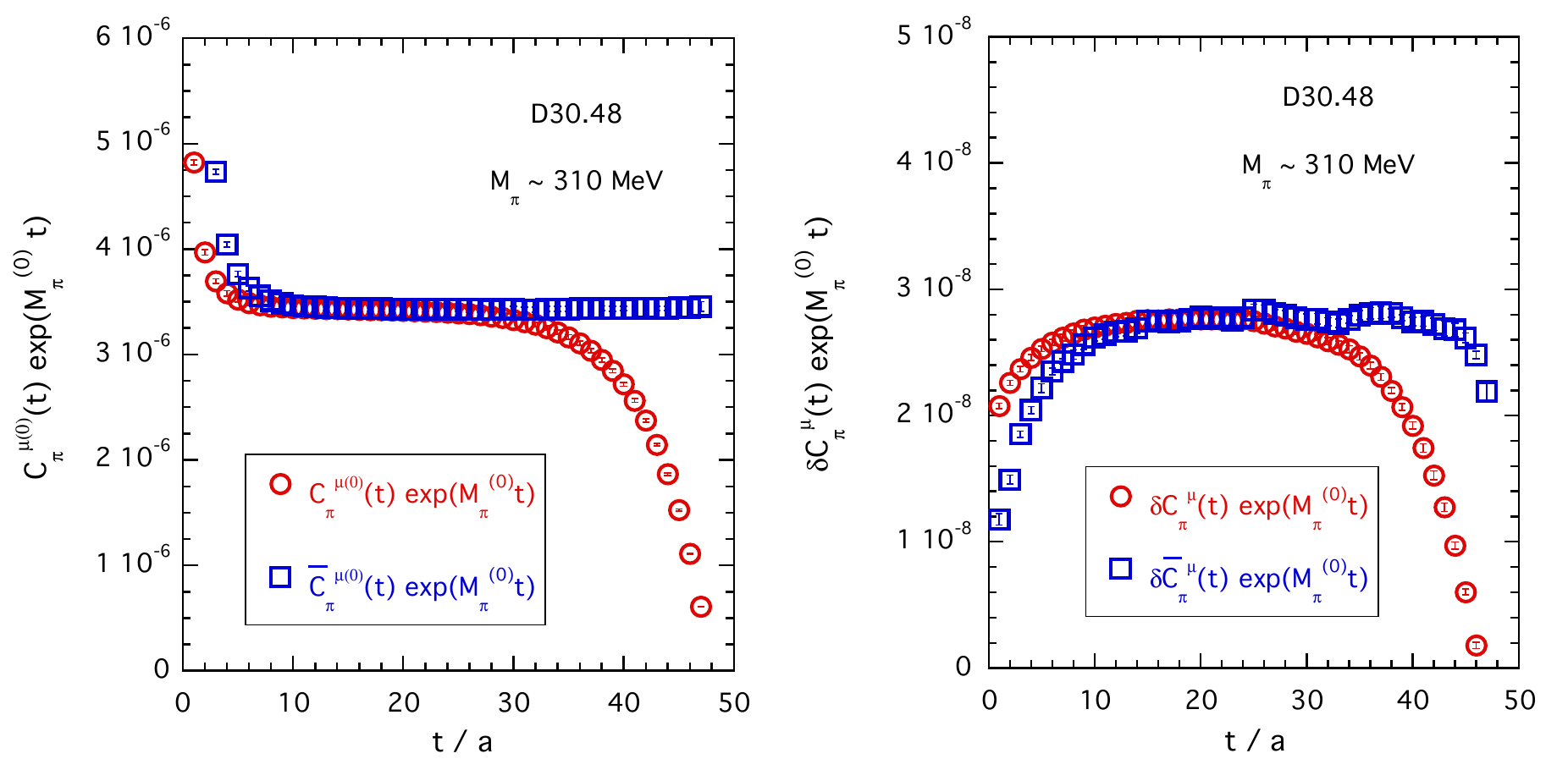}
\vspace{-0.5cm}
\caption{\it \footnotesize Time dependence of the correlators $C_\pi^{\mu(0)}(t)$ (left panels) and $\delta C_\pi^\mu(t)$ (right panels) for $\pi_{\mu 2}$ decays. These are given in lattice units and multiplied by the ground-state exponential, and were obtained from gauge ensemble A40.24 (top panels) and D30.48 (bottom panels). The blue squares represent the correlators $\delta \overline{C}_\pi^\mu(t)$ and $\overline{C}_\pi^{\mu(0)}(t)$ given by Eqs.\,(\ref{eq:backward_delta})\,-\,(\ref{eq:backward_tree}). Errors are statistical only. For details of the simulations see Appendix~\ref{sec:appA}.\hspace*{\fill}}
\label{fig:backward}
\end{figure}
The subtraction of the backward signals, needed for extracting directly the quantity $\delta A_P^\ell$ given by Eq.\,(\ref{eq:deltaAP_ell}), is beneficial also for extending the time region from which $\delta A_P^\ell$ (as well as the ratio $\delta A_P^\ell / A_P^{(0)}$) can be determined.

The quality of the signal for the ratio $\delta \overline{C}_P^\mu(t) / \overline{C}_P^{\mu(0)}(t)$ is illustrated in Fig.~\ref{fig:external} for charged kaon and pion decays into muons for the case of the ensembles B55.32 and D30.48.
\begin{figure}[htb!]
\centering
\includegraphics[scale=0.70]{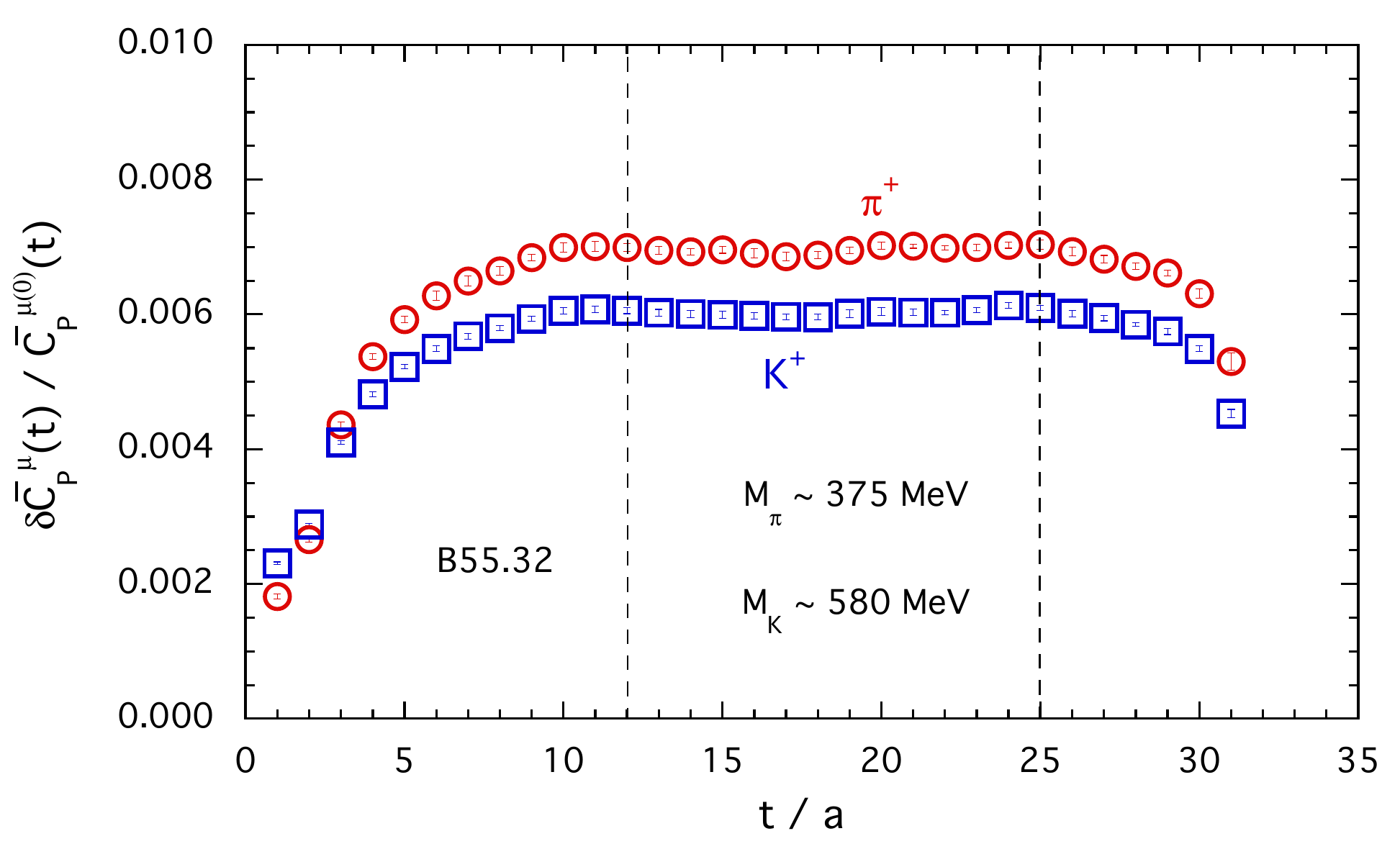}
\includegraphics[scale=0.70]{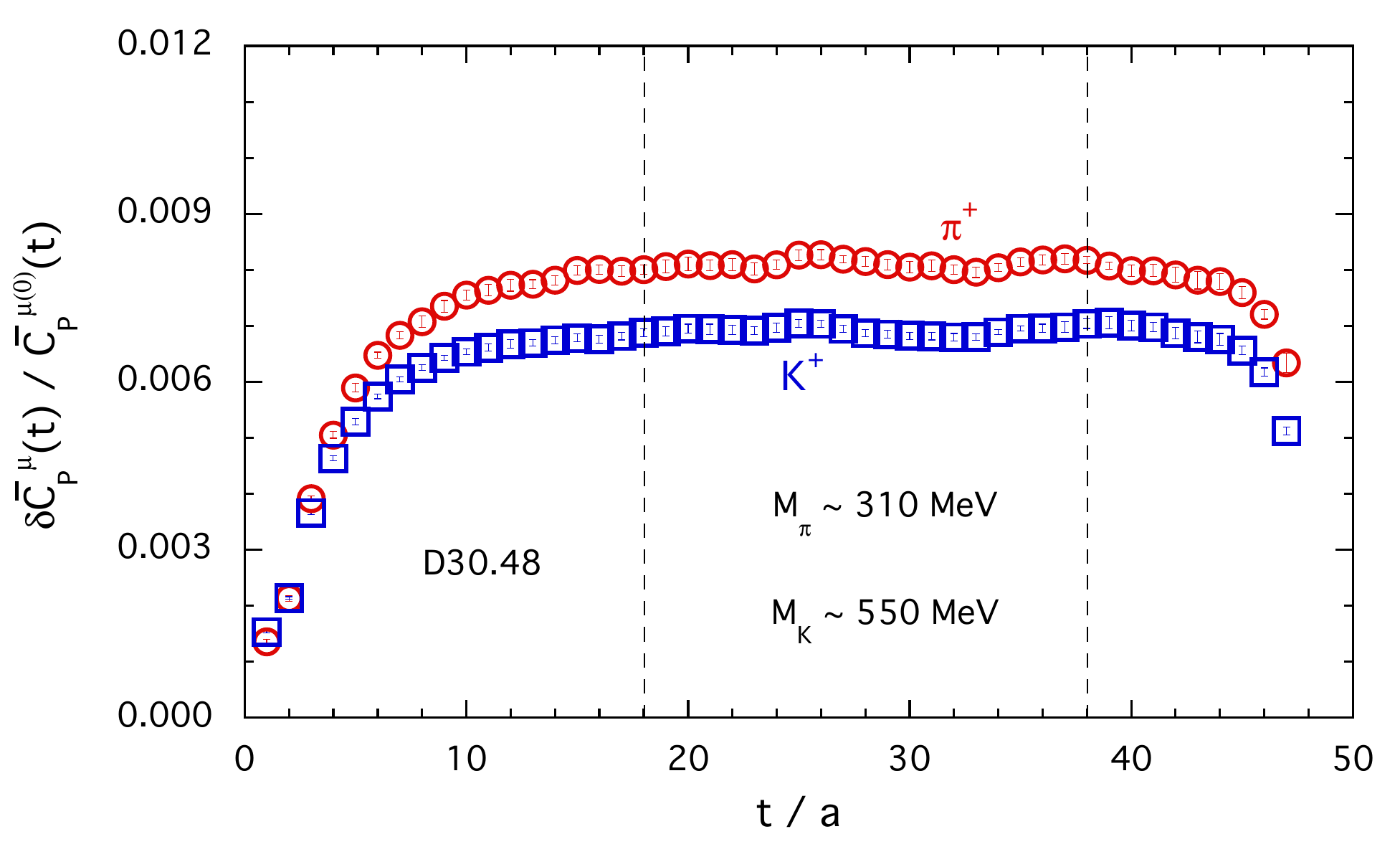}
\vspace{-0.25cm}
\caption{\it \footnotesize Results for the ratio $\delta \overline{C}_P^\mu(t) / \overline{C}_P^{\mu(0)}(t)$, given by Eq.\,(\ref{eq:deltaR}), for $K_{\mu 2}$ and $\pi_{\mu 2}$ decays obtained from the gauge ensembles B55.32 (top panel) and D30.48 (bottom panel). The vertical dashed lines indicate the time region used for the extraction of the ratio $\delta A_P^\mu / A_P^{(0)}$. Errors are statistical only.\hspace*{\fill}}
\label{fig:external}
\end{figure}

The calculation of the correction due to the diagram~\ref{fig:diagrams_4}(c) is straightforward, since it is obtained by simply multiplying the lowest order amplitude, $A_P^{(0)}$, by the one-loop lepton self-energy evaluated on the lattice.

\section{Renormalisation of the effective Hamiltonian and chirality mixing}
\label{sec:Wreg}

In this section we provide the basic formalism to derive the e.m.~corrections to the RCs non-perturbatively;  further details of the calculation will be presented in a forthcoming publication~\cite{dicarlo}. This procedure relates the bare lattice operators to those in the \RIMOMprime~(and similar) renormalisation schemes up to order ${\cal{O}}(\alphaem)$ and to all orders in $\alpha_s$. We also improve the precision of the matching of the weak operator $O_1$ (see Eq.\,(\ref{eq:O1bare})) renormalised in the \RIMOMprime~scheme to that in the W-regularisation by calculating the coefficient of the term proportional to $\alphaem \alpha_s\log(M_W^2/\mu^2)$ in the matching coefficient. Using the two-loop anomalous dimension thus determined,  we can evolve the operator to the renormalisation scale of $M_W$. Following this calculation the error due to renormalisation is reduced from order ${\cal{O}}(\alphaem \alpha_s(1/a))$ to order ${\cal{O}}(\alphaem \alpha_s(M_W))$.

The effective Hamiltonian, including the perturbative electroweak matching with the Standard Model~\cite{Sirlin:1981ie}, can be written in the form
\be
    {\cal H}_W = \frac{G_F}{\sqrt{2}}\,V_{q_1 q_2}^\ast \, \left[1 + \frac{\alphaem}{\pi}\, 
                         \mbox{log}\left( \frac{M_Z}{M_W} \right) \right] O_1^\textrm{W-reg}(M_W)  \, , 
    \label{eq:DeltaL0b}
\ee
where the term proportional to the logarithm has been already included in Eq.\,(\ref{eq:RPS}) and $O_1^\textrm{W-reg}(M_W) $ is the operator renormalised in the W-regularisation scheme, which is used to regularise the photon propagator. 
Since the W-boson mass is too large to be simulated on the lattice, a matching of the lattice weak operator $O_1$ to the W-regularisation scheme is necessary. In addition, for lattice formulations which break chiral symmetry, like the one used in the present study, the lattice weak operator $O_1$ mixes with other four-fermion operators of different chirality.

\subsection{The renormalised weak operator in the W-regularisation scheme}
\label{sec:OWreg}

In order to obtain the operator renormalised in the W-regularisation scheme, we start by renormalising the lattice four-fermion operator $O_1$ defined in Eq.\,(\ref{eq:O1bare})
in  the \RIMOMprime~scheme\,\cite{Martinelli:1994ty}, obtaining $O_1^{\text{\RIprime}}(\mu)$, and then perturbatively match the operator $O_1^{\text{\RIprime}}(\mu)$ to the one in the W-regularization\,\cite{Carrasco:2015xwa}
\be 
    O^{\textrm{W-reg}}_1(M_W) = Z^{\textrm{W-\RIprime}}\left(\frac{M_W}{\mu}, \alpha_s(\mu), \alpha_\mathrm{em} \right) \, 
                                                     O_1^{\text{\RIprime}}(\mu) \, .  
    \label{eq:O1Wreg}
\ee
The coefficient $Z^{\textrm{W-\RIprime}}\left(M_W/\mu, \alpha_s(\mu), \alpha_\mathrm{em} \right)$ can be computed by first  evolving  the operator in the \RIprime~scheme to the scale $M_W$ and then matching it to the corresponding operator in the W-scheme. The  coefficient can therefore be  written as the product of a matching coefficient and an evolution operator
\be
Z^{\textrm{W-\RIprime}}\left(\frac{M_W}{\mu}, \alpha_s(\mu), \alpha_\mathrm{em} \right)  = Z^{\textrm{W-\RIprime}}\left(1, \alpha_s(M_W), \alpha_\mathrm{em} \right) \, U^{\textrm{\RIprime}}\left(M_W, {\mu},  \alpha_\mathrm{em} \right)\, . 
\ee
Below  we will only consider terms of first order in $\alphaem$ and,  therefore we will consistently neglect the running of $\alphaem$.  

We note that the original bare lattice operators and $ O^{\textrm{W-reg}}_1(M_W)$ are gauge invariant, and thus the corresponding matching  coefficients are gauge invariant. This is not the case for $O_1^{\text{\RIprime}}(\mu)$ that instead depends not only on the external states chosen to define the renormalisation conditions, but also on the gauge. Consequently the matching  coefficient $Z^{\textrm{W-\RIprime}}\left(\frac{M_W}{\mu}, \alpha_s(\mu), \alpha_\mathrm{em} \right)$ and the evolution operator $U^{\textrm{\RIprime}}\left(M_W, {\mu},  \alpha_\mathrm{em} \right)$ are in general gauge dependent. However, at the order of perturbation theory to which we are working, the evolution operator turns out to be both scheme and gauge independent.

In the following, we  discuss in turn the matching coefficient, $Z^{\textrm{W-\RIprime}}\left(1, \alpha_s(M_W), \alpha_\mathrm{em} \right)$, the evolution operator  $U^{\textrm{\RIprime}}\left(M_W, {\mu},  \alphaem \right)$,  and the definition of the renormalised operator $O_1^{\mathrm{\RIprime}}(\mu) $, which will be obtained non-perturbatively. 

\vskip 0.5 cm
{\bf a) The matching coefficient.} 
At first order (one loop) in $\alphaem$ 
\be
     Z^{\textrm{W-\RIprime}}\left( 1,\alpha_s(M_W), \alphaem \right) = 1 + \frac{\alphaem}{4\pi} \,  C^\textrm{W-\RIprime} \, ,
    \label{eq:ZWRIprime}
\ee
where the strong interaction corrections for the \RIMOMprime~operator vanish, at this order, because of the Ward identities of the quark vector and axial vector currents appearing in the operator $O_1$  in the massless limit. We recall that we currently do not include terms of $O(\alpha_s(M_W)\alphaem)$ in the matching coefficient $Z^{\textrm{W-\RIprime}}$.

\vskip 0.5 cm
{\bf  b) The evolution operator.}
The evolution operator $U^{\textrm{RI}^\prime}\left(M_W, {\mu}, \alpha_\mathrm{em} \right)$ is the solution of the renormalisation group equation
\be
\label{eq:RGE}
\left[ \mu^2 \frac{\partial}{\partial \mu^2} + \beta(\alpha_s, \alphaem)\frac{\partial}{\partial \alpha_s} \right]U^{\textrm{\RIprime}}\left(M_W, {\mu},  \alpha_\mathrm{em} \right) =\gamma(\alpha_s, \alphaem)\,  U^\textrm{\RIprime}\left(M_W, {\mu},  \alpha_\mathrm{em} \right)\, , 
\ee
where $U^\textrm{\RIprime}\left(M_W, {\mu},  \alpha_\mathrm{em} \right)$ satisfies the initial condition $U^{\textrm{\RIprime}}\left(M_W, M_W,  \alpha_\mathrm{em} \right)=1$,  $\gamma(\alpha_s, \alphaem)$ is, in general, the anomalous dimension matrix\,\cite{Buras:1993dy,Ciuchini:1993vr}, although in our particular case it is actually a number (and not a matrix), and $\beta(\alpha_s,\alphaem)$ is the QCD $\beta$-function:
\be
\label{eq:beta(g)}
\beta(\alpha_s, \alphaem) = \frac{d\alpha_s}{d \log\mu^2}= - \betazero \, \frac{\alpha_s^2 }{4\pi}- \betaone\, \frac{\alpha_s^3}{(4\pi)^2}-\beta_1^{se}\,\frac{\alpha_s^2 \alphaem}{(4\pi)^2}~,
\ee
with 
\be
\betazero = 11 - \frac{2}{3} N_f ~ , \quad \betaone = 102 - \frac{38}{3}N_f ~, \quad \beta_1^{se} = -\frac{8}{9} \left( N_u + \frac{N_d}{4} \right )\, ,
\ee
where $N_f$ denotes the number of active flavours, and $N_u$ and  $N_d$ denote the number of up-like and down-like  active quarks respectively  so that $N_f = N_u + N_d$. We may expand $\gamma(\alpha_s, \alphaem)$ in powers of the couplings as follows 
\be
\gamma(\alpha_s,\alphaem) = \frac{\alpha_s}{4\pi} \, \gamma_s^{(0)} + \frac{\alpha_s^2}{(4\pi)^2} \, \gamma_s^{(1)} + \frac{\alphaem}{4\pi} \, \gamma_e^{(0)} +  \frac{\alpha_s \alphaem}{(4\pi)^2} \, \gamma_{se}^{(1)}~,
\ee
where $\gamma_{se}^{(1)}$ has been previously calculated in Ref.\,\cite{Brod:2008ss}.
In the case of the operator $O_1$  both $\gamma_s^{(0)}$ and $\gamma_s^{(1)}$ vanish whereas
\be
\gamma_e^{(0)} =  -2 ~ , \qquad  \gamma_{se}^{(1)} = +2 ~ \,. 
\label{eq:anom} 
\ee
It can be demonstrated that, in addition to the leading anomalous dimension $\gamma_e^{(0)}$, $\gamma_{se}^{(1)}$ is also independent of the renormalisation scheme, thus in particular it is the same in \RIprime~and in the W-regularisation schemes.  
It is then straightforward to derive $U^{\textrm{\RIprime}}\left(M_W, {\mu}, \alpha_\mathrm{em} \right)$
\bea
    U^{\textrm{\RIprime}}\left(M_W, {\mu}, \alpha_\mathrm{em} \right) & = & 1 - \frac{\alphaem}{4\pi} \gamma_e^{(0)}
         \log\left( \frac{M_W^2}{\mu^2}\right) - \frac{\alpha_s(\mu) \alphaem}{(4\pi)^2} \gamma_{se}^{(1)}
         \log\left( \frac{M_W^2}{\mu^2}\right) \nonumber \\[2mm]
        & = & 1 + \frac{\alphaem}{4\pi} \, 2 \left( 1 - \frac{\alpha_s(\mu)}{4\pi} \right) \log\left( \frac{M_W^2}{\mu^2}\right) ~ .
     \label{eq:URIprime}
\eea
Note that at this order the evolution operator is independent of the QCD $\beta$-function. This is a consequence of the fact that the QCD anomalous dimension vanishes for the operator $O_1$.

Combining Eqs.\,(\ref{eq:O1Wreg})\,-\,(\ref{eq:ZWRIprime}) and (\ref{eq:URIprime}) we obtain the relation between the operator $ O_1$ in the W-regularization scheme and the one in the \RIprime~ scheme,
\be
O^{\textrm{W-reg}}_1(M_W) = \left\{1 + \frac{\alphaem}{4\pi} \, \left[ 2 \left( 1 - \frac{\alpha_s(\mu)}{4\pi} \right) \log\left( \frac{M_W^2}{\mu^2}\right) + C^\textrm{W-\RIprime}  \right] \right\} \,  O_1^{\text{\RIprime}}(\mu) ~ ,
\label{eq:O1Wreg2}
\ee
which is valid at first order in $\alphaem$ and up to (and including) terms of  ${\cal{O}}(\alphaem \alpha_s(M_W))$ in the strong coupling constant.

\vskip 0.5 cm
{\bf c) The renormalised operator in the \RIMOMprime~scheme.}
When we include QCD and e.m.~corrections at ${\cal{O}}(\alphaem)$, the operator $O_1$ on the lattice with Wilson fermions mixes with a 
complete basis of operators with different chiralities. In addition to $O_1$, the mixing involves the following operators
\bea   
      O_2^\mathrm{bare} & = & \overline{q}_2 \gamma^\mu (1 +  \gamma_5) q_1 ~ \overline{\nu}_\ell \gamma_\mu (1 -  \gamma_5) \ell ~ , \nonumber \\   
      O_3^\mathrm{bare} & = & \overline{q}_2 (1 -  \gamma_5) q_1 ~ \overline{\nu}_\ell (1 +  \gamma_5) \ell ~ , \nonumber \\
      O_4^\mathrm{bare} & = & \overline{q}_2 (1 +  \gamma_5) q_1 ~ \overline{\nu}_\ell (1 +  \gamma_5) \ell ~ , \nonumber \\  
      O_5^\mathrm{bare} & = & \overline{q}_2 \sigma^{\mu \nu} (1 +  \gamma_5) q_1 ~ \overline{\nu}_\ell \sigma_{\mu \nu} (1 +  \gamma_5) \ell \, . 
     \label{eq:ops}
 \eea
The mixing is a consequence of the explicit chiral symmetry breaking of Wilson-like fermions on the lattice. Therefore, the renormalised operators in the \RIMOMprime~scheme, $\vec O^{\textrm{\RIprime}}(\mu)$, with $\vec O = (O_1, \ldots O_5)$, can be written in terms of bare lattice operators $\vec O^\mathrm{bare}(a)$ as
\be
 \vec O^\textrm{\RIprime}(\mu) = Z_O(\mu a)\,  \vec O^\mathrm{bare} (a)
 \label{eq:ORI}
\ee
where $Z_O(\mu a)$ is a $5\times5$ renormalisation matrix. We note that in pure QCD the operator $O_1$ mixes only with $O_2$, with scale independent coefficients, whereas the full $5\times5$ renormalisation matrix is necessary in general when e.m. corrections are included.

We find it particularly convenient to rewrite Eq.\,(\ref{eq:ORI}) in the form
\be
 \vec O^\textrm{\RIprime} =Z^{\QED} \left[ (Z^{\QED})^{-1} Z_O (Z^{\QCD})^{-1} \right] Z^\QCD \, \vec O^\mathrm{bare} = Z^{\QED}\,  {\cal R} \, Z^\QCD \, \vec O^\mathrm{bare}
 \label{eq:ORI2}
\ee
where $Z^\QCD$ is the mixing matrix in pure QCD (corresponding to $\alphaem = 0$), and
\be
    Z^\QED \equiv 1 + \frac{\alpha_\mathrm{em}}{4\pi} \Delta Z^{\QED}
    \label{eq:Z_QED}
\ee
is the pure, perturbative QED mixing matrix (corresponding to $\alpha_s = 0$). In Eq.\,(\ref{eq:ORI2}) we have introduced the ratio
\be
     {\cal{R}} =( Z^\QED)^{-1}  \, Z_O    (Z^\QCD)^{-1}   \equiv 1+ \frac{\alpha_\mathrm{em}}{4\pi} \; \eta \, ,
     \label{eq:R_QCD+QED}
\ee
so that, at first order in $\alphaem$, Eq.\,(\ref{eq:ORI2}) is written as
\be
 \vec O^\textrm{\RIprime} = \left[ 1 + \frac{\alpha_\mathrm{em}}{4\pi} \left( \Delta Z^{\QED} + \eta \right) \right] Z^\QCD \, \vec O^\mathrm{bare} ~.
 \label{eq:ORI3}
\ee

The ratio ${\cal R}$ encodes all the nonperturbative contributions of order $\mathcal{O}(\alphaem \alpha_s^n)$ with $n \ge 1$, other than the factorisable terms given by the product $Z^\QED \, Z^\QCD$. In other words if $Z_O$ were simply given by $Z_O = Z^\QED \, Z^\QCD$ at first order in $\alphaem$ then $\eta$ would be zero.  The case $\eta = 0$  thus corresponds to the {\it factorisation approximation} that was first introduced in Refs.\,\cite{Giusti:2017dmp,Giusti:2017jof}.

In this work, the ratio $ {\cal R}$ has been computed non-perturbatively on the lattice to all orders in $\alpha_s$ and up to first order in $\alphaem$. Introducing this ratio ${\cal{R}}$ in the non-perturbative calculation is useful 
since by using the same photon fields in the lattice calculation of $Z_O$ and $Z^\QED$ the statistical uncertainty due to the sampling of the photon field is significantly reduced. Note that the ratio is also free from cut-off effects of ${\cal{O}}(\alphaem a^n)$. The non-perturbative calculation of ${\cal{R}}$, in terms of the matrix $\eta$, is described in Appendix~\ref{sec:appC} and all the details and results will be presented in a forthcoming publication~\cite{dicarlo}.

As already mentioned, pure QCD corrections in Eq.\,(\ref{eq:ORI3}) only induce the mixing of the operator $O_1$ with the operator $O_2$. This mixing produces the renormalised QCD operators
\bea
O_1^\chi &\equiv&  (Z^\QCD \vec O^{\mathrm{bare}})_1 = \overline{q}_2 \gamma^\mu \left[ Z_V^{(0)} -  Z_A^{(0)} \gamma_5 \right]\!q_1 ~ 
                      \overline{\nu}_\ell \gamma_\mu (1 -  \gamma_5) \ell ~ , \nonumber \\  
O_2^\chi &\equiv&  (Z^\QCD \vec O^{\mathrm{bare}})_2 = \overline{q}_2 \gamma^\mu \left[ Z_V^{(0)} +  Z_A^{(0)} \gamma_5 \right]\!q_1 ~ 
                      \overline{\nu}_\ell \gamma_\mu (1 -  \gamma_5) \ell ~ ,
\label{eq:Ochi}
\eea
which, similarly to the corresponding continuum operators, belong respectively to the $(8, 1)$ and $(1,8)$ chiral representations with respect to a rotation of the quark fields~\cite{Bochicchio:1985xa}. These are the combinations entering on the r.h.s. of Eq.\,(\ref{eq:ORI3}).

When we include the e.m.~corrections at ${\cal{O}}(\alphaem)$, the matrices $\Delta Z^{\QED}$ and $\eta$ in Eq.\,(\ref{eq:ORI3}) induce, in general, the mixing of $O_1^\chi$ with the full basis of operators in Eq.\,(\ref{eq:ops}). As shown in Appendix\,\ref{sec:appD}, however, in the twisted-mass formulation used in this paper the only relevant chirality mixing is the one between the operators $O_1$ with $O_2$. Indeed, the mixing coefficients with the operators $O_3$ and $O_4$ are found to be odd in the parameter $\overline{r} \equiv r_1 r_\ell = - r_2 r_\ell$, defined by the product of the Wilson $r$-parameters of the valence quarks and the lepton (with $r_2 = - r_1$ in our procedure). Therefore, taking the average over the values of the parameter $\overline{r}$ (with $\overline{r} = \pm 1$) when computing the amplitude, eliminates the mixing with $O_3$ and $O_4$. Moreover,  the matrix element of the operator $O_5$ between a pseudoscalar meson and the vacuum vanishes, so that the mixing with the operator $O_5$ cannot contribute to the decay rate. Therefore, Eq.\,(\ref{eq:ORI3}) for the renormalised operator $O_1^\textrm{\RIprime}$ simplifies to 
\bea
O_1^\textrm{\RIprime}(\mu) &=& \left[ 1 + \frac{\alpha_\mathrm{em}}{4\pi} \left( \Delta Z^{\QED}(\mu a)_{11} + \eta(\mu a , \alpha_s(1/a))_{11}  \right) \right] O_1^\chi(a) + \nonumber \\
&+& \frac{\alpha_\mathrm{em}}{4\pi} \left( \Delta Z^{\QED}_{12} + \eta(\alpha_s(1/a))_{12}  \right)  O_2^\chi(a)\,,
\label{eq:O1RI}
\eea
where we have explicitly indicated the dependence of the various terms on $\alpha_s$ and the renormalisation scale. Since the mixing of the {\it bona fide} $(8,1)$ operator $O_1^\chi$ with $O_2^\chi$ is a consequence of the explicit chiral symmetry breaking of Wilson-like fermions on the lattice, the corresponding coefficient is due to lattice artefacts and can only be a function of the lattice bare coupling constant $\alpha_s(1/a)$~\cite{Bochicchio:1985xa}.

\subsection{Complete expression for the matching coefficients} 
\label{sec:finalmatch}

We are now in a position to collect the results of the previous subsection in order to provide the final expression relating the renormalised operator $O^{\textrm{W-reg}}_1$ in the W-regularization to the lattice bare operators $O_1$ and $O_2$ at first order in $\alphaem$. Combining Eqs.\,(\ref{eq:O1Wreg2}) and (\ref{eq:O1RI}) and choosing $\mu=1/a$ as renormalisation scale in the intermediate \RIMOMprime~scheme we obtain:
\bea
O^{\textrm{W-reg}}_1(M_W) &=& O_1^\chi(a) + \frac{\alphaem}{4\pi} \, \left[ 2 \left( 1 - \frac{\alpha_s(1/a)}{4\pi} \right) \log\left( a^2 M_W^2\right) + C^\textrm{W-\RIprime} + \Delta Z^{\QED}_{11}(1/a)  + \right. \nonumber \\
&+& \biggl.  \eta_{11}(\alpha_s(1/a)) \biggr] \,  O_1^\chi(a)
+  \frac{\alpha_\mathrm{em}}{4\pi} \biggl[ \Delta Z^{\QED}_{12} + \eta_{12}(\alpha_s(1/a)) \biggr]  O_2^\chi(a) ~ .
\label{eq:O1Wreg3}
\eea
Using the results of Ref.\,\cite{Carrasco:2015xwa}, obtained in perturbation theory at order ${\cal{O}}(\alpha_s^0)$, we have determined the values for the matching and mixing coefficients, 
\bea
&& C^\textrm{W-\RIprime} = - 5.7825 +  1.2373 \, \xi  ~ , \nonumber \\
&& \Delta Z^{\QED}_{11}(1/a) =  -  9.7565 -1.2373 \, \xi \quad , \quad \Delta Z^{\QED}_{12} = -0.5357 ~ ,
\label{eq:coeffpt} 
\eea
where $\xi$ is the photon gauge parameter ($\xi=0\, (1)$ in the Feynman (Landau) gauge).  It is worth noting that the renormalised operator in the W-regularization scheme is gauge independent, at any order of perturbation theory. In particular, as shown by Eq.\,(\ref{eq:coeffpt}), at first order in $\alphaem$ and at zero order in $\alpha_s$ the gauge dependence of the matching coefficient of $O_1^\chi$ cancels in the sum $C^\textrm{W-\RIprime}+\Delta Z^{\QED}_{11}=-15.539$. By contrast, for the matching coefficient of $O_2^\chi$, the two terms $\Delta Z^{\QED}_{12}$ and $\eta_{12}$ are separately gauge independent.

When inserted into the expression for amplitude for the decay $P \to \ell \nu$, the term of order $\alphaem$ of the renormalised operator $O^{\textrm{W-reg}}_1(M_W)$ of Eq.\,(\ref{eq:O1Wreg3}), namely $\delta O^{\textrm{W-reg}}_1(M_W)=O^{\textrm{W-reg}}_1(M_W) - O_1^\chi$, provides the contribution denoted as $\delta A_P^W$ in Eq.\,(\ref{eq:deltaAPS}) 
\be
    \delta A_P^W =  - \frac{\langle 0|  \, \mbox{Tr}\left\{ \delta O^{\textrm{W-reg}}_1(M_W) \overline{\ell} \gamma_0 (1 - \gamma_5) \nu \right\} \, |P^{(0)} \rangle}{X_P^{\ell,0}} ~ ,
\ee 
where $X_P^{\ell,0}$ is the leptonic trace defined in Eq.\,(\ref{eq:lepton_trace}). We then note that $O_1^\chi$ and $O_2^\chi$ entering in Eq.\,(\ref{eq:O1Wreg3}) give opposite contributions to the tree-level amplitude, i.e. 
\be
    \langle 0| \mbox{Tr}\left\{ O_1^\chi \, \overline{\ell} \gamma_0 (1 - \gamma_5) \nu \right\} |P^{(0)} \rangle  =  
        - \langle 0| \mbox{Tr}\left\{ O_2^\chi \, \overline{\ell} \gamma_0 (1 - \gamma_5) \nu \right\} |P^{(0)} \rangle = 
        - A_P^{(0)} \,X_P^{\ell,0} ~ ,
\ee
with $A_P^{(0)}$ given in Eq.\,(\ref{eq:AP0_ren}) . Therefore, after averaging the amplitude over the values of the parameter $\overline{r} = \pm 1$, in order to cancel out the contribution of the mixing with $O_3$ and $O_4$, one obtains
\be
    \label{eq:deltaAPW}
    \delta A_P^W = Z^\textrm{W-reg} A_P^{(0)}  ~ ,
\ee
with 
\be
    \label{eq:Z_Wreg}
     Z^\textrm{W-reg} = \frac{\alphaem}{4\pi} \left[2 \left( 1 - \frac{\alpha_s(1/a)}{4\pi} \right) \log\left(a^2M_W^2\right) -15.0032 + \eta_{11}(\alpha_s(1/a)) - \eta_{12}(\alpha_s(1/a)) \right] \, .
\ee

As already noted, the contribution $\delta A_P^W$ of the matching factor at order $\alphaem$ to the decay amplitude, expressed by Eqs.\,(\ref{eq:deltaAPW}) and (\ref{eq:Z_Wreg}), is gauge independent. It then follows that also the order $\alphaem$ contribution of the bare diagrams to the amplitude, expressed by the other terms in Eq.\,(\ref{eq:deltaAPS}), is by itself gauge independent. Therefore, we can numerically evaluate the two contributions separately by making different choices for the gluon and the photon gauge in the two cases\,\footnote{It should be noted, however, that while $Z^\textrm{W-reg}$ of Eq.\,(\ref{eq:Z_Wreg}) is gauge independent at any order of perturbation theory, its actual numerical value may display a residual gauge dependence due to higher order terms in the non-perturbative determination of $\eta_{11}$ which are neglected in the perturbatively evaluated matching coefficient.}. In particular, we have chosen to compute the matching factor $Z^\textrm{W-reg}$ of Eq.\,(\ref{eq:Z_Wreg}) in the Landau gauge for both gluons and photons, because this makes \RIprime~equivalent to RI up to higher orders in the perturbative expansions. On the other hand, in the calculation of the physical amplitudes described in Sec.\,\ref{sec:master} (and already computed in Ref.\,\cite{Giusti:2017dwk}) we have used a stochastic photon generated in the Feynman gauge, which has been adopted also in the calculation of $\Gamma^{{\rm pt}}_0(L)$ in Ref.\,\cite{Lubicz:2016xro}.

As discussed in Ref.\,\cite{Carrasco:2015xwa}, when we compute the difference $\Gamma_0(L) -\Gamma^{{\rm pt}}_0(L)$ in Eq.\,(\ref{eq:Gamma}) at leading order in $\alpha_\mathrm{em}$, the contribution from the lepton wave function RC cancels out provided, of course, it is evaluated in $\Gamma_0(L)$ and $\Gamma^{{\rm pt}}_0(L)$ in the same W-regularization scheme and in the same photon gauge. Since $\Gamma^{{\rm pt}}_0(L)$ has been computed in Ref.\,\cite{Lubicz:2016xro} by omitting the lepton wave function RC contribution in the Feynman gauge, we have to subtract the analogous contribution from Eq.\,(\ref{eq:Z_Wreg}) in the Feynman gauge. The QCD and QED corrections to the the lepton wave function RC at $O(\alpha_\mathrm{em})$ factorise, so that their contribution does not enter into the non-perturbative determination of the matrix $\eta$, which only contains, by its definition, non-factorisable QCD+QED contributions.  Therefore, as discussed in Ref.\,\cite{Carrasco:2015xwa}, the subtraction of the lepton wave function RC only requires the replacement of $Z^\textrm{W-reg}$ in Eq.\,(\ref{eq:Z_Wreg}) by the subtracted matching factor
\be
\tilde Z^\textrm{W-reg} = Z^\textrm{W-reg} - \frac 12 \Delta Z_\ell^\textrm{W-reg} \,,
\ee
where
\be
\Delta Z_\ell^\textrm{W-reg} = \frac{\alphaem}{4\pi} \left[ - \log\left(a^2M_W^2\right) -13.3524 \right] \,.
\ee
The final expression to be used in Eq.\,(\ref{eq:deltaAPS}) is therefore
\be
    \label{eq:deltaAPW2}
    \delta A_P^W = \tilde Z^\textrm{W-reg} A_P^{(0)} \,,
\ee
with 
\be
    \label{eq:Ztilde_Wreg}
    \tilde Z^\textrm{W-reg} =  \frac{\alphaem}{4\pi} \left[ \left( \frac{5}{2} - 2 \frac{\alpha_s(1/a)}{4\pi} \right)                                               
    \log\left(a^2M_W^2\right) - 8.3270 + \eta_{11}(\alpha_s(1/a)) - \eta_{12}(\alpha_s(1/a)) \right] \, .
\ee

To make contact with the factorisation approximation introduced in Refs.\,\cite{Giusti:2017dmp,Giusti:2017jof}, we rewrite Eq.\,(\ref{eq:Ztilde_Wreg})  as
\be
    \tilde Z^\textrm{W-reg}  \equiv Z^{\mathrm{\,fact}}\cdot Z^\textrm{W-reg}_{\eta=0}  
    \label{eq:Z_Wreg_prod}
\ee
where $Z^\textrm{W-reg}_{\eta=0}$ is the result in the factorisation approximation (i.e. with $\eta=0$) 
 \begin{equation}
    \label{eq:Z_Wreg_fact}
    Z^\textrm{W-reg}_{\eta=0} = \frac{\alphaem}{4\pi} \left[ \left( \frac{5}{2} - 2 \frac{\alpha_s(1/a)}{4\pi} \right) \log\left(a^2M_W^2\right) - 8.3270 \right]\,, 
\end{equation}
and $Z^{\mathrm{\,fact}}$ is the factor correcting the result for $\tilde Z^\textrm{W-reg}$ to include the entries of the matrix $\eta$ determined in Ref.\,\cite{dicarlo}
\begin{equation}
    \label{eq:Z_fact}
    Z^{\mathrm{\,fact}}  \equiv  1 +\frac{\alphaem}{4\pi} \,\frac{ \eta_{11}(\alpha_s(1/a)) - \eta_{12}(\alpha_s(1/a))}{Z^\textrm{W-reg}_{\eta=0}}\,.
 \end{equation}
The values of the coefficients $Z^\textrm{W-reg}_{\eta=0}$ and $Z^{\mathrm{fact}}$ are collected in Table~\ref{tab:factorisation} for the three values of the inverse coupling $\beta$ adopted in this work and for $\mu = 1 / a$. 
In the same Table we also include the values of the coefficient $Z_m^{\mathrm{fact}}$ corresponding to the non-factorisable e.m.~corrections to the mass RC (see Eq.\,(\ref{eq:Zf})), evaluated in Ref.\,\cite{dicarlo}. The two methods M1 and M2 correspond to different treatments of the $O(a^2\mu^2)$ discretisation effects and are described in Ref.\,\cite{Carrasco:2014cwa}. The difference of the results obtained with these two methods enters into the systematic uncertainty labelled as $()_{input}$ in Sec.\,\ref{sec:results} below.
The results in Table\,\ref{tab:factorisation} show that the non-factorisable corrections are significant, of O(12\,-\,25\%) for  $Z^\textrm{W-reg}$ and even larger, O(40\,-\,60\%), for ${Z}_m$.
\begin{table}[htb!]
\begin{center}
\begin{tabular}{||c||c||c|c||c|c||}
\hline
& & \multicolumn{2}{c||}{Method M1} &  \multicolumn{2}{c||}{Method M2}  \\
\hline 
$\beta$ & $Z^\textrm{W-reg}_{\eta=0}$ & $Z^{\mathrm{fact}}$ & $Z_m^{\mathrm{fact}}$ 
             & $Z^{\mathrm{fact}}$ & $Z_m^{\mathrm{fact}}$ \\
\hline \hline
$ ~ 1.90 ~ $ & ~ 0.00542 (11) ~ & ~ 1.184 (11) ~ & ~ 1.629 (41) ~ & ~ 1.126 (7) ~ & ~ 1.637 (14) ~ \\ 
\hline
$ ~ 1.95 ~ $ & ~ 0.00519 (10) ~ & ~ 1.172 \, (9) ~ & ~ 1.514 (33) ~ & ~ 1.123 (5) ~ & ~ 1.585 (12) ~ \\ 
\hline
$ ~ 2.10 ~ $ & ~ 0.00440 \, (7) ~ & ~ 1.160 \, (6) ~ & ~ 1.459 (17) ~ & ~ 1.136 (4) ~ & ~ 1.462 \, (6) ~ \\ 
\hline  
\end{tabular}
\end{center}
\vspace{-0.25cm}
\caption{\it \footnotesize Values of the coefficients $Z^\textrm{W-reg}_{\eta=0}$ (see Eq.\,(\ref{eq:Z_Wreg_fact})) and $Z^{\mathrm{fact}}$ (see Eq.\,(\ref{eq:Z_fact})) calculated for the three values of the inverse coupling $\beta$ adopted in this work and for $\mu = 1 / a$. In the fourth and sixth columns the values of the coefficient $Z_m^{\mathrm{fact}}$ corresponding to the non-factorisable e.m.~corrections to the mass RC in the $\overline{\rm MS}(2 \, \mbox{\rm GeV})$ (see Eq.\,(\ref{eq:Zf})) are shown. The evaluation of the RCs in the \RIMOMprime~scheme has been carried out in Ref.\,\cite{dicarlo} using the methods M1 and M2 of Ref.\,\cite{Carrasco:2014cwa} (see Appendix~\ref{sec:appA}).\hspace*{\fill}}
\label{tab:factorisation}
\end{table}

We close this section by noting that Eq.\,(\ref{eq:RPS}) implies that the contribution to $\delta R_P$  from the matching factor in Eq.\,(\ref{eq:deltaAPW2}) is 
$2 \tilde Z^\textrm{W-reg}$. Such a term is mass independent. Thus, as already pointed out in Ref.\,\cite{Giusti:2017dwk}, all the matching and mixing contributions to the axial amplitude in Eq.\,(\ref{eq:deltaAPS}) cancel exactly in the difference between the corrections corresponding to two different channels, e.g. in $\delta R_K - \delta R_\pi$. A similar cancelation also occurs in the difference between the corrections to the amplitudes corresponding to the meson P decaying into two different final-state leptonic channels.

\section{Finite volume effects at order ${\cal{O}}(\alphaem)$}
\label{sec:FVE}

The subtraction $\Gamma_0(L) - \Gamma_0^{\textrm{pt}}(L)$ in Eq.\,(\ref{eq:Gamma}) cancels both the IR divergences and the structure-independent FVEs, i.e.~those of order ${\cal{O}}(1/L)$. 
The point-like decay rate $ \Gamma_0^{\textrm{pt}}(L)$ is given by  
 \be
     \Gamma_0^{\textrm{pt}}(L) = \left( 1 + 2 \frac{\alphaem}{4\pi} ~ Y_P^\ell(L) \right) ~ \Gamma_P^{\textrm{tree}} ~ ,  
 \ee
where
 \be
    Y_P^\ell(L) = b_{\mathrm{IR}} ~ \mbox{log}(M_P L) + b_0 + \frac{b_1}{M_P L} + \frac{b_2}{(M_P L)^2} + \frac{b_3}{(M_P L)^3} +
                          {\cal{O}}(e^{-M_P L}) ~ 
       \label{eq:APS_pt}
 \ee
with the coefficients $b_j$ ($j = \mathrm{IR}, 0, 1, 2, 3$) depending on the dimensionless ratio $m_\ell / M_P$ and given explicitly in Eq.\,(98) of Ref.\,\cite{Lubicz:2016xro} (see also Ref.\,\cite{Tantalo:2016vxk}) after the subtraction of the lepton self-energy contribution in the Feynman gauge. 
An important result of Ref.\,\cite{Lubicz:2016xro} is that the structure-dependent FVEs start at order ${\cal{O}}(1 / (M_P L)^2)$. 
Consequently the coefficients $b_{\mathrm{IR},0,1}$ in the factor $Y_P^\ell(L)$ are ``universal", i.e.~they are the same as in the full theory when the structure of the meson $P$ is considered\footnote{Notice that the decay rate in the full theory, $\Gamma_0(L)$, can be affected also by non-universal FVEs of order ${\cal{O}}[1/(M_PL)^n]$ with $n \geq 4$ that do not appear in $\Gamma_0^{pt}(L)$.}.

Eq.\,(\ref{eq:deltaAPS}) is therefore replaced by
 \be
     \delta A_P = \delta A_P^W + \delta A_P^{\mathrm{SIB}} + \sum_{i = J,T,P,S} \delta A_P^i + \delta A_P^\ell  - 
                          Y_P^\ell(L) \, A_P^{(0)} ~ ,
     \label{eq:deltaAPS_sub}
 \ee 
where $\delta A_P^W$ is given by Eq.\,(\ref{eq:deltaAPW2}).

\begin{figure}[htb!]
\centering
\includegraphics[scale=0.75]{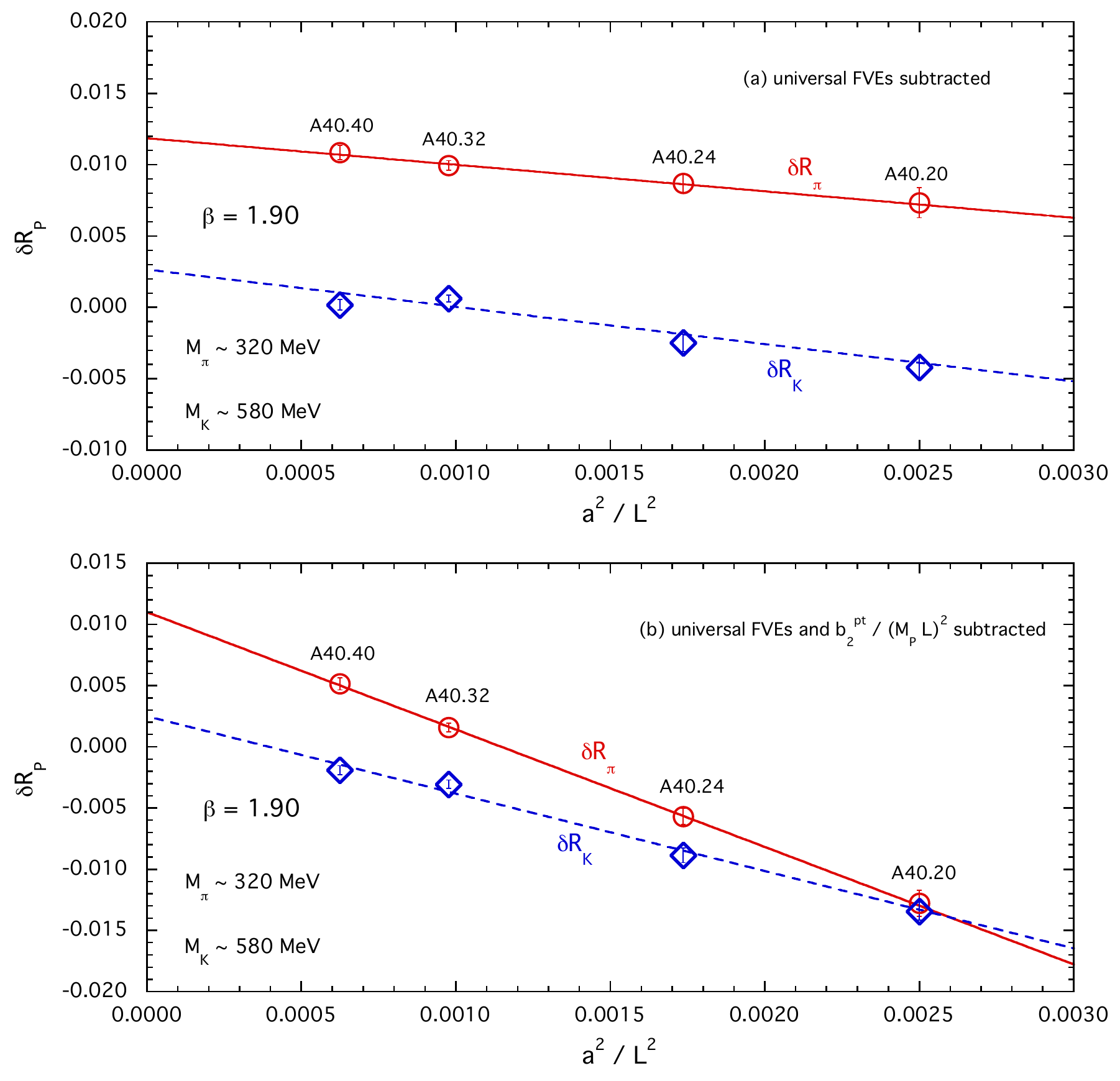}
\vspace{-0.4cm}
\caption{\it \footnotesize Results for the corrections $\delta R_\pi$ and $\delta R_K$ for the gauge ensembles A40.20, A40.24, A40.32 and A40.40 sharing the same lattice spacing, pion, kaon and muon masses, but with different lattice sizes (see Table \ref{tab:simudetails}). Top panel (a): the universal FVEs, i.e.~the terms up to order ${\cal{O}}(1/M_PL)$ in Eq.\,(\protect\ref{eq:APS_pt}), are subtracted for each quantity. Bottom panel (b): the same as in (a), but in addition to the subtraction of the universal terms, $b_2^{\mathrm{pt}}/(M_PL)^2$, where $b_2^{\mathrm{pt}}$ is the pointlike contribution to $b_2$ in Eq.\,(\protect\ref{eq:APS_pt}), is also removed. The solid and dashed lines are linear fits in $1 /L^2$. The maximum photon energy $\Delta E_\gamma$ corresponds to the fully inclusive case $\Delta E_\gamma = \Delta E_\gamma^{max, P} = M_{P} (1 - m_\mu^2 / M_P^2) / 2$.\hspace*{\fill}}
\label{fig:FVE}
\end{figure}

In order to study the FVEs in detail we consider four ensembles generated at the same values of $\beta$ and quark masses, but differing in the size of the lattice; these are the ensembles A40.40, A40.32, A40.24 and A40.20 (see Appendix~\ref{sec:appA}).
The residual FVEs after the subtraction of the universal terms as in Eq.\,(\ref{eq:deltaAPS_sub}) are illustrated in the plots in Fig.~\ref{fig:FVE} for $\delta R_\pi$ and $\delta R_K$ in the fully inclusive case, i.e.~where the energy of the final-state photon is integrated over the full phase space. In this case $\Delta E_\gamma = \Delta E_\gamma^{\mathrm{max}, P} = M_{P} (1 - m_\mu^2 / M_{P}^2) / 2$, which corresponds to $\Delta E_\gamma^{\mathrm{max}, K} \simeq 235$ MeV and $\Delta E_\gamma^{\mathrm{max}, \pi} \simeq 29$ MeV, respectively. With a muon as the final state lepton, the contribution from photons with energy greater than about 20\,MeV is negligible and hence the point-like approximation is valid. In the top plot the universal FV corrections have been subtracted and so we would expect the remaining effects to be of order ${\cal{O}}(1 / (M_P L)^2)$ and this is indeed what we see. 

In the bottom plot of Fig.~\ref{fig:FVE}, in addition to subtracting the universal FVEs, we also subtract the contribution to the order ${\cal{O}}(1 / (M_P L)^2)$ corrections from the point-like contribution to $b_2$, which can be found in Eq.\,(3.2) of Ref.\,\cite{Tantalo:2016vxk}. We observe that this additional subtraction does not reduce the ${\cal{O}}(1 / (M_P L)^2)$ effects, underlining the expectation that these effects are indeed structure dependent.

It can be seen that after subtraction of the universal terms the residual structure-dependent FVEs are almost linear in $1 / L^2$, which implies that the FVEs of order ${\cal{O}}(1 / (M_P L)^3)$ are quite small; indeed they are too small to be resolved with the present statistics.
Nevertheless, since the QED$_\mathrm{L}$ formulation of QED on a finite box, which is adopted in this work, violates locality\,\cite{Patella:2017fgk}, we may expect that there are also FVEs of order ${\cal{O}}(a^3 / L^3)$\,\cite{Tantalo:2016vxk}.
We have checked explicitly that the addition of such a term in fitting the results shown in Fig.~\ref{fig:FVE} changes the extrapolated value at infinite volume well within the statistical errors.

A more detailed description of the full analysis, including the continuum and chiral extrapolations, is given in the following section. As far as the FVEs are concerned, the central value is obtained by subtracting the universal terms and fitting the residual ${\cal{O}}(1/L^2)$ corrections to 
\begin{equation}
\frac{K_P}{(M_PL)^2}+ \frac{K_P^\ell}{(E^\ell_PL)^2}\,,\label{eq:FVansatz}
\end{equation}
where $K_P$ and $K_P^\ell$ are constant fitting parameters and $E_P^\ell$ is the energy of the charged lepton in the rest frame of the pseudoscalar $P$ (see Eq.\,(\ref{eq:RPS_fit}) below). Such an ansatz is introduced to model the unknown dependance of $b_2$ on the ratio $m_\ell / M_P$. For the four points in each of the plots of Fig.~\ref{fig:FVE} $m_\ell/M_P$ takes the same value, but this is not true for all the ensembles used in the analysis.  We estimate the uncertainty due to the use of the ansatz in Eq.\,(\ref{eq:FVansatz}) by repeating the same analysis, but on the data in which, in addition to subtracting the universal terms in Eq.\,(\ref{eq:APS_pt}), we also subtract the term $b_2^{\mathrm{pt}}/(M_P L)^2$, where $b_2^{\mathrm{pt}}$ is contribution to $b_2$ from a point-like meson\,\cite{Tantalo:2016vxk}. Since $b_2^{\mathrm{pt}}$ depends on $m_\ell / M_P$, the result obtained with this additional subtraction is a little different from that obtained with only the universal terms removed and we take the difference as an estimate of the residual FV uncertainty.

\section{Results for charged pion and kaon decays into muons}
\label{sec:results}

We now insert the various ingredients described in the previous sections into the master formula in Eq.\,(\ref{eq:RPS}) for the decays $\pi^+ \to \mu^+ \nu [\gamma]$ and $K^+ \to \mu^+ \nu [\gamma]$.

The results for the corrections $\delta R_\pi$ and $\delta R_K$ are shown in Fig.~\ref{fig:PSplus}, where the ``universal'' FSEs up to order ${\cal{O}}(1/L)$ have been subtracted from the lattice data (see the empty symbols) and all photon energies (i.e.~$\Delta E_\gamma = \Delta E_\gamma^{\mathrm{max}, P} = M_P (1 - m_\mu^2 / M_P^2) / 2$) are included, since the experimental data on $\pi_{\ell 2}$ and $K_{\ell 2}$ decays are fully inclusive.
As already pointed out in section \ref{sec:intro}, structure dependent contributions to  real photon emission should be included.
According to the ChPT predictions of Ref.\,\cite{Cirigliano:2007ga}, however, these contributions are negligible in for both kaon and pion decays into muons, while the same does not hold as well for decays into final-state electrons (see Ref.\,\cite{Carrasco:2015xwa}).
This important conclusion needs to be explicitly validated 
by an ongoing dedicated lattice study of the real photon emission amplitudes in light and heavy P-meson leptonic decays.

The combined chiral, continuum and infinite-volume extrapolations are performed using the following SU(2)-inspired fitting function
 \bea
     \delta R_P & = & R_P^{(0)} + R_P^{(\chi)} ~ \mbox{log}(m_{ud}) + R_P^{(1)} m_{ud} + R_P^{(2)} m_{ud}^2 + D_P\,a^2 \nonumber \\[2mm]
                       & + & \frac{K_P}{M_P^2 L^2} + \frac{K_P^\ell}{(E_P^\ell)^2 L^2} + \delta \Gamma^{\textrm{pt}}(\Delta E_\gamma^{max, P})~ ,
     \label{eq:RPS_fit}
 \eea
where $m_{ud} = \mu_{ud} / Z_P$ and $\mu_{ud}$ is the bare (twisted) mass (see Table\,\ref{tab:simudetails} in Appendix\,\ref{sec:appA} below), $E_P^\ell$ is the lepton energy in the P-meson rest frame, $R_P^{(0), (1), (2)}$, $D_P$, $K_P$ and $K_P^\ell$ are free parameters.
In Eq.\,(\ref{eq:RPS_fit}) the chiral coefficient $R_P^{(\chi)}$ is known for both pion and kaon decays from Ref.\,\cite{Bijnens:2006mk}; in QED the coefficients are
 \be
      R_\pi^{(\chi)} = \frac{\alphaem}{4 \pi} \left( 3 - 2 X \right) ~ , \qquad
      R_K^{(\chi)} = - \frac{\alphaem}{4 \pi} X ~ ,
       \label{eq:Rchi_full}
 \ee
 while in qQED they are
  \be
      R_\pi^{(\chi)} = \frac{\alphaem}{4 \pi} \left( 3 - \frac{10}{9} X \right) ~ , \qquad
      R_K^{(\chi)} = - \frac{\alphaem}{4 \pi} \frac{8}{9} X  ~ ,
      \label{eq:Rchi_qQED}
  \ee
where $X$ is obtained from the chiral limit of the ${\cal{O}}(\alphaem)$ correction to $M_{\pi^\pm}^2$ (i.e.~$\delta M_{\pi^\pm}^2 = 4 \pi \alphaem X f_0^2 + {\cal{O}}(m_{ud})$).
In Ref.\,\cite{Giusti:2017dmp} we found $X = 0.658(40)$.

\begin{figure}[htb!]
\begin{center}
\includegraphics[scale=0.85]{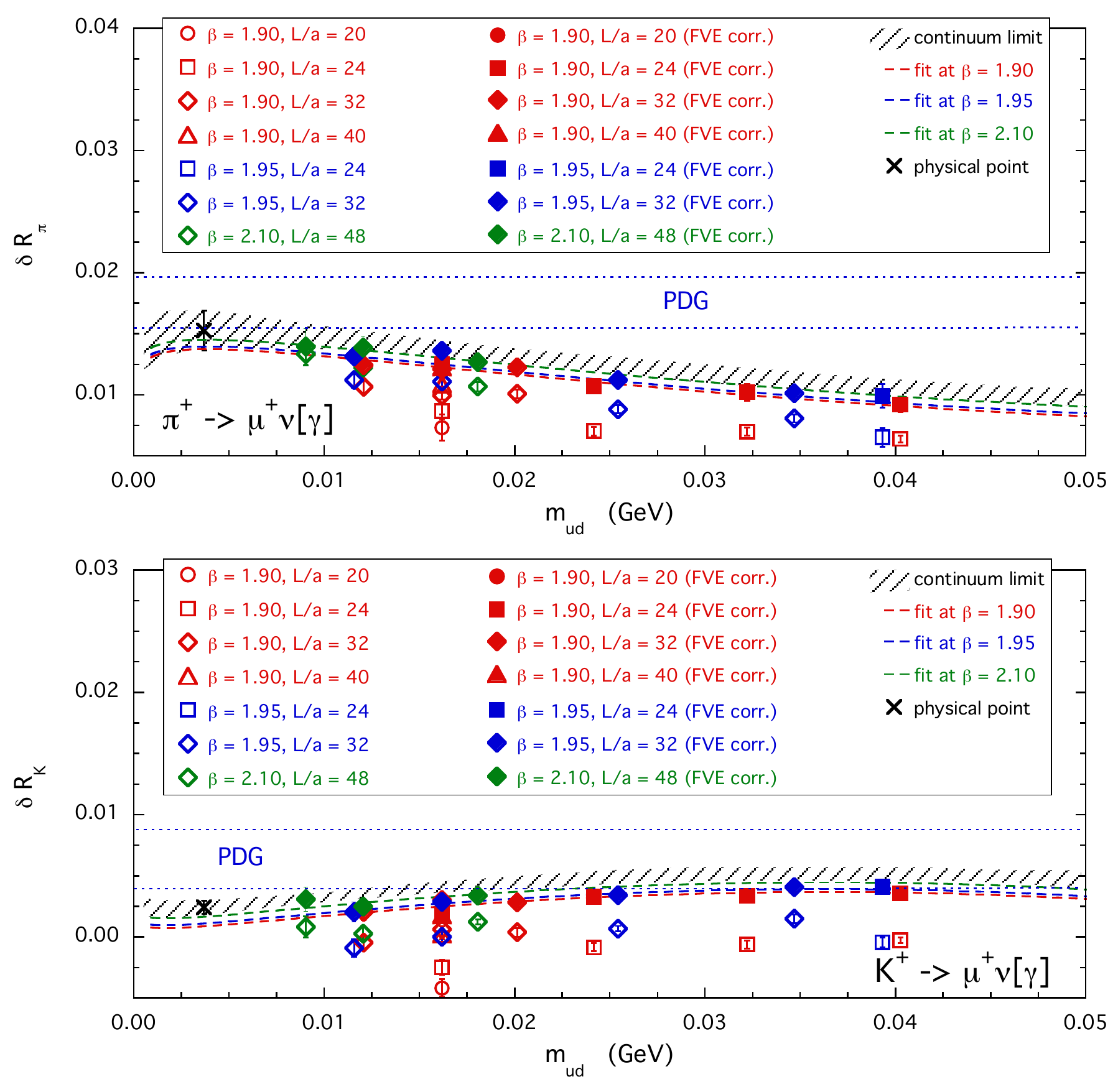}
\end{center}
\vspace{-0.75cm}
\caption{\it \footnotesize Results for the corrections $\delta R_\pi$ (top panel) and $\delta R_K$ (bottom panel) obtained after the subtraction of the ``universal'' FSE terms up to order ${\cal{O}}(1/L)$ in Eq.\,(\protect\ref{eq:APS_pt}) (empty markers). The full markers correspond to the lattice data corrected by the residual FSEs obtained in the case of the fitting function (\protect\ref{eq:RPS_fit}) including the chiral log. The dashed lines are the (central) results in the infinite volume limit at each value of the lattice spacing, while the shaded areas identify the results in the continuum limit at the level of one standard deviation. The crosses represent the values $\delta R_\pi^{\textrm{phys}}$ and $\delta R_K^{\textrm{phys}}$ extrapolated at the physical point $m_{ud}^{\textrm{phys}}(\overline{\rm MS}, 2\,{\rm GeV}) = 3.70~(17)$ MeV~\protect\cite{Carrasco:2014cwa}. The blue dotted lines correspond to the values $\delta R_\pi^{\textrm{phys}} = 0.0176~(21)$ and $\delta R_K^{\textrm{phys}} = 0.0064~(24)$, obtained using ChPT \protect\cite{Cirigliano:2011tm} and adopted by the PDG\,\protect\cite{Rosner:2015wva}.\hspace*{\fill}}
\label{fig:PSplus}
\end{figure}

Using Eq.\,(\ref{eq:RPS_fit}) we have fitted the data for $\delta R_\pi$ and $\delta R_K$ using a $\chi^2$-minimization procedure with an uncorrelated $\chi^2$, obtaining values of $\chi^2 / \mbox{d.o.f.}$ always around $0.9$.
The uncertainties on the fitting parameters do not depend on the $\chi^2$-value, because they are obtained using the bootstrap samplings of Ref.\,\cite{Carrasco:2014cwa} (see Appendix~\ref{sec:appA}).
This guarantees that all the correlations among the data points and among the fitting parameters are properly taken into account.

The quality of our fits is illustrated in Fig.~\ref{fig:PSplus}.
It can be seen that the residual SD FVEs are still visible in the data and well reproduced by our fitting ansatz in Eq.\,(\ref{eq:RPS_fit}).
Discretisation effects on the other hand, only play a minor role.

At the physical pion mass in the continuum and infinite-volume limits we obtain
 \bea
      \label{eq:Rpi_phys}
      \delta R_\pi^{\textrm{phys}} & = & + 0.0153 ~ (16)_{stat + fit} ~ (4)_{input} ~ (3)_{chiral} ~ (6)_{FVE} ~ (2)_{disc} 
                                                              ~ (6)_{qQED} \nonumber \\
                                                  & = & + 0.0153 ~ (19) ~ , \\
      \label{eq:RK_phys}
      \delta R_K^{\textrm{phys}} & = & + 0.0024 ~ (6)_{stat + fit} ~ (3)_{input} ~ (1)_{chiral} ~ (3)_{FVE} ~ (2)_{disc} 
                                                            ~ (6)_{qQED} \nonumber \\
                                                 & = & + 0.0024 ~ (10) ~ ,
 \eea
where
\begin{itemize} 

\item $()_{stat+fit}$ indicates the uncertainty induced by the statistical Monte Carlo errors of the simulations and its propagation in the fitting procedure;

\item$()_{input}$ is the error coming from the uncertainties of the input parameters of the quark-mass analysis of Ref.\,\cite{Carrasco:2014cwa}; 

\item $()_{chiral}$ is the difference between including or excluding the chiral logarithm in Eq.\,(\ref{eq:RPS_fit}), i.e.~taking $R_\chi \neq 0$ or $R_\chi = 0$; 

\item $()_{FVE}$ is the difference between the analyses of the data corresponding to the FVE subtractions up to the order ${\cal{O}}(1/L)$ alone or by also subtracting the term proportional to $b_2^{\mathrm{pt}}/(M_P L)^2$ (see Fig.~\ref{fig:FVE} and the discussion towards the end of Sec.\,\ref{sec:FVE}); 

\item $()_{disc}$ is the uncertainty coming from including ($D \neq 0$) or excluding (setting $D = 0$) the discretisation term proportional to $a^2$ in Eq.\,(\ref{eq:RPS_fit}); 

\item $()_{qQED}$ is our estimate of the uncertainty of the QED quenching. This is obtained using the ansatz (\ref{eq:RPS_fit}) with the coefficient $R_\chi$ of the chiral log fixed either at the value (\ref{eq:Rchi_qQED}), which corresponds to the qQED approximation, or at the value (\ref{eq:Rchi_full}), which includes the effects of the up, down and strange sea-quark charges\,\cite{Bijnens:2006mk}. The change both in $\delta R_\pi^{\textrm{phys}}$ and in $\delta R_K^{\textrm{phys}}$ is $\simeq 0.0003$, which has been already added in the central values given by Eqs.\,(\ref{eq:Rpi_phys}) and (\ref{eq:RK_phys}). To be conservative, we use twice this value for our estimate of the qQED uncertainty.

\end{itemize}

Our results in Eqs.\,(\ref{eq:Rpi_phys})\,-\,(\ref{eq:RK_phys}) can be compared with the ChPT predictions $\delta R_\pi^{\textrm{phys}} = 0.0176(21)$ and $\delta R_K^{\textrm{phys}} =  0.0064(24)$ obtained in Ref.\,\cite{Cirigliano:2011tm} and adopted by the PDG\, \cite{PDG,Rosner:2015wva}.
The difference is within one standard deviation for $\delta R_\pi^{\textrm{phys}}$, while it is larger for $\delta R_K^{\textrm{phys}}$.
Note that the precision of our determination of $\delta R_\pi^{\textrm{phys}}$ is comparable to the one obtained in ChPT, while our determination of $\delta R_K^{\textrm{phys}}$ has a much better accuracy compared to that obtained using ChPT; the improvement in precision is a factor of about 2.2.
We stress that the level of precision of our pion and kaon results depends crucially on the non-perturbative determination of the chirality mixing, carried out in section \ref{sec:Wreg} by including simultaneously QED at first order and QCD at all orders.

As already stressed, the correction $\delta R_P$ and the QCD quantity $f_P^{(0)}$ separately depend on the prescription used for the separation between QED and QCD corrections\,\cite{Gasser:2010wz}. 
Only the product $f_P^{(0)} \sqrt{1 + \delta R_P}$ is independent of the prescription and its value, multiplied by the relevant CKM matrix element, yields the P-meson decay rate.
We remind the reader that our results (\ref{eq:Rpi_phys})\,-\,(\ref{eq:RK_phys}) are given in the GRS prescription (see the dedicated discussion in sections\,\ref{sec:QCDQCD} and~\ref{sec:master}) in which the renormalised couplings and quark masses in the full theory and in isosymmetric QCD coincide in the $\overline{\rm MS}$ scheme at a scale of 2\,GeV\,\cite{Gasser:2003hk}. We remind the reader that, to the current level of precision, this GRS scheme can be considered equivalent to the FLAG scheme.
 
Taking the experimental values $\Gamma(\pi^- \to \mu^- \bar{\nu}_\mu [\gamma]) = 3.8408 (7) \cdot 10^7$ s$^{-1}$ and $\Gamma(K^- \to \mu^- \bar{\nu}_\mu [\gamma]) = 5.134 (11) \cdot 10^7$ s$^{-1}$ from the PDG\,\cite{PDG} and using our results (\ref{eq:Rpi_phys})\,-\,(\ref{eq:RK_phys}), we obtain
 \bea
       \label{eq:fpi0Vud}
       f_\pi^{(0)} |V_{ud}| & = & 127.28 ~ (2)_{\mathrm{exp}} ~ (12)_{\mathrm{th}} \, \mbox{MeV} = 127.28 \, (12) \, \mbox{MeV} ~ , ~ \\
       \label{eq:fK0Vus}
       f_K^{(0)} |V_{us}| & = & 35.23 ~ (4)_{\mathrm{exp}} ~ (2)_{\mathrm{th}} \, \mbox{MeV} = 35.23 \, (5) \, \mbox{MeV} ~ ,
 \eea 
where the first error is the experimental uncertainty  and the second is that from our theoretical calculations.
The result for the pion in Eq.\,(\ref{eq:fpi0Vud}) agrees within the errors with the updated value $f_\pi^{(0)} |V_{ud}| = 127.12(13)$\,MeV\,\cite{PDG}, obtained by the PDG and based on the model-dependent ChPT estimate of the e.m.~corrections from Ref.\,\cite{Cirigliano:2011tm}. Our result for the kaon in Eq.\,(\ref{eq:fK0Vus}) however, is larger than the corresponding PDG value $f_K^{(0)} |V_{us}| = 35.09(5)$\,MeV~\cite{PDG}, based on the ChPT calculation of Ref.\,\cite{Cirigliano:2011tm}, by about $2$ standard deviations. 

As anticipated in the Introduction and discussed in detail in Sec.\,\ref{sec:master}, we cannot use the result (\ref{eq:fpi0Vud}) to determine the CKM matrix element $|V_{ud}|$, since the pion decay constant was used by ETMC~\cite{Carrasco:2014cwa} to set the lattice scale in isosymmetric QCD and its value, $f_\pi^{(0)} = 130.41(20)$\,MeV, was based on the determination of $|V_{ud}|$ obtained from super-allowed $\beta$-decays in Ref.\,\cite{Hardy:2014qxa}.
On the other hand, adopting the best lattice determination of the QCD kaon decay constant, $f_K^{(0)} = 156.11(21)$\,MeV~\cite{FLAG,Dowdall:2013rya,Carrasco:2014poa,Bazavov:2017lyh}\,\footnote{The average value of $f_{K^\pm}$ quoted by FLAG\,\cite{FLAG} includes the strong IB corrections. In order to obtain $f_K^{(0)}$ therefore, we have subtracted this correction which is given explicitly in Refs.\,\cite{Dowdall:2013rya,Carrasco:2014poa,Bazavov:2017lyh}.}, we find that Eq.\,(\ref{eq:fK0Vus}) implies 
\be
      \label{eq:Vus_K}
      |V_{us}| = 0.22567(26)_{\mathrm{exp}}\,(33)_{\mathrm{th}} = 0.22567 \, (42)\,,
\ee
which is a result with the excellent precision of $\simeq 0.2 \%$.

Since the non-factorisable e.m.~corrections to the mass RC (see the coefficient $Z_m^{\mathrm{fact}}$ in Table~\ref{tab:factorisation}) were not included in Ref.\,\cite{Giusti:2017dwk}, we update our estimate of the ratio of the kaon and pion decay rates:
\be
     \delta R_{K\pi}^{\textrm{phys}} = \delta R_K^{\textrm{phys}} - \delta R_\pi^{\textrm{phys}} = -0.0126 \, (14) ~ .
     \label{eq:RKpi_phys}
\ee
Using the pion and kaon experimental decay rates we get
\be  
    \frac{|V_{us}|}{|V_{ud}|} \frac{f_K^{(0)}}{ f_\pi^{(0)}} = 0.27683 \,(29)_{\mathrm{exp}} \, (20)_{\mathrm{th}} = 0.27683 \, (35) \, .
    \label{eq:ratioVf}
\ee
Using the best $N_f = 2+1+1$ lattice determination of the ratio of the QCD kaon and pion decay constants, $f_K^{(0)} / f_\pi^{(0)} = 1.1966~(18)$\,\cite{FLAG,Dowdall:2013rya,Carrasco:2014poa,Bazavov:2017lyh}, we find
\be  
    \frac{|V_{us}|}{|V_{ud}|} = 0.23135 \, (24)_{\mathrm{exp}} \, (39)_{\mathrm{th}} = 0.23135 \, (46) \, . 
    \label{eq:VusVud}
\ee
Taking the updated value $|V_{ud}| = 0.97420\,(21)$ from super-allowed nuclear beta decays\,\cite{Hardy:2016vhg}, Eq.\,(\ref{eq:VusVud}) yields the following value for the CKM element $|V_{us}|$:
\be  
    \label{eq:Vus}
    |V_{us}| = 0.22538 \,(24)_{\mathrm{exp}} \, (39)_{\mathrm{th}} = 0.22538 \, (46) \, ,
\ee
which agrees with our result (\ref{eq:Vus_K}) within the errors. 
Note that our result (\ref{eq:Vus}) agrees with the latest estimate $|V_{us}| = 0.2253(7)$, recently updated by the PDG\,\cite{PDG}, but it improves the error by a factor of approximately 1.5.

Taking the values $|V_{ub}| = 0.00413(49)$\,\cite{PDG} and $|V_{ud}| = 0.97420(21)$\,\cite{Hardy:2016vhg} our result in Eq.\,(\ref{eq:Vus}) implies that the unitarity of the first-row of the CKM matrix is confirmed to better than
the per-mille level
\be
     \label{eq:unitarity}
     |V_{ud}|^2 + |V_{us}|^2 + |V_{ub}|^2 = 0.99988 \, (46) \, .
\ee

With the same value $|V_{ud}| = 0.97420(21)$ from super-allowed nuclear beta decays~\cite{Hardy:2016vhg}, our result (\ref{eq:fpi0Vud}) implies for the QCD pion decay constant (in the GRS prescription) the following value 
 \be
       \label{eq:fpi0}
       f_\pi^{(0)} = 130.65 \, (12)_{\mathrm{exp+th}} \, (3)_{V_{ud}} \, \mbox{MeV} = 130.65\,(12) \, \mbox{MeV} ~ ,
 \ee 
which, as anticipated in Sec.\,\ref{sec:master}, agrees within the errors with the value $f_\pi^{(0)} = 130.41\,(20)$\,MeV adopted in Ref.\,\cite{Carrasco:2014cwa} to set the lattice scale in the isosymmetric QCD theory.
This demonstrates the equivalence of the GRS and PDG schemes within the precision of our simulation.

In a recent paper~\cite{Seng:2018yzq}  the hadronic contribution to the electroweak radiative corrections to neutron and super-allowed nuclear $\beta$ decays has been analyzed in terms of dispersion relations and neutrino scattering data. 
With respect to the result $V_{ud} = 0.97420 (21)$ from Ref.~\cite{Hardy:2016vhg} a significant shift in the central value and a reduction of the uncertainty have been obtained, namely $V_{ud} = 0.97370 (14)$~\cite{Seng:2018yzq}.
The impact of the new value of $V_{ud}$ on our determinations of $V_{us}$ and $f_\pi^{(0)}$ is $V_{us} = 0.22526\,(46)$ and $f_\pi^{(0)} = 130.72 \,(12)$ MeV, i.e.~well within the uncertainties shown in Eqs.\,(\ref{eq:Vus}) and (\ref{eq:fpi0}), respectively. On the contrary, the first-row CKM unitarity (\ref{eq:unitarity}) will be significantly modified into
\be
\label{eq:unitarity_alt}
|V_{ud}|^2 + |V_{us}|^2 + |V_{ub}|^2 = 0.99885 (34) ~ ,
\ee
which would imply a $\simeq 3.4 \sigma$ tension with unitarity.
A confirmation of the new calculation of the radiative corrections made in Ref.~\cite{Seng:2018yzq} is therefore urgently called for.

Before closing this section, we comment briefly about the comparison between our result $\delta R_K^{\mathrm{phys}} = 0.0024(10)$ and the corresponding model-dependent ChPT prediction $\delta R_K^{\mathrm{phys}} = 0.0064(24)$ from Ref.\,\cite{Cirigliano:2011tm}.
The latter is obtained by adding a model-dependent QED correction of $0.0107(21)$ and a model-independent next-to-leading strong IB contribution equal to $-0.0043(12)$.
Our result on the other hand, obtained in the GRS prescription, stems from a QED correction equal to $0.0088(9)$ and a strong IB term equal to $-0.0064(7)$ (see also Ref.\,\cite{Giusti:2017xrv}).
The difference between our result and the ChPT prediction of Ref.\,\cite{Cirigliano:2011tm} appears to be mainly due to a different strong IB contribution.
Thus, in the present $N_f = 2 + 1 + 1$ study, we confirm for the strong IB term a discrepancy at the level of about 2 standard deviations, which was already observed at $N_f = 2$ in Ref.\,\cite{deDivitiis:2013xla}.

\section{Conclusions}
\label{sec:conclusions}

In this paper we have presented the details of the first lattice computation of the leading e.m.~and strong IB corrections to the $\pi^+ \to \mu^+ \nu$ and $K^+ \to \mu^+ \nu$ leptonic decay rates, following a method recently proposed in Ref.\,\cite{Carrasco:2015xwa}. This expands significantly on the discussion of Ref.\,\cite{Giusti:2017dwk}, where the results and a brief outline of the calculation had been presented.
The results were obtained using the gauge ensembles produced by the European Twisted Mass Collaboration with $N_f = 2 + 1 + 1$ dynamical quarks. 
Systematics effects are evaluated and the impact of the quenched QED approximation is estimated. 

The effective weak Hamiltonian in the W-regularisation scheme appropriate for this calculation is obtained from the bare lattice operators in two stages. First of all, the lattice operators are renormalised non-perturbatively in the RI$^\prime$-MOM scheme at $O(\alphaem)$ and to all orders in the strong coupling $\alpha_s$. Because of the breaking of chiral symmetry in the twisted mass formulation we have adopted this renormalisation includes the mixing with other four-fermion operators of different chirality. In the second step we perform the matching from the RI$^\prime$-MOM scheme to the W-regularisation scheme perturbatively. By calculating and including the two-loop anomalous dimension at $O(\alphaem\alpha_s)$\,\cite{Brod:2008ss}, the residual truncation error of this matching is of $O(\alphaem\alpha_s(M_W))$, reduced from $O(\alphaem\alpha_s(1/a))$ in our earlier work\,\cite{Giusti:2017dwk,Carrasco:2015xwa}.

The evaluation of isospin breaking (IB) ``corrections" raises the question of how QCD without these corrections is defined. Since IB corrections change hadronic masses and other physical quantities a prescription is needed to define QCD, whether isosymmetric or not, and in Sec.\,\ref{sec:QEDQCD} and Appendix\,\ref{sec:appB} we discuss this issue in detail. In particular, the correction $\delta R_P$ and the QCD quantity $f_P^{(0)}$ separately depend on the prescription used for the definition of QCD\,\cite{Gasser:2010wz}. 
Only the product $f_P^{(0)} \sqrt{1 + \delta R_P}$ is independent of the prescription and its value, multiplied by the relevant CKM matrix element, yields the P-meson decay rate.
In this paper we chose to follow the conventionally used GRS prescription (see the dedicated discussion in sections\,\ref{sec:QCDQCD} and~\ref{sec:master}) in which the renormalised couplings and quark masses in the full QCD+QED theory and in isosymmetric QCD coincide in the $\overline{\rm MS}$ scheme at a scale of 2\,GeV\,\cite{Gasser:2003hk}. For future studies however, we advocate the use of ``hadronic schemes" in which QCD is defined by requiring that a set of hadronic quantities (for example a set of hadronic masses) take their physical values in QCD and in QCD+QED.

The main results of the calculation are presented in Sec.\,\ref{sec:results} together with a detailed discussion of their implications.
In summary, after extrapolation of the data to the physical pion mass, and to the continuum and infinite-volume limits, the isospin-breaking corrections to the leptonic decay rates can be written in the form:
\bea
     \Gamma(\pi^\pm \to \mu^\pm \nu_\ell [\gamma]) & \equiv & (1+\delta R_\pi^\mathrm{phys})\,\Gamma^{(0)}(\pi^\pm \to \mu^\pm \nu_\ell)\nonumber\\
     &=&(1.0153 \pm  0.0019)\,\Gamma^{(0)}(\pi^\pm \to \mu^\pm \nu_\ell), \label{eq:pion_result_final}\\[2mm]
      \Gamma(K^\pm \to \mu^\pm \nu_\ell [\gamma])  & \equiv& (1+\delta R_K^\mathrm{phys})\,\Gamma^{(0)}(\pi^\pm \to \mu^\pm \nu_\ell)\nonumber\\
      &=&(1.0024 \pm 0.0010)\, \Gamma^{(0)}(K^\pm \to \mu^\pm \nu_\ell)\,,
      \label{eq:kaon_result_final}
 \eea
where $\Gamma^{(0)}$ is the leptonic decay rate at tree level in the GRS scheme (see Eqs.\,(\ref{eq:Rpi_phys}) and (\ref{eq:RK_phys})). These results can be compared with the ChPT predictions $\delta R_\pi^{\textrm{phys}} = 0.0176(21)$ and $\delta R_K^{\textrm{phys}} =  0.0064(24)$ obtained in Ref.\,\cite{Cirigliano:2011tm} and adopted by the PDG\, \cite{PDG,Rosner:2015wva}.
The difference is within one standard deviation for $\delta R_\pi^{\textrm{phys}}$, while it is larger for $\delta R_K^{\textrm{phys}}$. We also underline that our result $|V_{us}|=0.22538(46)$ in Eq.\,(\ref{eq:Vus}), together with the value of $V_{ud}$ determined in Ref.\,\cite{Hardy:2016vhg} and $|V_{ub}|$ from the PDG\,\cite{PDG}, implies that the unitarity of the first row of the CKM matrix is satisfied at the per-mille level (see Eq.\,(\ref{eq:unitarity})).

\section*{Acknowledgments}
We gratefully acknowledge the CPU time provided by the Partnership for Advanced Computing in Europe (PRACE) under the project Pra10-2693 {\em ``QED corrections to meson decay rates in Lattice QCD''} and by CINECA under the specific initiative INFN-LQCD123 on the BG/Q system Fermi at CINECA (Italy).
V.L., G.M.~and S.S.~thank MIUR (Italy) for partial support under the contract PRIN 2015. 
G.M.~also acknowledges partial support from the European Research Council Ideas Advanced Grant n. 267985 ``DaMeSyFla''.
C.T.S.~was partially supported by the UK Science and Technology Facilities Council (STFC) grant No. ST/P000711/1 and by an Emeritus Fellowship from the Leverhulme Trust.
N.T.~gratefully acknowledges the University of Rome Tor Vergata for the support granted to the project PLNUGAMMA.

\appendix

\section{Details of the simulation}
\label{sec:appA}

The gauge ensembles used in this work are those generated by ETMC with $N_f = 2 + 1 + 1$ dynamical quarks and used in Ref.\,\cite{Carrasco:2014cwa} to determine the up, down, strange and charm quark masses. 
We use the Iwasaki action \cite{Iwasaki:1985we} for the gluons and the Wilson Twisted Mass Action \cite{Frezzotti:2000nk, Frezzotti:2003xj, Frezzotti:2003ni} for the sea quarks. 
In the valence sector we adopt a non-unitary setup \cite{Frezzotti:2004wz} in which the strange quark is regularized as an Osterwalder-Seiler fermion \cite{Osterwalder:1977pc}, while the up and down quarks have the same action as the sea.
Working at maximal twist such a setup guarantees an automatic ${\cal{O}}(a)$-improvement \cite{Frezzotti:2003ni, Frezzotti:2004wz}.

We have performed simulations at three values of the inverse bare lattice coupling $\beta$ and at several different lattice volumes as shown in Table \ref{tab:simudetails}. We allow a separation of 20 trajectories between each of the $N_{\mathrm{cfg}}$ analysed configurations.
For the earlier investigation of finite-volume effects (FVEs) ETMC had produced three dedicated ensembles, A40.20, A40.24 and A40.32, which share the same quark masses and lattice spacing and differ only in the lattice size $L$.
To improve such an investigation, which is crucial in the present work,  we have generated a further gauge ensemble, A40.40, at a larger value of the lattice size $L$.

\begin{table}[htb!]
\begin{center}
\begin{tabular}{||c|c|c|c||c|c|c|c||c|c|c||}
\hline
ensemble & $\beta$ & $V / a^4$ &$N_{\mathrm{cfg}}$&$a\mu_{sea}=a\mu_{ud}$&$a\mu_\sigma$&$a\mu_\delta$& $a\mu_s$ & $M_\pi {\rm (MeV)}$ & $M_K {\rm (MeV)}$ & $M_\pi L$ \\
\hline \hline
$A40.40$ & $1.90$ & $40^{3}\times 80$ &$100$ &$0.0040$ &$0.15$ &$0.19$ & $0.02363$ & 317 (12) & 576 (22) & 5.7 \\
\cline{1-1} \cline{3-5} \cline{9-11}
$A30.32$ & & $32^{3}\times 64$ &$150$ &$0.0030$ & & & & 275 (10) & 568 (22) & 3.9 \\
$A40.32$ & & & $100$ & $0.0040$ & & & & 316 (12) & 578 (22) & 4.5 \\
$A50.32$ & & & $150$ & $0.0050$ & & & & 350 (13) & 586 (22) & 5.0 \\
\cline{1-1} \cline{3-5} \cline{9-11}
$A40.24$ & & $24^{3}\times 48 $ & $150$ & $0.0040$ & & & & 322 (13) & 582 (23) & 3.5 \\
$A60.24$ & & & $150$ & $0.0060$ & & & & 386 (15) & 599 (23) & 4.2 \\
$A80.24$ & & & $150$ & $0.0080$ & & & & 442 (17) & 618 (14) & 4.8 \\
$A100.24$ & & & $150$ & $0.0100$ & & & & 495 (19) & 639 (24) & 5.3 \\
\cline{1-1} \cline{3-5} \cline{9-11}
$A40.20$ & & $20^{3}\times 48 $ & $150$ & $0.0040$ & & & & 330 (13) & 586 (23) & 3.0 \\
\hline \hline
$B25.32$ & $1.95$ & $32^{3}\times 64$ & $150$ &$0.0025$&$0.135$ &$0.170$ & $0.02094$ & 259 ~(9) & 546 (19) & 3.4 \\
$B35.32$ & & & $150$ & $0.0035$ & & & & 302 (10) & 555 (19) & 4.0 \\
$B55.32$ & & & $150$ & $0.0055$ & & & & 375 (13) & 578 (20) & 5.0 \\
$B75.32$ & & & $~80$ & $0.0075$ & & & & 436 (15) & 599 (21) & 5.8 \\
\cline{1-1}\cline{3-5} \cline{9-11}
$B85.24$ & & $24^{3}\times 48 $ & $150$ & $0.0085$ & & & & 468 (16) & 613 (21) & 4.6 \\
\hline \hline
$D15.48$ & $2.10$ & $48^{3}\times 96$ & $100$ &$0.0015$&$0.1200$ &$0.1385 $& $0.01612$ & 223 ~(6) & 529 (14) & 3.4 \\ 
$D20.48$ & & & $100$ & $0.0020$ & & & & 256 ~(7) & 535 (14) & 3.9 \\
$D30.48$ & & & $100$ & $0.0030$ & & & & 312 ~(8) & 550 (14) & 4.7 \\
 \hline   
\end{tabular}
\end{center}
\vspace{-0.25cm}
\caption{\it \footnotesize Values of the valence and sea bare quark masses (in lattice units), of the pion and kaon masses for the $N_f = 2+1+1$ ETMC gauge ensembles used in Ref.\,\cite{Carrasco:2014cwa} and for the gauge ensemble, A40.40 added to improve the investigation of FVEs. 
A separation of $20$ trajectories between each of the $N_{\mathrm{cfg}}$ analysed configurations. The bare twisted masses $\mu_\sigma$ and $\mu_\delta$ describe the strange and charm sea doublet as in to Ref.\,\cite{Frezzotti:2003xj}. The values of the strange quark bare mass $a \mu_s$, given for each $\beta$, correspond to the physical strange quark mass $m_s^{\textrm{phys}}(\overline{\rm MS}, 2\,\mbox{{\rm GeV}}) = 99.6 (4.3)$ MeV and to the mass RCs determined in Ref.\,\cite{Carrasco:2014cwa}. The central values and errors of pion and kaon masses are evaluated using the bootstrap procedure of Ref.\,\cite{Carrasco:2014cwa}.\hspace*{\fill}}
\label{tab:simudetails}
\end{table}

At each lattice spacing, different values of the light sea-quark masses have been considered. 
The light valence and sea quark masses are always taken to be degenerate. 
The bare mass of the valence strange quark ($a\mu_s$) is obtained, at each $\beta$, using the physical strange mass and the mass RCs determined in Ref.\,\cite{Carrasco:2014cwa}. 
There the ``FLAG" hadronic scheme was adopted in which the pion and kaon masses in isosymmetric QCD are equal to $M_\pi^{(0),\textrm{FLAG}} = 134.98$ MeV and $M_K^{(0),\textrm{FLAG}} = 494.2$ MeV and the lattice scale is fixed by the value $f_\pi^{(0),\textrm{FLAG}} = 130.41~(20)$ MeV for the physical pion decay constant. 
In the charm sector instead, the $D_s$-meson mass $M_{D_s}^{(0)}$ was chosen to be equal to its experimental value $M_{D_s^+} = 1969.0(1.4)$\,MeV\,\cite{PDG}.
The values of the lattice spacing are found to be $a = 0.0885(36)$, $0.0815(30)$, $0.0619(18)$\,fm at $\beta = 1.90$, $1.95$ and $2.10$, respectively.

The two valence quarks $q_1$ and $q_2$ in the P meson are regularized with opposite values of the Wilson $r$-parameter ($r_2 = - r_1$) in order to guarantee that discretisation effects on the P meson mass are of order ${\cal{O}}(a^2 \mu ~ \Lambda_{\textrm{QCD}})$.
The lepton is a free twisted-mass fermion with Wilson parameter $r_\ell$ and its mass is taken fixed at the physical muon value $m_\ell = m_\mu = 105.66$\,MeV \,\cite{PDG}.
The regularization of the (massless) neutrino is irrelevant and it is taken to be a free fermion field.

In this work we made use of the bootstrap samples generated for the input parameters of the quark mass analysis of Ref.\,\cite{Carrasco:2014cwa}.
There, eight branches of the analysis were adopted differing in: 
\begin{itemize}
\item the continuum extrapolation adopting for the matching of the lattice scale either the Sommer parameter $r_0$ or the mass of a fictitious P-meson made up of two valence strange(charm)-like quarks; 
\item the chiral extrapolation performed with fitting functions chosen to be either a polynomial expansion or a Chiral Perturbation Theory (ChPT) Ansatz in the light-quark mass;
\item the choice between the methods M1 and M2, which differ by ${\cal{O}}(a^2)$ effects, used to determine the mass RC $Z_m = 1 / Z_P$ in the RI$^\prime$-MOM scheme . 
\end{itemize}

\section{Relating observables in the full theory and in QCD}
\label{sec:appB}

In this appendix we provide the detailed derivation of the relation between observables calculated in the full theory (QCD+QED) and in QCD (in the absence of QED). We start by a discussion of the separation of the QCD action from that in the full theory.

\subsection{Actions of the full theory and of QCD} \label{subsec:actions}
The lattice action in the full theory given in Eq.\,(\ref{eq:fullaction}) can be written as
\be
    \label{eq:Sfullsplit}
     S^\textrm{full} = S^\textrm{QCD} + \sum_\ell S_{\ell,0} + S^A + S^\textrm{ct} + \Delta S \, ,
\ee
where $S_{\ell, 0} = S^\textrm{kin}_{\ell, 0} + m_\ell S_\ell^m$ and the counterterm $S^\textrm{ct}$ and the QED vertices $\Delta S$ are given by 
\bea
    S^\textrm{ct} & = & \left\{ \frac1{g_s^2} - \frac{1}{g_0^2} \right\} \, S^\textrm{YM} + \sum_f \big\{(m_f^\textrm{cr} - m_0^\textrm{cr})
                                   S^\textrm{cr}_f + (m_f - m_{f, 0}) S_f^m \big\} \, , \\
    \Delta S & = & \sum_\ell \left( S^\textrm{kin}_\ell - S^\textrm{kin}_{\ell, 0} \right)+\sum_f\left( S_f^\textrm{kin} - S_{f, 0}^\textrm{kin} \right) \, .
\eea

We now consider these terms in detail using Wilson fermions for illustration. 
The kinetic term for the quark with flavour $f$, $S_f^\textrm{kin}$, is given by:
\be
    S_f^\textrm{kin} = \sum_x ~ \bar\psi_f(x)\left\{\gamma_\mu\frac{\nabla_\mu[UV_f] + \nabla_\mu^\ast[UV_f]}{2} - 
                                 \frac{\nabla_\mu[UV_f]\nabla_\mu^\ast[UV_f]}{2}\right\}\psi_f(x) \, ,
    \label{eq:Sfkin}
\ee
where $\psi_f$ is the quark field, while $U_\mu$ and $V_{f,\mu}$ are the QCD and QED gauge links, respectively. 
Specifically
\be
    V_{f,\mu}(x) = e^{-ie_f e A_\mu(x)} \, ,
\ee
where $e_f$ is the charge of the quark with flavour $f$ in units of the positron charge.
The forward and backward derivatives are given by 
\bea
\label{eq:forwardfull}
\nabla_\mu[UV_f] \psi_f(x) &=& U_\mu(x) V_{f,\mu}(x) \psi_f(x + \hat\mu) - \psi_f(x)\quad\textrm{and}\\
\label{eq:backwardfull}
    \nabla^\ast_\mu[UV_f] \psi_f(x) &=& \psi_f(x) - U^\dagger_\mu(x - \hat\mu) V_{f,\mu}^\dagger(x - \hat\mu) \psi_f(x - \hat\mu) \, .
\eea
The leptonic action is given by
\be
    S_\ell^\textrm{kin} + S^m_\ell = \sum_{x,ell} ~ \bar\psi_\ell(x) \left\{ \gamma_\mu \frac{\nabla_\mu[V_\ell] + \nabla_\mu^\ast[V_\ell]}{2} - 
                                                       \frac{\nabla_\mu[V_\ell]\nabla_\mu^\ast[V_\ell]}{2} + m_\ell \right\} \psi_\ell(x)
\ee
with $\psi_\ell$ being the lepton field.
The renormalisation of the lepton masses is performed perturbatively, by requiring that the on-shell masses correspond to the physical ones.

In QCD the kinetic term  only includes the gluon links so that for Wilson fermions
\be
    S_{f,0}^\textrm{kin} = \sum_x ~ \bar\psi_f(x) \left\{ \gamma_\mu\frac{\nabla_\mu[U] + \nabla_\mu^\ast[U]}{2} - 
                                       \frac{\nabla_\mu[U] \nabla_\mu^\ast[U]}{2} \right\} \psi_f(x) \, ,
    \label{eq:Sf0kin}
\ee
and the derivatives are defined in Eqs.\,(\ref{eq:forwardfull})\,-\,(\ref{eq:backwardfull}) with the $V_f = 1$. Since for leptonic and semileptonic decays, leptonic spinors are present even in the absence of electromagnetism, it is also convenient to define the kinetic action for free leptons:
\be
    S_{\ell,0}^\textrm{kin} = \sum_x ~ \bar\psi_\ell(x) \left\{ \gamma_\mu\frac{\nabla_\mu[1] + \nabla_\mu^\ast[1]}{2} - 
                                          \frac{\nabla_\mu[1]\nabla_\mu^\ast[1]}{2} \right\} \psi_\ell(x) \, .
    \label{eq:Sl0kin}
\ee

\subsection{Relation between observables in the full theory and in QCD}
Physical observables are determined from correlation functions evaluated from lattice computations in the full theory. For a generic observable ${\cal{O}}$ evaluated in the full theory up to ${\cal{O}}(\alphaem)$ we write:
\bea
    \langle O \rangle & = & \frac{\int_{U, A, \psi_f, \psi_\ell} e^{-S^\textrm{full}} O[\psi_f, \psi_\ell, U, A]}{\int_{U, A, \psi_f, \psi_\ell} e^{-S^\textrm{full}}} 
                                         \nonumber \\ 
                               & = & \frac{\int_{U, \psi_f} e^{-S^\textrm{QCD}} \int_{A, \psi_\ell} e^{-S^A - \sum_\ell S_{\ell, 0}} \, \left\{1 - S^\textrm{ct} - \Delta S +
                                         \frac{(\Delta S)^2}{2} \right\} O[\psi_f, \psi_\ell,U,A]} {\int_{U, \psi_f} e^{-S^\textrm{QCD}} \int_{A, \psi_\ell} 
                                         e^{-S^A - \sum_\ell S_{\ell, 0}} \, \left\{1 - S^\textrm{ct} - \Delta S + \frac{(\Delta S)^2}{2} \right\}} \, ,
   \label{eq:Of_full} 
\eea
where in the integrand $O$ is a multilocal composite operator.
For a given choice of the strong coupling $g_s$, the parameters of the action, the bare quark masses and the lattice spacing, are determined by imposing that a set of physical quantities take their experimental values as explained in Sec.\,\ref{subsec:renormalisationfull}. Physical quantities other than those used for the calibration can now be determined unambiguously up to lattice artefacts, which are removed by taking the continuum limit.

In general, the determination of physical observables requires the processing of correlation functions of the form of Eq.\,(\ref{eq:Of_full}). 
Hadronic masses, for example are obtained from the behaviour in the time separation of two interpolating operators and the determination of hadronic matrix elements may
require the cancelation of interpolating operators at the source and/or sink using a combination of three- and two-point functions. The discussion in this appendix concerns the evaluation of a generic correlation function.

We now turn to the definition of correlation functions in QCD defined in a generic scheme. For a generic observable ${\cal{O}}$ we define its value in QCD by:
\begin{equation}
\langle {\cal O}\rangle^\textrm{QCD}\equiv\frac{\int_{U, \psi_f} e^{-S^\textrm{QCD}} \int_{A, \psi_\ell} e^{-S^A - \sum_\ell S_{\ell, 0}} \, O[\psi_f, \psi_\ell, U,A]} {\int_{U, \psi_f} e^{-S^\textrm{QCD}} \int_{A, \psi_\ell} e^{-S^A - \sum_\ell S_{\ell, 0}}} \,,\label{eq:OQCDdef}
\end{equation}
where the bare quark masses and the lattice spacing are defined as discussed in Sec.\,\ref{sec:QCDQCD}. The free QED action is included in the numerator and denominator of 
Eq.\,(\ref{eq:OQCDdef}) since even without radiative corrections the physical quantities such as $\Gamma(K_{\ell 2})$ and $\Gamma(\pi_{\ell 2})$ studied in this paper are obtained by combining the results for hadronic matrix elements obtained from QCD simulations with leptonic spinors. Moreover, for other quantities, for example the long-distance contributions to the amplitude for the rare kaon decay $K^+\to\pi^+\nu\bar{\nu}$, there are internal free lepton propagators even in the absence of isospin breaking\,\cite{Christ:2016eae}.

Comparing Eqs.\,(\ref{eq:Of_full}) and (\ref{eq:OQCDdef}) we arrive at  
\be
    \langle O\rangle^\textrm{full} = \left\langle  O \right\rangle^\textrm{QCD} - \left\langle O \, S^\textrm{ct}
                                                      \right\rangle^\textrm{QCD}_\textrm{conn} - \left\langle O \, 
                                                     \left\{ \Delta S - \frac{(\Delta S)^2}{2}\right\}  \right\rangle^\textrm{QCD}_\textrm{conn} \equiv \left\langle  O \right\rangle^\textrm{QCD}+\left\langle \delta O\right\rangle^\textrm{QCD}\, ,
    \label{eq:Ofexpansion}
\ee
where the subscript ``conn'' reminds that only connected Feynman diagrams contribute: $\langle O_1 O_2\rangle_\textrm{conn} = \langle O_1 O_2\rangle - \langle O_1\rangle \, \langle O_2\rangle$.

There is one final subtlety which we must account for. We need to convert the results obtained from simulations in lattice units (i.e. in units of the lattice spacing) into values given in physical units such as MeV. Eq.\,(\ref{eq:Ofexpansion}) is also written in lattice units. Imagine that the observable $O$ has mass dimension $n$ and rewrite Eq.\,(\ref{eq:Ofexpansion}) with the lattice spacing included explicitly:
\begin{equation}
\langle a^n O\rangle^\textrm{full}=\left\langle  a_0^n O \right\rangle^\textrm{QCD}+\left\langle a_0^n\,\delta O\right\rangle^\textrm{QCD}
\end{equation}
where, since we are working to first order in isospin breaking, in the second term on the right-hand side we do not need to distinguish between the lattice spacing in the full theory ($a$) and that obtained in QCD  ($a_0$).
The quantity which we wish to determine, $\langle O\rangle^\textrm{full}$ in physical units, is therefore given by
\begin{equation}\label{eq:Ofinal}
\langle O\rangle^\textrm{full}=\frac{\left\langle  a_0^n O \right\rangle^\textrm{QCD}}{a_0^n}+\frac{\left\langle a_0^n\,\delta O\right\rangle^\textrm{QCD}}{a_0^n}-\frac{n\,\delta a}{a_0^{n+1}}\left\langle  a_0^n O \right\rangle^\textrm{QCD}\,,
\end{equation}
where $\delta a=a-a_0$. The three expectation values on the right-hand side of (\ref{eq:Ofinal}) are directly computed in QCD simulations.

\section{Non-perturbative renormalisation in the  \RIMOMprime~scheme}
\label{sec:appC}

In this paper, as explained in Sec.\,\ref{sec:Wreg}, we have renormalised the weak four-fermion operator $O_1$ non-perturbatively on the lattice to all orders in $\alpha_s$ and up to first order in $\alpha_\mathrm{em}$. In this appendix we describe the main steps of the non-perturbative renormalisation procedure at ${\cal O}(\alpha_\mathrm{em})$ and we refer the reader to a forthcoming publication\,\cite{dicarlo} for further details and results.

Given the amputated Green function, $\LO$, of an operator $O$ computed in a given gauge between external states with momentum $p$ and a suitable projector on the relevant Dirac structure, $P_O$, we define the projected Green function as 
\be
\GO(pa) = \tr \left[ \LO(pa) \, P_O \right] \, .
\ee 
In the \RIMOMprime~scheme, the renormalisation constant (RC) $Z_O(\mu a)$ is found by imposing the condition\,\cite{Martinelli:1994ty}
\be
Z_{\Gamma_O} (\mu a) \Gamma_O(pa) \vert_{p^2=\mu^2} = 1 ~, 
\label{eq:rimom}
\ee
where
\be
Z_{\Gamma_O}(\mu a)  = Z_O(\mu a)  \, \prod_f Z_f^{-1/2}(\mu a) ~.
\label{eq:ZGammaO}
\ee
The $Z_f$ are the RCs of the external fields and the index $f$ runs over all external fields entering the expression of the composite operator $O$. For the four-fermion operators considered in this work, the RCs $Z_O(\mu a)$ and the projected Green functions $\Gamma_O(pa)$ are $5\times 5$ matrices, the latter with elements $(\Gamma_O)_{ij}=\tr \left[ \Lambda_{O_i} \, P_{O_j} \right]$. In QCD+QED the RCs $Z_O$ and $Z_f$ depend both on the strong and the e.m.~coupling constants. 
 
Following the discussion of Sec.\,\ref{sec:Wreg} (see Eqs.\,(\ref{eq:ORI2})\,-\,(\ref{eq:ORI3})), we write the RCs of any composite operator, and in particular of the fields, bilinear and four-fermion operators, in the generic decomposition
\bea
 Z_O &=& Z_O^{\QED} \left[ (Z_O^{\QED})^{-1} Z_O (Z_O^{\QCD})^{-1} \right] Z_O^\QCD = 
    \left[1 +   \frac{\alphaem}{4\pi} \left( \Delta Z_O^{\QED} + \eta_O \right) \right] Z_O^\QCD  \nonumber \\
   &=&  \left(1 +   \frac{\alphaem}{4\pi} \Delta Z_O \right) Z_O^\QCD ~ ,
\label{eq:ZO}
\eea
where $Z_O^\QCD$ and $Z_O^{\QED}$ are the RCs of the operator $O$ in pure QCD and pure QED respectively and we have put 
\be
\Delta Z_O = \Delta Z_O^{\QED} + \eta_O~ .
\label{eq:DeltaZO}
\ee
The first term, $\Delta Z_O^{\QED}$, in Eq.\,(\ref{eq:DeltaZO}) represents the pure QED contribution to the RC at ${\cal O}(\alphaem)$, whereas $\eta_O$ contains the ${\cal O}(\alphaem)$ non-factorisable QCD+QED correction.

In terms of the QCD renormalised operators $O^\chi$, as those in Eq.\,(\ref{eq:Ochi}), we define the QCD renormalised projected Green function $\Gamma_O^\chi$, and expand it at first order in $\alphaem$
 \bea
 \Gamma_O^\chi(\mu a)  &=& Z_{\Gamma_O}^\QCD(\mu a)  \Gamma_O(pa) \vert_{p^2=\mu^2}  = Z_{\Gamma_O}^\QCD(\mu a) \left[\Gamma^\QCD_O(\mu a)   +  \frac{ \alphaem}{4\pi} \Delta \Gamma_O(\mu a)  \right] = \nonumber \\
&=& 1 + \frac{ \alphaem}{4\pi} \Delta \Gamma_O^\chi(\mu a) \, ,
    \label{eq:GammaOchi}
 \eea
where we have used the \RIMOMprime~renormalisation condition $Z_{\Gamma_O}^\QCD(\mu a) \Gamma^\QCD_O(\mu a)  =1$ applied in the pure QCD theory and defined
\be
\Delta \Gamma_O^\chi(\mu a) = Z_{\Gamma_O}^\QCD(\mu a)  \, \Delta \Gamma_O(\mu a) ~.
\ee

Using Eqs.\,(\ref{eq:ZO}) and (\ref{eq:GammaOchi}), we can rewrite Eq.\,(\ref{eq:rimom}) at first order in $\alphaem$ as
\be
1=Z_{\Gamma_O}(\mu a) \Gamma_O(\mu a) = 1 +  \frac{\alphaem}{4\pi}  \left(\Delta Z_{\Gamma_O}(\mu a) + 
\Delta \Gamma_O^\chi(\mu a)  \right)
\ee
which provides, in turn, the RI-MOM renormalisation condition at order $\alphaem$
\be
\Delta Z_{\Gamma_O}(\mu a) = - \Delta \Gamma_O^\chi(\mu a) \, .
\label{eq:masterRIg}
\ee
Using the expression of $Z_{\Gamma_O}$ in Eq.\,(\ref{eq:ZGammaO}) in terms of $Z_O$ and the external fields RCs one also obtains
\be
\Delta Z_O(\mu a)= - \Delta \Gamma_O^\chi(\mu a) + \frac 12 \sum_f \Delta Z_f(\mu a) \, .
\label{eq:masterRI}
\ee
Thus, $\Delta Z_O$ is expressed directly in terms of the ${\cal O}(\alphaem)$ contribution to the QCD renormalised projected Green function $\Delta \Gamma_O^\chi=  Z_{\Gamma_O}^\QCD \Delta \Gamma_O$ evaluated at $p^2=\mu^2$.

In the following we describe a completely non-perturbative determination of the RCs $\Delta Z_O(\mu a)$ to all orders in $\alpha_s$. We will assume that all the relevant RCs of fields and composite operators in pure QCD have been already determined, by following the standard \RIMOMprime~renormalisation procedure. With appropriate modifications to the kinematical conditions and projectors, the discussion can readily be adapted to similar schemes, such as the Symmetric Momentum subtraction one\,\cite{Sturm:2009kb}.

In addition to the renormalisation of the four-fermion operator appearing in the Hamiltonian, the e.m.~shift of the quark masses (see Sec.\,\ref{sec:noncrossed}) requires the knowledge of the RC of the pseudoscalar density\,\cite{deDivitiis:2013xla}. We therefore start by discussing the non-perturbative renormalisation of quark bilinear operators.

\subsection{Renormalisation of the quark field and bilinear operators}
\label{sec:bilinear_cond_with_em}

 
We start with the renormalisation of the quark fields. The e.m.~corrections to a quark propagator can be represented schematically in the form
\bea
    && \frac{ \alpha_\mathrm{em}}{4\pi} \, S^\QCD(p) \Delta S_q (p) S^\QCD(p)  = \nonumber \\ 
    && \golself + \golltad   - [m_f-m_f^0] \goi \mp [m^{cr}_f-m^{cr}_0] \goip \, , 
    \label{eq:prop_corrections}
\eea
where the last two diagrams represent the mass and critical Wilson parameter counter-terms\,\cite{deDivitiis:2013xla}. 

The amputated one-particle irreducible two-point function is then given by
\be 
    \label{eq:Delta_em_propagator}
    \Delta\Sigma_q (p)  = - \, \< S^\QCD(p) \>^{-1} \, \< S^\QCD(p) \Delta S_q (p) S^\QCD(p) \>\, \< S^\QCD(p)\>^{-1} \, 
\ee
and the correction to the quark field RC in the \RIMOMprime~scheme is obtained, according to Eq.\,(\ref{eq:masterRIg}), as
\be
    \Delta Z_q = - \frac{i}{12} \tr \left[ \frac{\pslash \, \Delta\Sigma^\chi_q(p)}{p^2}\right]_{p^2=\mu^2} =
    - \frac{i}{12} (Z_q^\QCD)^{-1} \, \tr \left[ \frac{\pslash \, \Delta\Sigma_q(p)}{p^2}\right]_{p^2=\mu^2} ~ .
\ee

The e.m.~correction to the RC $Z_O$ of a generic bilinear operator $O_\Gamma = \bar{q}_2 \, \Gamma \, q_1$, where $\Gamma$ is one of the Dirac matrices (${\Gamma=1,\gamma^5,\gamma^\mu,\gamma^\mu\gamma^5,\sigma^{\mu\nu}}$), is given by Eq.\,(\ref{eq:masterRI}), which in this case reads
\be
    \label{eq:Delta_ZO}
    \Delta \ZO  = - \Delta \Gamma_O^\chi + \frac{1}{2} \left(\Delta Z_{q_1} + \Delta Z_{q_2} \right) ~ .
\ee
Two kinds of corrections contribute to the amputated Green function: either the QCD Green function is amputated with the e.m.~corrections on the inverse propagators, or the correction to the Green function itself is amputated with QCD propagators. Thus we have
\be
\Delta \Gamma_O^\chi = (Z_{q_1}^\QCD)^{-1/2} (Z_{q_2}^\QCD)^{-1/2} Z_O^\QCD \, \tr \left[ \Delta\LO \, P_O \right] 
\ee
with
\bea
 \Delta \LO \, & = & \, \Delta\Sigma_{q_2}(p) \, G_O^\QCD(p) \, \gamma_5 \, \< {S^\QCD}^\dagger(p) \>^{-1} \, \gamma_5  +  \< S^\QCD(p) \>^{-1} \, G_O^\QCD(p) \, \gamma_5 \, \Delta\Sigma_{q_1}^\dagger(p) \, \gamma_5 \, +  \nn \\
 && + \  \< S^\QCD(p) \>^{-1} \, \Delta G_O(p) \, \gamma_5 \< {S^\QCD}^\dagger( p) \>^{-1} \, \gamma_5~, 
\eea
where $G_O$ is the non-amputated Green function and $\Delta G_O$ is given diagrammatically by
\be
    \label{eq:DeltaG_O}
    \Delta G_O (p) = \left\langle\ \gexc \  +\ \gin \ + \ \gout \ \right\rangle .
\ee

In this work we have used an improved method to compute the first diagram in Eq.\,(\ref{eq:DeltaG_O}), as well as all the diagrams containing a photon propagator connecting different points. In this method, some of the sequential propagators introduced in Ref.\,\cite{Giusti:2017dmp} are summed in order to reduce the number of inversions of the Dirac matrix. All details of the calculation will be given in the forthcoming publication \cite{dicarlo}.

Before closing this subsection, we stress that in the calculation of $Z_P$ and its e.m.~correction $\Delta Z_P$, the Goldstone pole contamination has been taken into account and subtracted. In pure QCD, at each $p^2$ and for each combination of valence quark masses, $\mu_1$ and $\mu_2$, the amputated Green function has been fitted to the ansatz
\be
    \label{eq:GB_QCD}
    \Gamma_P^\QCD  = A_0  + B_0  \, M^2_P + \frac{C_0}{M^2_P} ~ ,
\ee
where $M_P \equiv M_P(\mu_1,\mu_2)$ is the mass of the pseudoscalar meson composed of valence quarks of mass $\mu_1$ and $\mu_2$. When including QED in the calculation, Eq.\,(\ref{eq:GB_QCD}) has to be modified to take into account the e.m.~correction to the meson mass. By considering the ansatz in Eq.\,(\ref{eq:GB_QCD}) in QCD+QED and expanding it in terms of $\alpha_\mathrm{em}$ one finds
\be\label{eq:doublepole}
    \Delta \Gamma_P  = A_1 + B_1 \, M^2_P + \frac{C_1}{M^2_P} + B_0 \, \Delta M_P^2 - C_0 \, \frac{\Delta M_P^2}{M_P^4}\,,
\ee
where $\Delta M_P^2$ is the correction to $M_P^2$ evaluated in Ref.\,\cite{Giusti:2017dmp}. Note, in particular, that $\Delta \Gamma_P$ also receives the contribution of a double pole. In Eq.\,(\ref{eq:doublepole}) only the coefficients $A_1$, $B_1$, $C_1$ need to be fitted, since the values of $B_0$ and $C_0$ are already obtained from the QCD fit in Eq.\,(\ref{eq:GB_QCD}). 

\subsection{Renormalisation of the four-fermions operators}
\label{sec:4f_cond_with_em}

We conclude this section by describing the calculation of the RCs of the complete basis of four-fermion operators $O_i$ ($i = 1,\ldots,5$), in the \RIMOMprime~scheme. In this case, the renormalisation condition~(\ref{eq:masterRI}) for the renormalisation matrix at ${\cal{O}}(\alphaem)$ reads:
\be
    \label{eq:deltaZij}
    \Delta  Z_O = - \Delta \Gamma_O^\chi + \frac{1}{2} \left( \Delta Z_{q_1} + \Delta Z_{q_2} + 
                           \Delta Z_{\ell}\right) ~,
\ee
where $\Delta Z_{\ell}$ is only e.m.~and can be computed in perturbation theory. We remind the reader that this term is omitted in the actual calculation since its contribution cancels out in the difference $\Gamma_0(L) - \Gamma_0^{\textrm{pt}}(L)$. 

In Eq.\,\eqref{eq:deltaZij} $\Delta \Gamma_O^\chi$ is a matrix expressed by
\be
(\Delta \Gamma_O^\chi)_{ij} = (Z_{q_1}^\QCD)^{-1/2} (Z_{q_2}^\QCD)^{-1/2} \sum_{k=1,\ldots,5}(Z_O^\QCD)_{ik}  \, \tr \left[ \Delta\Lambda_{O_k} \, P_{O_j} \right] ~ .
\ee
As in the case of bilinear operators, the correction to the amputated Green function gets two kind of contributions,
\bea
    \label{eq:DeltaLambda}
    \Delta \Lambda_{O_i} \, &=& \, \Delta\Sigma_{q_2}(p) \, G_{O_i}^\QCD(p) \gamma_5 \, \< {S^\QCD}^\dagger(p) \>^{-1} \, \gamma_5 +  
    \< S^\QCD(p) \>^{-1} \, G_{O_i}^\QCD(p) \gamma_5 \, \Delta\Sigma_{q_1}^\dagger(p) \, \gamma_5 \, + \nn \\
    && \ +   \< S^\QCD(p) \>^{-1} \, \Delta G_{O_i}(p) \, \gamma_5 \< {S^\QCD}^\dagger(p) \>^{-1} \, \gamma_5 ~ ,
\eea
and in this case ${\Delta G}_{O_i}$ is given by
\bea
    \label{eq:DeltaG4f}
    \Delta G_{O_i} (p) & = & \left\< \quad \Gselfin + \Gselfout + \right. \\
                          && \ \ \: \left. + \Gexch + \Gexchin + \Gexchout \ \quad \right\> . \nonumber
\eea
The fermionic lines on the left-hand side of the diagrams in Eq.\,(\ref{eq:DeltaG4f}) represent the ingoing and outgoing light quarks. On the right-hand side, the external charged anti-lepton and the neutrino propagators are drawn for illustration but not actually included in the calculation. For this reason their amputation is neglected in Eq.\,(\ref{eq:DeltaLambda}). The lepton self-energy is not reported in Eq.\,\eqref{eq:DeltaG4f} since its contribution cancels out in the amputation.

\section{Matching, chirality mixing and fermion operators in the twisted mass regularisation}
\label{sec:appD}

In the main text and in Appendix~\ref{sec:appC} we have described the renormalisation of the relevant operators in the {\it physical basis}. This discussion is valid for a generic Wilson-like fermion regularisation. In this appendix we address instead some important aspects peculiar to the twisted mass fermions used in our numerical calculation. We derive, in particular, the relations between RCs in the so-called physical and twisted basis, for the bilinear and four-fermion operators considered in this work. 

The relevant observation is that the lattice action for twisted mass fermions at maximal twist in the twisted basis only differs from the standard Wilson fermion lattice action for the twisted rotation of the fermion mass term. The two actions become identical in the chiral limit. It then follows that, in any mass-independent renormalisation scheme, the RCs for twisted mass operators in the twisted basis are the same as those of the corresponding operators with standard Wilson fermions. It is customary to denote these RCs, for a generic operator $O$, as $Z_O$. They are valid for both standard Wilson and twisted mass operators in the twisted basis and differ, in general, from the RCs for twisted mass operators in the physical basis, that we denote here as $Z_O^{(0)}$. 

At maximal twist the rotation from the twisted to the physical basis for both quark and lepton fields is given by
\be 
    q_{\rm twisted} = \frac{1}{\sqrt{2}} \left( 1 + i \gamma_5 r_q \right) q ~ , \qquad 
    \ell_{\rm twisted} = \frac{1}{\sqrt{2}} \left(1 + i \gamma_5 r_\ell \right) \ell ~ ,
\ee
where $q$ and $\ell$ are the quark and lepton fields in the physical basis and $r_q$ and $r_\ell$ are the  corresponding $r$-parameters.
In our simulations we use opposite values of the $r$-parameter for the two valence quarks, ${r_2 = - r_1}$ ($r_i=\pm1$). The quark and lepton bilinears then transform as 
\bea
    \label{eq:twistandshout}
    \left[ \overline{q}_2 \gamma_\mu (1 \pm \gamma_5) q_1 \right]_{\rm twisted} & = & \pm i r_1 \left[ \overline{q}_2 \gamma_\mu 
          (1 \pm \gamma_5) q_1 \right] ~ , \nonumber \\ 
    \left[ \overline{q}_2 (1 \pm \gamma_5) q_1 \right]_{\rm twisted} & = & \left[ \overline{q}_2 (1 \pm \gamma_5) q_1 \right]~ , \nonumber \\ 
    \left[ \overline{q}_2 \sigma_{\mu\nu} (1 + \gamma_5) q_1 \right]_{\rm twisted} & = & \left[ \overline{q}_2 \sigma_{\mu\nu} (1 + \gamma_5) q_1 \right] \nonumber \\
    \left[ \overline{\nu} \gamma_\mu (1 - \gamma_5) \ell \right]_{\rm twisted} & = & \frac{1}{\sqrt{2}} \left( 1 - i r_\ell \right) 
        \left[ \overline{\nu} \gamma_\mu (1 - \gamma_5) \ell \right] ~ ,  \\ 
    \left[ \overline{\nu} (1 + \gamma_5) \ell \right]_{\rm twisted} & = & \frac{1}{\sqrt{2}} \left(1 + i r_\ell \right) \left[ \overline{\nu} 
        (1 + \gamma_5) \ell \right] \nonumber \\
    \left[ \overline{\nu} \sigma_{\mu\nu} (1 + \gamma_5) \ell \right]_{\rm twisted} & = & \frac{1}{\sqrt{2}} \left(1 + i r_\ell \right) 
        \left[ \overline{\nu} \sigma_{\mu\nu} (1 + \gamma_5) \ell \right]  ~ . \nonumber 
\eea

From Eqs.\,(\ref{eq:twistandshout}) one readily derives the relations between the quark vector and axial vector current in the two basis,
\bea
&& (V_\mu)_{\rm twisted} = \left[ \overline{q}_2 \gamma_\mu q_1 \right]_{\rm twisted}  = 
i \, r_1 \, \left[ \overline{q}_2 \gamma_\mu  \gamma_5 q_1 \right] =  i \, r_1 \, A_\mu  ~ , \nonumber \\
&& (A_\mu)_{\rm twisted} = \left[ \overline{q}_2 \gamma_\mu \gamma_5 q_1 \right]_{\rm twisted}  = 
i \, r_1 \, \left[ \overline{q}_2 \gamma_\mu  q_1 \right] =  i \, r_1 \, V_\mu  ~ , 
\eea
which, in turn, determine the relation between the RCs in the two basis
\bea
&& \hat V_\mu = Z_V^{(0)} \, V_\mu = - i \, r_1 \, (\hat A_\mu)_{\rm twisted} = - i \, r_1 \, Z_A \, (A_\mu)_{\rm twisted} = Z_A \, V_\mu ~ , 
\nonumber \\
&& \hat A_\mu = Z_A^{(0)} \, A_\mu  = - i \, r_1 \, (\hat V_\mu)_{\rm twisted} = - i \, r_1 \, Z_V \, (V_\mu)_{\rm twisted} = Z_V \, A_\mu ~,
\label{eq:ZVA}
\eea
where $\hat O$ denotes the generic renormalised operator. One then sees from Eq.\,(\ref{eq:ZVA}) that the RC $Z_V^{(0)}$ of the vector current in the physical basis, with $r_1 = -r_2$,  is simply the RC of the axial current in the twisted basis, which in turn is just $Z_A$ computed with  Wilson fermions in the chiral limit. Analogously, $Z_A^{(0)}$ in the physical basis, with $r_1 = -r_2$, corresponds to $Z_V$ computed with Wilson fermions in the chiral limit.

From the transformations (\ref{eq:twistandshout}) one can also derive the relations between the four-fermion operators $O_1-O_5$ of Eq.\,(\ref{eq:O1bare}) and (\ref{eq:ops}) in the physical and twisted basis
\begin{align}
& (O_1)_{\rm twisted} = -\frac{i}{\sqrt{2}} \, r_1 \left(1 - i r_\ell \right) O_1  ~ , &
&  O_1 = + \frac{i}{\sqrt{2}} \, r_1 \left(1 + i r_\ell \right) (O_1)_{\rm twisted}  ~ , 
\nonumber \\
& (O_2)_{\rm twisted} = + \frac{i}{\sqrt{2}} \, r_1 \left(1 - i r_\ell \right) O_2  ~ , &
&  O_2 = - \frac{i}{\sqrt{2}} \, r_1 \left(1 + i r_\ell \right) (O_2)_{\rm twisted}   ~ , 
\nonumber \\
& (O_3)_{\rm twisted} = \frac{1}{\sqrt{2}} \left(1 + i r_\ell \right) O_3  ~ , &
& O_3 = \frac{1}{\sqrt{2}} \left(1 - i r_\ell \right) (O_3)_{\rm twisted}    ~ , 
 \\
& (O_4)_{\rm twisted} = \frac{1}{\sqrt{2}} \left(1 + i r_\ell \right) O_4  ~ , &
&  O_4 = \frac{1}{\sqrt{2}} \left(1 - i r_\ell \right) (O_4)_{\rm twisted}  ~ , 
\nonumber \\
& (O_5)_{\rm twisted} = \frac{1}{\sqrt{2}} \left(1 + i r_\ell \right) O_5  ~ , &
&  O_5 = \frac{1}{\sqrt{2}} \left(1 - i r_\ell \right) (O_5)_{\rm twisted}  ~ . 
\nonumber
\end{align}
We can then obtain the relation between the renormalisation matrix in the physical basis, $Z^{(0)}$, and the corresponding matrix $Z$ for standard Wilson fermions. In particular, for the weak operator $O_1$ one finds
\bea
\hat O_1 &=&  \sum_{j = 1,\dots , 5} Z^{(0)}_{1j} \, O_j =  \frac{i}{\sqrt{2}} \, r_1 \left(1 + i r_\ell \right) (\hat O_1)_{\rm twisted} =
 \frac{i}{\sqrt{2}} \, r_1 \left(1 + i r_\ell \right) \sum_{j = 1,\dots , 5} Z_{1j} \, (O_j)_{\rm twisted} = \nonumber \\
&=& \frac{i}{\sqrt{2}} \, r_1 \left(1 + i r_\ell \right) \left[ -\frac{i}{\sqrt{2}} \, r_1 \left(1 - i r_\ell \right) \left( Z_{11} \, O_1 -  Z_{12} \, O_2\right)
+ \frac{1}{\sqrt{2}} \left(1 + i r_\ell \right) \sum_{j = 3,4, 5} Z_{1j} \, O_j  \right] = \nonumber \\
&=& Z_{11} \, O_1 -  Z_{12} \, O_2 - \overline{r} \sum_{j = 3,4, 5} Z_{1j} \, O_j ~,
\eea
with $\overline{r} \equiv r_1 r_\ell$. Therefore
\be
Z^{(0)}_{11} = Z_{11} \, , \quad
Z^{(0)}_{12} = -Z_{12} \, , \quad
Z^{(0)}_{13} =  - \overline{r} \, Z_{13} \, , \quad
Z^{(0)}_{14} =  - \overline{r} \, Z_{14} \, , \quad
Z^{(0)}_{15} =  - \overline{r} \, Z_{15} \, .
\label{eq:Zi1}
\ee
Eq.\,(\ref{eq:Zi1}) shows in particular that the mixing coefficients $Z_{13,14,15}$ for the operators $O_{3,4,5}$ are proportional to $\overline{r} \equiv r_1 r_\ell$. Thus, we can eliminate the mixing with these operators by simply averaging the numerical results over the two possible values $\overline{r}=\pm 1$.

In order to illustrate the above point, using the results of Ref.\,\cite{Carrasco:2015xwa} obtained in perturbation theory at order ${\cal{O}}(\alpha_s^0)$, the coefficients $\Delta Z^{QED}_{1j} = Z^{(0)}_{1j}/(\alphaem/4\pi) $ are explicitly given in the physical basis by
\be 
    \Delta Z^{\mathrm{QED}}_{12}  = - 0.5357 \, , \quad   
    \Delta Z^{\mathrm{QED}}_{13}  = - 1.6072 \, \overline{r} \, , \quad  
    \Delta Z^{\mathrm{QED}}_{14}  = 3.2143 \, \overline{r} \, , \quad 
    \Delta Z^{\mathrm{QED}}_{15}  = 0.8036 \, \overline{r} \, .
    \label{eq:Zi2}
\ee

\end{document}